\documentclass[useAMS,usenatbib,linenumbers]{aastex631}
\newcommand{\bz}{$\langle B_z \rangle$}

\newcommand{\vsini}{$v \sin i$}
\newcommand{\kms}{km\,s$^{-1}$}

\newcommand{\msun}{M$_\odot$}

\newcommand{\teff}{$T_{\rm eff}$}

\RequirePackage{color}

\definecolor{adu}{rgb}{0.0, 0.1, 0.7}

\shorttitle{UV Spectropolarimetry of Hot Star Magnetospheres}
\shortauthors{Shultz et al.}

\begin{document}
\nolinenumbers


\title{Ultraviolet Spectropolarimetry With Polstar: Hot Star Magnetospheres}

\author{M. E. Shultz}
\affiliation{Department of Physics and Astronomy, University of Delaware, 217 Sharp Lab, Newark, Delaware, 19716, USA}
\author{R. Casini}
\affiliation{High Altitude Observatory, National Center for Atmospheric Research, P.O. Box 3000, Boulder CO 80307-3000, USA}

\author{M. C. M. Cheung}
\affiliation{Lockheed Martin Solar and Astrophysics Laboratory, 3251 Hanover St, Palo Alto, CA 94304, USA}

\author{A. David-Uraz}
\affiliation{Department of Physics and Astronomy, Howard University, Washington, DC 20059, USA}
\affiliation{Center for Research and Exploration in Space Science and Technology, and X-ray Astrophysics Laboratory, NASA/GSFC, Greenbelt, MD 20771, USA}
\author{T. del Pino Alem\'an}
\affiliation{Instituto de Astrof\'isica de Canarias, E-38205 La Laguna, Tenerife, Spain}
\affiliation{Departamento de Astrof\'isica, Universidad de La Laguna, E-38206 La Laguna, Tenerife, Spain}

\author{C. Erba}
\affiliation{Department of Physics \& Astronomy, East Tennessee State University, Johnson City, TN 37614, USA}
\author{C.\ P.\ Folsom}
\affiliation{Tartu Observatory, University of Tartu, Observatooriumi 1, T\~{o}ravere, 61602, Estonia}

\author{K. Gayley}
\affiliation{Department of Physics \& Astronomy, University of Iowa, 203 Van Allen Hall, Iowa City, IA, 52242, USA}

\author{R.\ Ignace}
\affiliation{Department of Physics \& Astronomy, East Tennessee State University, Johnson City, TN 37614, USA}

\author{Z. Keszthelyi}
\affiliation{Anton Pannekoek Institute for Astronomy, University of Amsterdam, Science Park 904, 1098 XH, Amsterdam, The Netherlands}

\author{O. Kochukhov}
\affiliation{Department of Physics and Astronomy, Uppsala University, Box 516, 75120 Uppsala, Sweden}

\author{Y. Naz\'e}
\affiliation{GAPHE, Universit\'e de Li\`ege, All\'ee du 6 Ao\^ut 19c (B5C), B-4000 Sart Tilman, Li\`ege, Belgium}
\author{C. Neiner}
\affiliation{LESIA, Paris Observatory, PSL University, CNRS, Sorbonne Universit\'e, Univ. Paris Diderot, Sorbonne Paris Cit\'e, 5 place\\ Jules Janssen, 92195 Meudon, France}
\author{M. Oksala}
\affiliation{Department of Physics, California Lutheran University, 60 West Olsen Road 3700, Thousand Oaks, CA, 91360, USA}

\author{V. Petit}
\affiliation{Department of Physics and Astronomy, University of Delaware, 217 Sharp Lab, Newark, Delaware, 19716, USA}

\author{P. A. Scowen}
\affiliation{NASA Goddard Space Flight Center, 8800 Greenbelt Rd., Greenbelt, MD 20771}

\author{N. Sudnik}
\affiliation{Nicolaus Copernicus Astronomical Centre of the Polish Academy of Sciences, Bartycka 18, 00-716 Warsaw, Poland}

\author{A. ud-Doula}
\affiliation{Penn State Scranton, 120 Ridge View Drive, Dunmore, PA 18512, US}

\author{J. S. Vink}
\affiliation{Armagh Observatory and Planetarium, College Hill, BT61 9DG Armagh, Northern Ireland}

\author{G.\ A.\ Wade}
\affiliation{Department of Physics and Space Science, Royal Military College of Canada, PO Box 17000, Station Forces, Kingston, ON, K7K 7B4}









\begin{abstract}
Polstar is a proposed NASA MIDEX space telescope that will provide high-resolution, simultaneous full-Stokes spectropolarimetry in the far ultraviolet, together with low-resolution linear polarimetry in the near ultraviolet. In this white paper, we describe the unprecedented capabilities this observatory would offer in order to obtain unique information on the magnetic and plasma properties of the magnetospheres of hot stars. This would enable a test of the fundamental hypothesis that magnetospheres should 
act to rapidly drain angular momentum, thereby spinning the star down, whilst simultaneously reducing the net mass-loss rate. Both effects are expected to lead to dramatic differences in the evolution of magnetic vs. non-magnetic stars. 
\end{abstract}




\section{Introduction}\label{sec:intro}

\subsection{Background}\label{subsec:background}

The hot, massive stars of the upper main sequence are cosmic creation engines. While much less numerous than cooler stars, they exert a wide-ranging influence on galactic structure and stellar ecology. The majority of the periodic table -- all elements up to iron \citep{johnson_science} -- is forged in their cores, with heavier elements being synthesized in their deaths in Type II supernovae, at which point the enriched material is returned to the interstellar medium (ISM). With tens to hundreds of thousands of times the luminosity of the Sun, and much higher temperatures with spectra peaking in the far ultraviolet (FUV), they contribute most of the ISM's ionizing radiation. Their high luminosities launch powerful, radiatively driven winds with mass loss rates thousands to millions of times the Sun's to terminal velocities of thousands of km/s, injecting substantial matter and momentum into the ISM. 

The combination of ionizing radiation and powerful stellar winds hollows out star-forming material, quenching star formation. Upon supernova detonation at the end of a hot star's life, the resulting shock wave contributes a final burst of rapidly moving material that can trigger star formation by initiating the gravitational collapse of nearby molecular clouds. Following the supernova, the stellar remnant begins an extended afterlife as a neutron star or black hole, long-lived objects that -- in the deep cosmic future -- will be the sole inhabitants of the cosmos. 

The structure and evolution of single stars are largely functions of the stellar mass, therefore any process that changes the mass will change the evolution of the star. For O-type stars, which have mass-loss rates of around 10$^{-6}~{\rm M_\odot~yr^{-1}}$, main-sequence lifetimes on the order of 10 Myr, and initial masses of around $50~{\rm M_\odot}$, mass loss via stellar winds can result in considerable reductions in mass. The evolution of such stars therefore cannot be understood in isolation from the effects of their wind, since differences in the terminal, pre-supernova state of a star have consequences for the type of supernova as well as the type of remnant that emerges. 

Many massive stars are rapid rotators, with equatorial surface rotational velocities of hundreds of km/s. In the most extreme cases, stars may rotate near their critical or breakup velocities, leading to deformation of their shape from a spherical to an oblate form, bulged around the equator, with much cooler equatorial temperatures due to gravity darkening \citep{1924MNRAS..84..665V}. Even in less extreme cases, rotation leads to mixing, replenishing the nuclear-burning core with fresh material from the envelope, and thereby having a strong effect on the evolution of the star \citep[e.g.][]{2011A&A...530A.115B}. 

Due to the dominant role played by massive stars in terms of mass and energy input via winds, ionizing radiation, and supernovae, understanding the evolution of galaxies requires understanding the evolution of massive stars, which in turn requires that we understand their winds and their rotation along with all phenomena that can modify these key parameters. 
It should also be emphasized that there is no such thing in nature as a really `normal' or `standard' massive star. Instead, there is a diversity of special cases, such as e.g. classical Be stars or interacting binaries, which collectively comprise the hot star population. It is only via understanding of the properties of these individual groups that the properties of the massive stars can be properly accounted for in population synthesis models, which in turn are key inputs for models of galactic chemical and structural evolution. 

Polstar is a proposed NASA MIDEX mission that will utilize high-resolution spectropolarimetry in the far ultraviolet as a sensitive probe of the circumstellar environments of hot stars \citep{2021SPIE11819E..08S}. In this white paper we describe the broad utility of Polstar for the study of magnetic hot stars and their magnetospheres. In the remainder of this section, general background is provided on hot star magnetic fields, magnetospheres, magnetospheric diagnostics, and the effects of magnetospheres on stellar evolution, culminating with the motivation for exploring these effects with Polstar. Experimental design is described in Sect. \ref{sec:experiment}. Sect. \ref{sec:lsd_magnetometry} compares the expected results of photospheric magnetometry in the ultraviolet and the visible regions of the spectrum. The known properties of ultraviolet magnetospheric diagnostics, a comparison with visible and X-ray diagnostics, and an overview of the simulations and models used to interpret them, are given in Sect. \ref{sec:uv_spectroscopy}. Sect.\ \ref{sec:circumstellar_magnetometry} presents predictions for the amplitude and morphology of circularly polarized Zeeman signatures originating in the magnetosphere. Expectations for linear spectropolarimetry arising from scattering in the circumstellar environment, drawing on both models and observations acquired with high-resolution visible instrumentation, are described in Sect. \ref{sec:lin_specpol}. Linearly polarized broadband magnetospheric signatures, again drawing on both models and observations, are given in Sect. \ref{sec:continuum_linpol}. Measurement of weak magnetic fields in the circumstellar environment via the Hanle effect is described in Sect. \ref{sec:hanle}. The properties of the primary sample, and the expected results of a Polstar observing campaign, are outlined in Sect. \ref{sec:sample}. Enabled science falling under the purview of magnetic and magnetospheric activity in hot stars is described in Sect.\ref{sec:enabled}. Synergies with other Polstar scientific objectives related to massive stars, the interstellar medium, and protoplanetary disks are outlined in Sect. \ref{sec:other_objectives}. The white paper is summarized in Sect. \ref{sec:summary}.

\subsection{Massive star magnetism}\label{subsec:magnetism}

Magnetic fields are a crucial factor that leads to drastic modifications in both mass-loss rates and rotation of hot stars. 
They are found in approximately 10\% of the OBA population \citep{grunhut2017,2017A&A...599A..66S,2019MNRAS.483.3127S}. The magnetic fields of stars with radiative envelopes share similar properties from approximately spectral type A5 to the top of the main sequence. They are strong \citep[ranging from hundreds of G to tens of kG;][]{2019MNRAS.490..274S}, they are stable over timescales of at least decades \citep{2018MNRAS.475.5144S}, and they are globally organized and, with few exceptions, geometrically simple \citep[being well-described by tilted dipoles with most of the magnetic energy in low-order poloidal field components;][]{2019A&A...621A..47K}. Magnetohydrodynamic simulations have demonstrated that such fields, once established, can remain stable over evolutionary timescales \citep[e.g.][]{2004Natur.431..819B}, leading to the hypothesis that they are of `fossil' origin i.e.\ remnants of some previous event in the star's life \citep{2009ARA&A..47..333D}.
Since there is no correlation of magnetic field strengths with rotational or stellar properties, as expected and observed for the dynamo fields of cool stars \citep[e.g.][]{2016MNRAS.457..580F,2017NatAs...1E.184S}, and as there is furthermore no known dynamo mechanism capable of sustaining globally organized magnetic fields in the radiative atmospheres of hot stars, it is generally believed that these magnetic fields are fossils arising in some earlier phase of the star's life \citep{2009ARA&A..47..333D}, e.g.\ due to amplification of ISM seed fields during formation and/or convective phases during pre-main sequence evolution \citep[e.g.][]{2019A&A...622A..72V}, or in dynamos powered by binary mergers \citep[e.g.][]{
2019Natur.574..211S}. An improved understanding of the formation mechanism of fossil fields therefore offers the promise of improving our general understanding of the formation or evolution of hot stars.

While the 10\% incidence rate of strong magnetic fields in massive stars may seem like a minor issue, the detection of `ultra-weak' ($\sim$0.1-10 G) magnetic fields in a number of bright A-type stars suggests that magnetic phenomena may be ubiquitous on the upper main sequence \citep{2020MNRAS.492.5794B}. Weak magnetic fields have also been detected in numerous blue supergiants, indicating that they may play a role in post-main sequence evolution as well \citep{2017MNRAS.471.1926N,2018MNRAS.475.1521M,2021mobs.confE..47O}. These weak fields are expected to arise from small-scale dynamos powered by embedded helium and iron opacity-bump convection zones \citep{2020ApJ...900..113J}, which are revealed as the stellar wind strips away the outer layers of the star \citep{2021arXiv211003695J}. 

\subsection{Massive star magnetospheres}\label{subsec:magnetospheres}

Strong magnetic fields are easily capable of trapping the ionized wind of a massive star \citep{udDoula2002}. The surface mass flux from opposite magnetic colatitudes is channeled by the magnetic field and collides in the magnetic equatorial plane, leading to accumulation of a torus of high-density plasma surrounding the magnetic equator, with X-rays being produced in the cooling shocks \citep{2014MNRAS.441.3600U}. 

Magnetic confinement extends out to the Alfv\'en surface, the distance from the star at which the magnetic energy density and the kinetic energy density of the wind equalize, with the Alfv\'en radius $R_{\rm A}$ defined in the magnetic equatorial plane. Alfv\'en radii commonly range from a few stellar radii (as is typical for an O-type star, where the powerful wind rapidly overpowers the magnetic field) to tens of stellar radii (as is typical for a strong magnetic field trapping the much weaker wind of a B-type star). 

Beyond the Alfv\'en surface, the wind is magnetically unconfined and escapes from the star. Within the Alfv\'en surface, plasma is forced into corotation with the stellar magnetic field by the Lorentz force. If a star is slowly rotating, gravity pulls the confined plasma back to the star. This can dramatically reduce the net mass-loss rate of the star \citep{udDoula2002,2017MNRAS.466.1052P}. If a star is rapidly rotating, such that the Kepler corotation radius $R_{\rm K}$ (at which gravity is balanced by the centrifugal force imparted by corotation) is inside the Alfv\'en radius, gravitational infall is prevented in the part of the magnetosphere above $R_{\rm K}$. The former case is referred to as a {\em Dynamical Magntosphere} (DM) since mass-balancing occurs on dynamical timescales; the latter case is referred to as a {\em Centrifugal Magnetosphere} \citep[CM;][]{petit2013}. Importantly, all magnetic stars have DMs in at least the inner part of the magnetosphere\footnote{Unless the star is critically rotating, in which case $R_{\rm K}$ becomes identical to the equatorial radius; while theoretically possible, no such object has been found, and in any case such a phase would be extremely short-lived due to strong magnetic braking.}.

Plasma inside a CM is unable to fall back to the star, and magnetohydrodynamic simulations indicate that it will instead build up until it reaches a critical density beyond which the magnetic field is unable to confine it, at which point it is explosively ejected outwards -- a phenomenon referred to as centrifugal breakout \citep{udDoula2008}. This mass-balancing mechanism has been verified via analysis of the H$\alpha$ emission properties of stars with CMs \citep{2020MNRAS.499.5379S,2020MNRAS.499.5366O}. Thus, plasma accumulated in the DM is trapped and returned to the star, whereas plasma trapped in the CM will eventually be lost; only the DM reduces the net mass-loss rate. 

Poynting stresses in the magnetosphere lead to rapid angular momentum loss \citep{2009MNRAS.392.1022U}, as a result of which magnetic hot stars are systematically more slowly rotating than non-magnetic stars of comparable spectral type and luminosity class \citep{2018MNRAS.475.5144S}. In some cases, the rotational period change can be directly measured \citep{2010ApJ...714L.318T}. Rotational spindown can be so extreme that rotational periods of up to decades in length have been identified \citep[e.g.][]{2010A&A...520A..59N,2017MNRAS.468.3985S,2017MNRAS.471.2286S,2021MNRAS.506.2296E}. An important consequence of this is that the CM quickly shrinks as a star ages, and is detectable only during the initial phase of the main sequence \citep{2019MNRAS.490..274S}.

\subsection{Multiwavelength magnetospheric diagnostics}\label{subsec:diagnostics}

In addition to their important consequences for stellar evolution, magnetospheres have a number of observational consequences. There are available diagnostics across the electromagnetic spectrum, each of which probes a different magnetospheric component, and is detectable in a different part of stellar and magnetospheric parameter space. 

X-rays are emitted due to magnetically confined wind shocks, making magnetic stars much more luminous in X-rays than non-magnetic stars \citep[e.g.][]{2014ApJS..215...10N,2014MNRAS.441.3600U}. While X-rays are detectable for most magnetic OB stars, time series data are difficult to acquire. 
Velocity-resolved information from line emission is furthermore difficult to obtain for all but the brightest X-ray sources. However, in 2031 the launch of the {\em Athena} mission will in fact provide velocity-resolved X-ray information \citep{2013arXiv1306.2307N}, which may provide a powerful complement to the data gathered by Polstar (although the science case presented here is independent of any results from Athena). 

Rapidly rotating magnetic AB stars\footnote{This diagnostic is unavailable for O-type stars, since the large radio photospheres produced by their dense winds swallow any gyrosynchrotron radiation produced by their magnetospheres \citep[e.g.][]{2015MNRAS.452.1245C}.} can exhibit radio gyrosynchrotron emission originating at high magnetic latitudes \citep[e.g.][]{2021MNRAS.507.1979L}, beamed auroral radio emission via the electron cyclotron mechanism \citep[e.g.][]{2021arXiv210904043D}, and line emission in near infrared and visible Bracket, Paschen, and Balmer series H lines, most prominently in H$\alpha$ \citep[e.g.][]{grun2012,2015A&A...578A.112O}. 
Since the weak winds of B-type stars are unable to fill their DMs to sufficient density to become optically thick in H$\alpha$ \citep{petit2013}, their magnetospheres are detectable in visible data only around young, rapidly rotating, strongly magnetic stars with large CMs, which due to rapid magnetic braking disappear after less than 1/3 of the stars' main sequence lifetime \citep{2019MNRAS.490..274S}. 

By contrast, ultraviolet signatures associated with magnetospheres have been detected across the Hertzsprung-Russell diagram, 
making FUV resonance lines the most versatile and ubiquitous diagnostic by far. These lines probe both the free, unconfined stellar wind, and the dynamical magnetosphere. In other words, UV diagnostics provide information on that part of the wind which is escaping, and that part which is returned, to the star. UV resonance line variability is almost universally detectable in magnetic OB stars. It therefore provides the only diagnostic that a) can be used for magnetic stars across the full range of magnetic, rotational, and stellar parameters, b) provides simultaneous information on the wind and the magnetosphere, and c) provides velocity-resolved information with which the detailed magnetospheric structure can be examined. 

The application of ultraviolet spectroscopy to the study of hot star magnetospheres is conceptually similar to its utility in the study of the solar magnetosphere. One recent example is the IRIS mission, which has been doing high resolution FUV spectroscopy of solar plasmas since 2013. The FUV band of IRIS regularly observes the same Si~{\sc IV} doublet that is used to probe the winds of hot stars. In the solar case, Si~{\sc iv} line profiles are used to probe Doppler flows of plasmas at $T \sim 80$~kK \citep[e.g.][]{2015ApJ...801...83C}.

\subsection{Magnetic massive star evolution}\label{subsec:evolution}

   \begin{figure}[t]
   \centering
   \includegraphics[width=0.6\textwidth, clip]{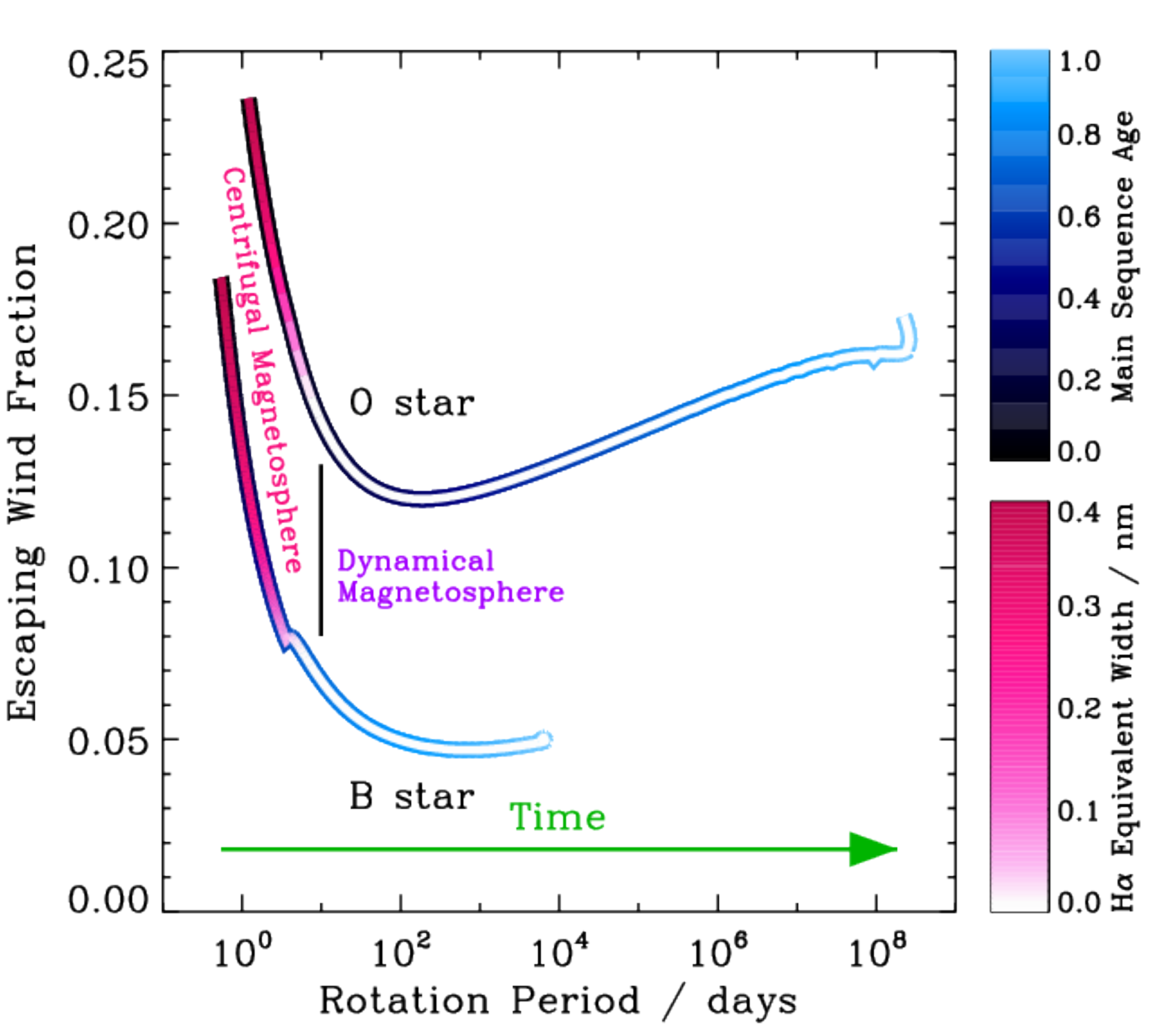}
      \caption{MESA evolutionary models \citep{2020MNRAS.493..518K} showing the change in the escaping wind fraction as a function of the rotation period for an O- and B-type star, each with an initial magnetic field strength of 6 kG and an initial critical rotation fraction of 0.5. The H$\alpha$ equivalent width is calculated following an empirically guided centrifugal breakout formalism \citep{2020MNRAS.499.5366O,2020MNRAS.499.5379S}; note that centrifugal magnetospheres are only detectable via H$\alpha$ for the first third of the main sequence; in contrast, dynamical magnetospheres are detectable in the ultraviolet at all ages. The escaping wind fraction initially decreases as the centrifugal magnetosphere shrinks and centrifugal breakout declines in importance, and then begins to increase as the star's increasing radius and luminosity drive an increased mass-loss rate, decreasing surface magnetic field strength \citep[e.g.][]{2019MNRAS.490..274S}, and therefore the total size of the magnetosphere decreases.}
         \label{fb_prot}
   \end{figure}

Stellar structure and evolution are both profoundly influenced by mass-loss and rotation. Mass-loss during the main sequence influences the pre-supernova mass of the star, as well as the mass of the remnant. Since material trapped in a dynamical magnetosphere is returned to the star via graviational infall, magnetic fields have the effect of reducing the net mass-loss rate \citep{udDoula2002}. Building on this phenomenon, \cite{2017MNRAS.466.1052P} demonstrated that magnetospheric mass-loss quenching can reduce mass-loss rates by amounts comparable to a reduction of metallicity to that prevailing in the early, unenriched universe \citep[since radiative winds are accelerated via line driving, mass-loss rates are a strong function of metallicity, e.g.][]{vink2001}. Thus, magnetic stars are potential progenitors to the formation of the heavy stellar-mass black holes such as those found by gravitational wave observations \citep{2016PhRvL.116f1102A,2017MNRAS.466.1052P}, i.e.\ the formation of such objects is not limited to the early universe or to low-metallicity environments such as the Magellanic Clouds.  

Rapid rotation can lead to mixing in the convective core, replenishing the material available for nuclear fusion and thereby extending the main sequence lifetime; the rapid spindown induced by magnetic fields can therefore be expected to hasten main sequence evolution. Evolutionary models incorporating rotational spindown and mass-loss reduction have successfully reproduced the qualitative evolution of magnetic stars \citep[e.g.][]{2019MNRAS.485.5843K,2020MNRAS.493..518K,2021A&A...646A..19T,2021arXiv210813734S}, despite systematic uncertainty regarding factors such as the internal rotational profile. These models firmly predict that the evolutionary tracks of magnetic stars should differ considerably from those of stars without magnetic fields.

Combining everything together, the expected evolutionary scenario for magnetic hot stars -- illustrated in Fig.\ \ref{fb_prot} -- is as follows: 1) young, rapid rotators with strong magnetic fields lose the majority of the trapped plasma from their CMs via centrifugal breakout \citep{udDoula2008}; 2) angular momentum loss quickly shrinks the CM, decreasing the net mass-loss rate as the DM grows from the inside out \citep{2009MNRAS.392.1022U}, until; 3) the CM disappears, the DM slams shut on the wind, and the star evolves as an essentially non-rotating object with a nearly constant mass. As can be seen in Fig.\ \ref{fb_prot}, centrifugal magnetospheres disappear early in a magnetic star's main sequence evolution, and become undetectable in H$\alpha$ at an even earlier phase \citep[e.g.][]{2019MNRAS.490..274S,2020MNRAS.499.5379S}. In contrast, dynamical magnetospheres are detectable in the ultraviolet throughout a star's evolution.

\subsection{Polstar and motivation for this study}\label{subsec:motivation}

   \begin{figure}[ht]
   \centering
   \includegraphics[width=0.95\textwidth]{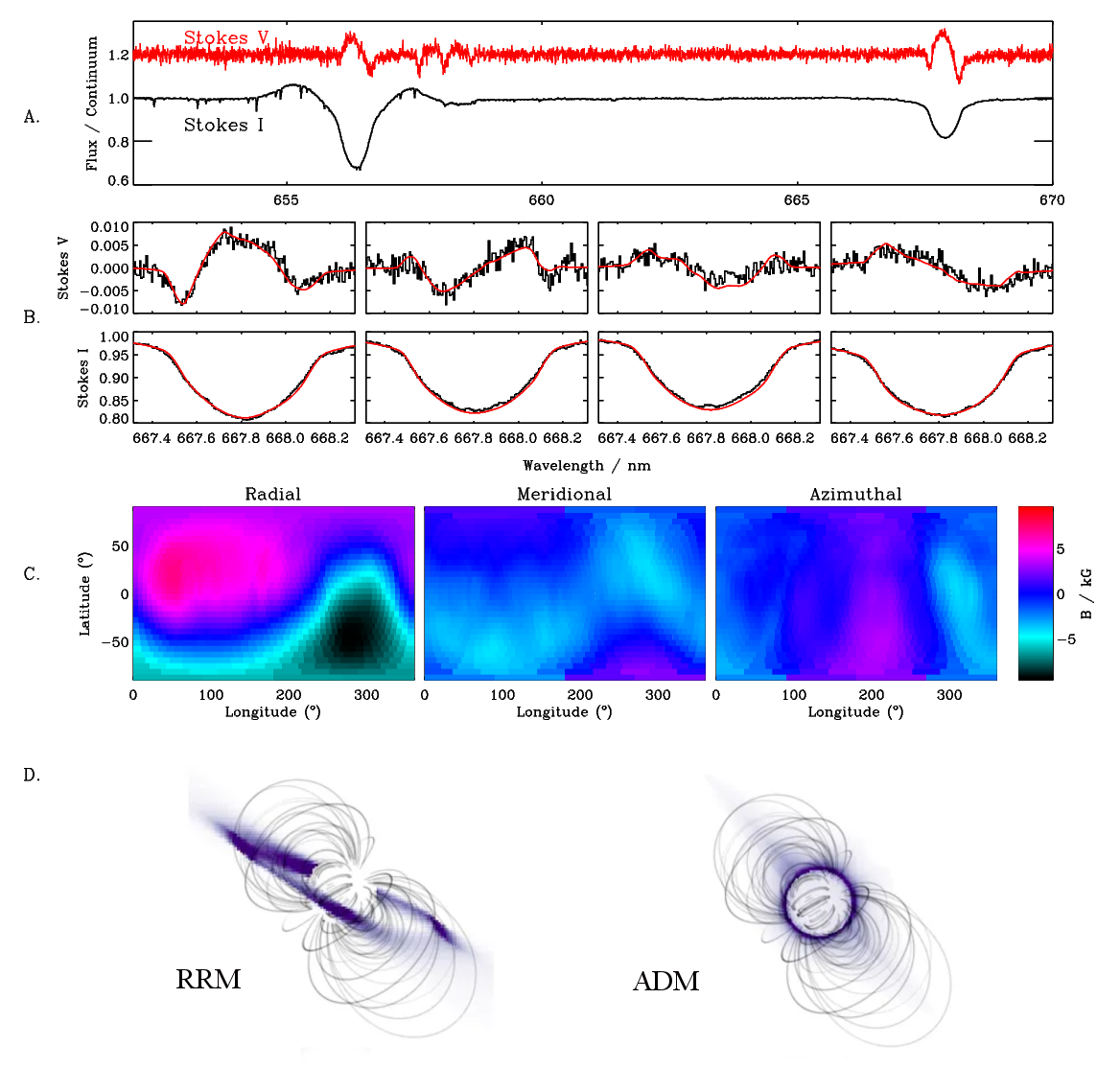}
      \caption{Illustration of the data analysis flow. (A.) Polarized spectra of the target are acquired. (B.) Information from multiple rotational phases is combined in order to obtain (C.) a model of the surface magnetic field. Magnetospheric signatures (e.g. the H$\alpha$ line at 656 nm in (A.)) are then compared to predictions from (D.) Rigidly Rotating Magnetosphere (RRM) or Analytic Dynamical Magnetosphere (ADM) models obtained via extrapolation of the surface field into the circumstellar environment. The data and models shown in this figure are adapted from those presented by \cite{2015MNRAS.451.2015O} for $\sigma$ Ori E (with the exception of the ADM model, which is based on the same field structure).}
         \label{sigorie_pol_to_model}
   \end{figure}

Polstar is a proposed NASA MIDEX space mission equipped with a 60-cm telescope and a full-Stokes (IQUV) spectropolarimeter divided in 2 channels in the ultraviolet \citep{2021SPIE11819E..08S}. The first channel provides spectropolarimetry at high spectral resolution of R$\sim$33000 over the 122-200 nm far-UV bandpass. The second channel provides spectropolarimetry over the 180-320 nm NUV band with low- to mid-resolution (R$\sim$30 to 250). These wavelength ranges include particularly interesting resonance lines sensitive to the winds of hot stars, such as N~\textsc{v}~123.9, 124.3 nm, Si~\textsc{iv}~139.4, 140.3 nm, and C~\textsc{iv}~154.8, 155.1 nm, as well as a large quantity of photospheric lines. Therefore Polstar is very well suited to study hot stars and their circumstellar environments. 

High-resolution UV spectroscopy is relatively sparsely available for magnetic massive stars. A handful of objects have extensive time series, predominantly acquired with the International Ultraviolet Explorer (IUE) space telescope, with which the rotational modulation of their resonance lines can be examined; for other stars, only snapshot observations with IUE or the Hubble STIS or COS instruments are available, as the high time pressure on the HST makes it impractical to obtain high-cadence time series. 

Even for stars with existing high-resolution UV spectroscopy, Channel 1 Polstar spectroscopy would provide several important advantages over existing data. First, the spectral resolution is higher than either IUE (about 8000) or HST/COS (about 15,000). Second, Polstar will be able to obtain a significantly higher signal-to-noise $S/N$: whereas a typical IUE spectrum has a $S/N \sim 10$, Polstar spectroscopy will easily reach values on the order of 100, and for some targets on the order of 1000. This will enable stellar rotation to be resolved in spectral lines, and will furthermore enable the detection of subtle features associated with magnetospheric activity. 

While UV spectroscopy is available for some stars, UV spectropolarimetry and polarimetry is not. These capabilities will enable Polstar data to detect and measure circumstellar magnetic fields, both in Stokes $V$ via the Zeeman effect in wind-sensitive UV resonance lines, and via the Hanle effect in Stokes $QU$ (which only works in the UV). Linear spectropolarimety and broadband polarimetry will further offer unique information on the circumstellar geometry. 



In this white paper, we describe how Polstar can be used to test the fundamental hypothesis that magnetic fields lead to angular momentum loss in the early part of a magnetic star's evolution, while trapping material and dramatically reducing stellar mass-loss rates throughout the entirety of its main sequence lifetime. While the primary focus of this white paper pertains to the utility of Polstar to conduct such an experiment, the considerations developed here are of relevance to other proposed UV spectropolarimeters such as Arago \citep{2019arXiv190801545M}, Pollux on LUVOIR \citep{2018SPIE10699E..3BB}, or any other similar mission that may be launched in the future. 






\section{Experimental design}\label{sec:experiment}

\begin{table}
\label{exptable}
\begin{center}
\caption{Summary of experimental design. From left to right, the columns give: the physical property of the target star being measured; the observational correlate of that property; the Polstar capability utilized; the requirements for the measurement to detect the feature of interest; the post-processing method used to reach the necessary signal-to-noise ratio; and the Section in the current white paper in which the measurements and methodologies are described in detail.} 
\begin{tabular}{llllll}
\hline\hline 
Physical feature & Measurement & Mission Capability & Requirements & Method & Section \\
\hline
Surface magnetic field & Zeeman effect      & Channel 1   & $R \sim 30,000$ & LSD & \ref{sec:lsd_magnetometry} \\
                       & in photospheric lines & Stokes IQUV & pol. precision $\sim 10^{-4}$ & & \\
\hline
Circumstellar magnetic field & 1) Zeeman effect & 1) Channel 1 &  1) $R \sim 200-1500$ & 1,2) Wavelength & 1) \ref{sec:circumstellar_magnetometry} \\
1) $\gtrsim 100$~G                             & in resonance lines & Stokes IV & pol. precision $\sim 10^{-4}$ & binning, co-addition & \\
2) $\lesssim 100$~G                             & 2) Hanle effect & 2) Channel 1 & 2) $R \sim 200-1500$ & of spectra & 2) \ref{sec:hanle} \\
                             & in resonance lines & Stokes IQU &  pol. precision $\sim 10^{-4}$ & & \\
\hline
Magnetospheric velocity, & velocity-resolved flux & Channel 1 & $R \sim 30,000$ & N/A & \ref{sec:uv_spectroscopy} \\
column density           & in resonance lines                 & Stokes I  & $S/N \sim 100$ & & \\
\hline
Magnetospheric geometry & 1) scattering in & 1) Channel 1 & 1) $R \sim 200-1500$, & 1,2) Wavelength & 1) \ref{sec:lin_specpol} \\
                        & resonance lines  & Stokes IQU   & pol. precision $\sim 10.^{-3}$  & binning, co-addition & \\
                        & 2) scattering in & 2) Channel 2 & 2) pol. precision & of spectra & 2) \ref{sec:continuum_linpol} \\
                        & continuum        & Stokes QU    & $\sim 10^{-5}$ & & \\
\hline 
\hline
\end{tabular}
\end{center}
\end{table}

The goal of Polstar observation of magnetic massive stars is to test the theoretical prediction that magnetic confinement in the early, rapidly rotating evolutionary phase leads to rapid angular momentum loss accompanied by mass escape from the centrifugal magnetosphere via breakout, switching to a mass-trapping phase when the CM disappears and the magnetosphere locks down on the stellar wind, as illustrated in Fig. \ref{fb_prot} \citep[see also e.g.][]{udDoula2008,2009MNRAS.392.1022U}. This requires observation of magnetic hot stars across the full range of stellar parameters, evolutionary phases, rotational periods, and surface magnetic field strengths and geometries. Moreover, since magnetic wind confinement leads to rotational modulation of all signatures associated with the surface magnetic field, observations must be acquired sampling the full rotational phase curve -- indeed, doing so enables magnetic and magnetospheric models to be inferred. Crucially, ultraviolet polarimetry will also enable the different components of the magnetosphere -- the outflowing wind, and the trapped downflow -- to be separately identified. 

Uniquely amongst Polstar science objectives, the observation of magnetic stars will utilize the full range of the observatory's capabilities: high-resolution spectroscopy, circular spectropolarimetry, linear spectropolarimetry, and broadband linear polarization. This comprehensive usage is summarized in Table \ref{exptable}. Each will provide key constraints that can be combined to obtain detailed models of the three-dimensional density, velocity, and magnetic structure in the  circumstellar environment, and linking this directly to the photospheric magnetic field and surface mass flux. 

The initial steps of the analytic flow from observations to models is well illustrated by the case of the prototyoical magnetosphere host star $\sigma$ Ori E, for which a detailed magnetospheric analysis was performed by \cite{2015MNRAS.451.2015O}. This flow is illustrated in Fig.\ \ref{sigorie_pol_to_model}.

\noindent 1) Polarized spectra is obtained for a target. The $S/N$ can be boosted with mean line profiles extracted via least-squares deconvolution \citep[LSD;][]{1997MNRAS.291..658D}. The surface magnetic field of the star is measured via the Zeeman effect.

\noindent 2) Direct analysis of the Stokes $V$ profiles enables detailed maps of the surface magnetic field via Zeeman Doppler Imaging \citep[ZDI;][]{Piskunov2002-ZDItechnique}. 

\noindent 3) The surface magnetic field is extrapolated into the circumstellar environment via potential field extrapolation. This is then used to guide Analytic Dynamical and Rigidly Rotating Magnetosphere models, providing predictions for the magnetospheric structure in the innermost region (where rotation is not important) and the outermost region (where rotation is key). 

While steps 1) to 3) are possible with ground-based data, ultraviolet polarimetry will enable the following key steps:

\noindent 4) Stokes $V$ profiles in UV resonance lines will directy measure the circumstellar magnetic field, testing the expected decline in magnetic field strength with increasing distance from the star. 

\noindent 5) High-resolution spectroscopy and linear spectropolarimetry will provide information on the velocity and density structure in the magnetosphere and wind. 

\noindent 6) Broadband polarimetry will enable the magnetospheric geometry to be inferred via integrated light. 

For each target, 10 high-resolution spectropolarimetric sequences will be obtained with Channel 1, and 30 low-resolution Channel 2 observations, with each dataset evenly sampling the rotational phase curve. The larger number of Channel 2 observations is necessitated by the complex behaviour of broadband linear polarimetry phase curves.

\section{Stellar magnetometry in the UV}\label{sec:lsd_magnetometry}

   \begin{figure}[t]
   \centering
   \includegraphics[width=0.95\textwidth, clip]{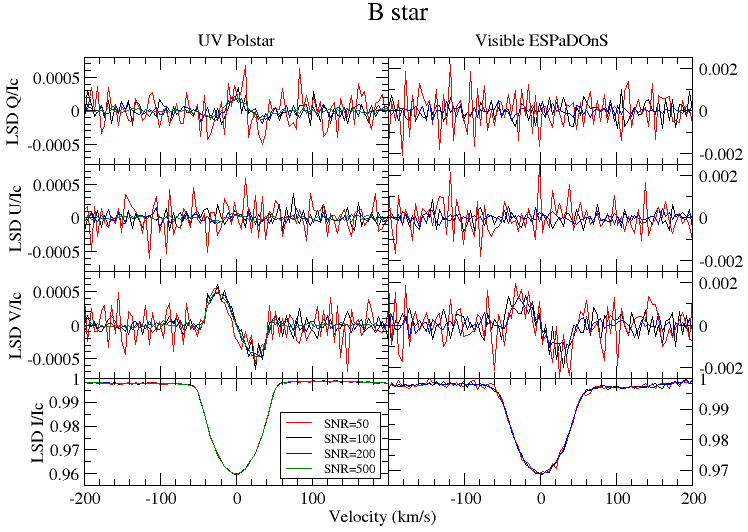}
      \caption{LSD Stokes I (bottom) and QUV (top 3 rows) profiles computed from simulated UV spectra over the 122-213 nm wavelength range with $R=33000$, i.e. corresponding to Polstar channel 1 (left) and over the 370-990 nm range with $R=65000$, i.e. corresponding to the ESPaDOnS spectropolarimeter, for a B star with T$_{\rm eff}=20000$ K and $\log g =4$,  with $v\sin i = 50$ km/s, a polar field strength $B_{\rm pol}=3$ kG, and S/N~=~50, 100, 200, and 500 at 150 and 450 nm respectively for the UV and Visible range.}
         \label{LSD_UVVis_Bstar}
   \end{figure}

   \begin{figure}[t]
   \centering
   \includegraphics[width=0.95\textwidth, clip]{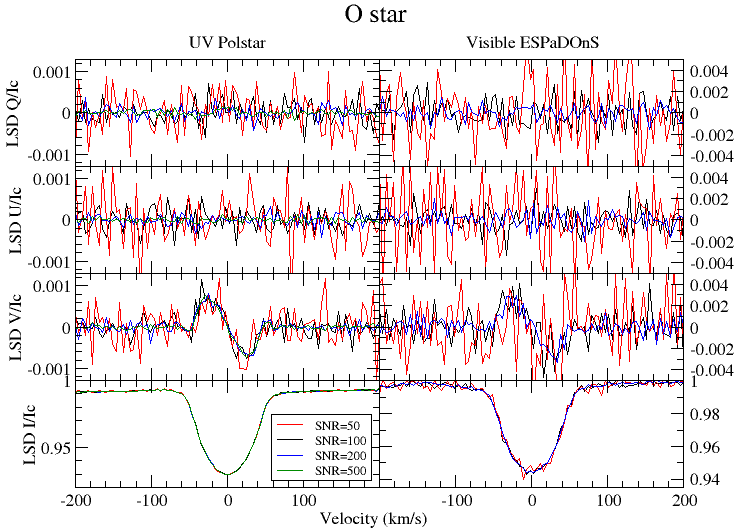}
      \caption{Same as Fig.~\ref{LSD_UVVis_Bstar} but for an O star with T$_{\rm eff}$ = 40000 K and $\log g =4.5$.}
         \label{LSD_UVVis_Ostar}
   \end{figure}

The Zeeman effect in spectral lines is commonly used to detect and measure magnetic fields at the surface of stars. If the magnetic field is strong enough, a Zeeman splitting of the lines can be directly observed in the spectra. If the field is weaker however, this splitting only appears as a small broadening of the line, which can be difficult to disentangle from other types of line broadening. However, each Zeeman component of the spectral lines is polarized in a different way. Therefore, measuring the polarization of spectral lines, i.e. measuring their Stokes parameters with spectropolarimetry, provides a very sensitive way to detect and characterize magnetic fields.

The Least-Squares Deconvolution (LSD) technique \citep{1997MNRAS.291..658D} 
is a powerful method to increase the S/N of Zeeman signatures. It combines the intensity and polarization signal from all photospheric lines available in the spectrum to produce mean Stokes profiles and therefore allows the detection of weak magnetic fields that would otherwise remain hidden in the noise. This technique has been successfully applied for Visible \citep[e.g.][]{1997MNRAS.291..658D, Wade2000-LSD-4Stokes} and IR \citep[e.g.][]{Martioli2020-LSD-in-IR, Moutou2020-LSD-in-IR, Petit2021A&A-LSD-in-vis-IR} spectropolarimetry.

The large number of photospheric lines available in the UV domain makes LSD even more powerful in this waveband. Figures~\ref{LSD_UVVis_Bstar} and \ref{LSD_UVVis_Ostar} compare model IQUV LSD profiles obtained for an O and B star in channel 1 of Polstar and in the Visible. For Polstar we used a resolution of R=33000 and for the Visible we used R=65000, which corresponds to the resolution of ESPaDOnS. ESPaDOnS is the best high-resolution spectropolarimeter currently available for Visible light and is installed at the Canada-France-Hawaii Telescope. 

For these tests, model spectra were calculated using the {\sc Zeeman} spectrum synthesis code \citep{Landstreet1988-Zeeman, Wade2001-Zeeman}, including a 3 kG dipole magnetic field with a stellar inclination angle $i=90^\circ$ and a magnetic obliquity angle $\beta=90^\circ$. The models were computed in local thermodynamic equilibrium (LTE), although non-LTE effects may be important at T$_{\rm eff} > 30000$ K, and they neglect wind lines.
Synthetic noise was added as a function of the flux per detector pixel, for several realistic S/N levels, to produce model observations. The S/N was scaled as the square root of the flux per detector pixel (assuming detector pixel sizes are constant in velocity), however this does not account for the wavelength dependence of instrumental efficiency, or a wavelength dependence in interstellar extinction or atmospheric transmission. The S/N values are those obtained at 150 and 450 nm respectively. 

LSD was applied to the model observations, generating LSD profiles, and allowing us to assess the detectability of signals in Stokes V, as well as Q and U. Only photospheric lines have been included in the LSD profiles. Exceptionally broad features, and any lines blended with them, were removed (H lines, some He lines, some other resonance lines in the UV). Lines in the Visible that would be blended with telluric lines were also removed from the mask.

\begin{figure*}[t]
   \centering
   \includegraphics[width=0.95\textwidth, clip]{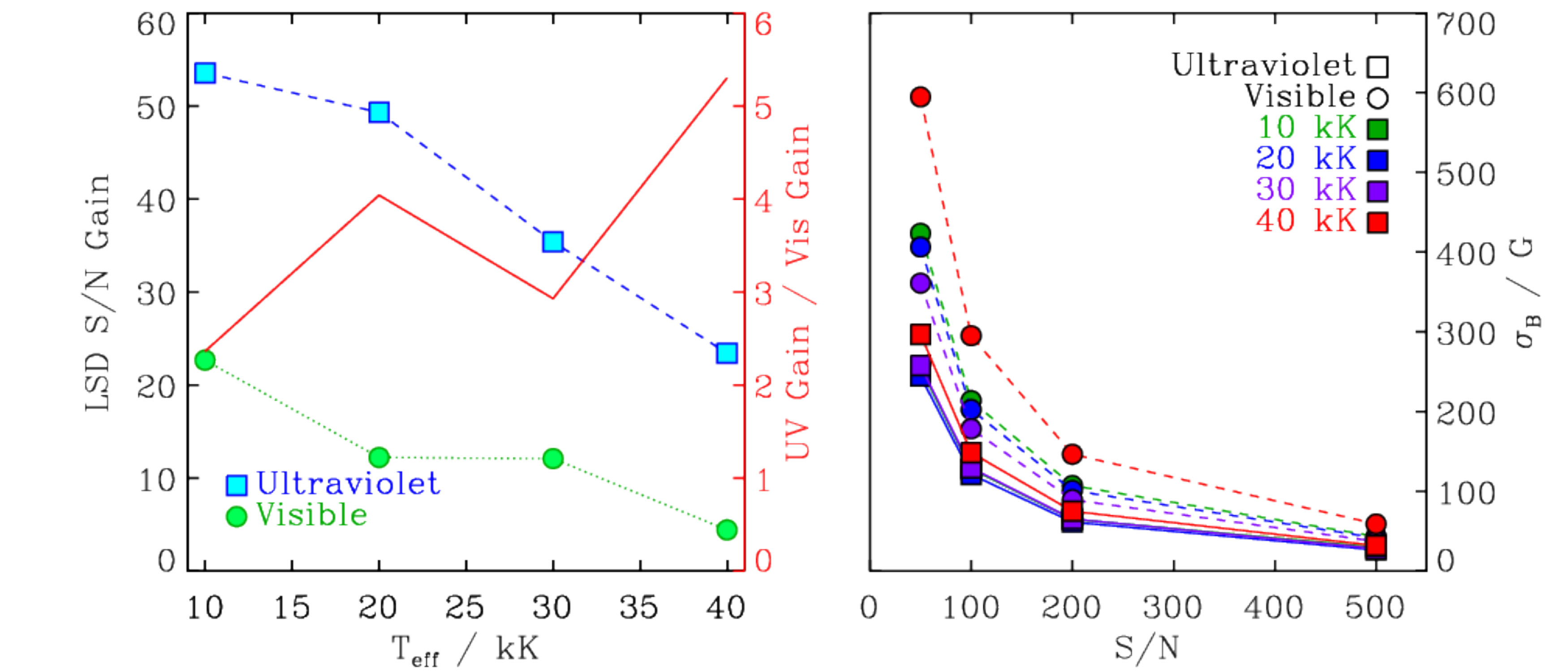}
      \caption{{\em Left}: $S/N$ gain obtained from ultraviolet and visible spectra as a function of $T_{\rm eff}$. The solid red line shows the relative gain achieved in the UV vs.\ the visible. {\em Right:} \bz~error bar obtained from visible and ultraviolet spectra for the 4 different \teff~models, as a function of $S/N$.}
         \label{uv_vis_lsd_snr_gain}
   \end{figure*}

The gain in S/N through the LSD process can be estimated from these results.  The S/N in a model observation is already specified, and the S/N of the LSD profile is approximated by the inverse of the error bars from the continuum normalized profiles.  In these calculations the same scaling parameters are used for all LSD line masks/profiles, and the visible and UV profiles are extracted with 1.8 and 4.5 \kms\ pixels, respectively, matching the mean pixel size of their model observations.  This allows one to examine the efficiency of the LSD process as a function of T$_{\rm eff}$, and compare the relative efficiencies of LSD in the visible (ESPaDOnS) and UV (Polstar) domains, shown in Figure \ref{uv_vis_lsd_snr_gain}.  

It is worth considering that Zeeman effect scales with wavelength.  For weaker magnetic fields the amplitude of the polarimetric signal is proportional to wavelength.  This leads to lower amplitude Stokes Q, U, and V signals in the UV than in the visible, by a factor of 2-4 depending on the wavelengths considered. However, there are many more strong lines available in the UV, thus the gain provided by multi-line techniques like LSD outweighs the loss in the amplitude of the polarimetric signal, which leads to an improved sensitivity to magnetic fields of hotter stars in the UV.
This is illustrated in Figs.~\ref{LSD_UVVis_Bstar} and \ref{LSD_UVVis_Ostar}, where the amplitudes in Stokes Q, U and V are lower for the UV profiles than the visible, but the noise is decreased even further, leading to more significant detections in the UV.

The relative magnetic performance of UV vs.\ visible spectra is demonstrated in Fig.\ \ref{uv_vis_lsd_snr_gain}. The left panel shows the LSD $S/N$ gain achieved in the two spectral regions as a function of \teff. Due to the much larger number of lines available in the UV, UV data provides a consistently higher $S/N$ gain, with an improvement over visible data range from about a factor of 2 at 10 kK to over 5 at 40 kK. The right panel shows the \bz~error bar $\sigma_B$ as a function of $S/N$, and demonstrates that the lower amplitude of the Zeeman effect in the UV is more than compensated for by the higher LSD $S/N$ gain, with error bars on average a factor of 2 smaller in the UV. While \bz~is measured only using circular polarization, the same improvement in sensitivity is also expected in Stokes $QU$, as qualitatively demonstrated in Figs.\ \ref{LSD_UVVis_Bstar} and \ref{LSD_UVVis_Ostar}. Thus, ultraviolet spectropolarimetry is optimally suited to hot star magnetometry.

These results demonstrate the increased effectiveness of LSD in the UV compared to the Visible for O and B stars, even though the Polstar wavelength range is much shorter than the one of ESPaDOnS, due to an increase in the number and strength of the available spectral lines. This leads to an improved ability to detect and precisely characterize the magnetic fields of O and B stars.

The magnetic and rotation axes of hot stars are usually not aligned. Therefore, as the star rotates, we observe the magnetic field at various orientations, which produces variability in the observed line profiles. From LSD profiles taken at different rotational phases, it is thus possible to reconstruct the photospheric and magnetic maps of the star through Zeeman-Doppler Imaging (ZDI). ZDI has been successfully applied to Visible spectropolarimetry \citep[e.g.][]{Donati-Brown1997-ZDI, Piskunov2002-ZDItechnique, Kochukhov2002-ZDI-alpha2CVn, 2006MNRAS.370..629D}, with a focus primarily on low- and intermediate-mass stars. Additionally, ZDI maps using full-Stokes (IQUV) information are more detailed than maps using only Stokes IV \citep[e.g.][]{Kochukhov2004-ZDI-4Stokes, 2010A&A...513A..13K, Rosen2015-ZDI-4Stokes-cool}. Figs.~\ref{LSD_UVVis_Bstar} and \ref{LSD_UVVis_Ostar} show that linear polarization (Stokes QU) becomes accessible at reasonable S/N values in the UV while they are not easily detectable in the Visible. Therefore the full-Stokes capability of PolStar is potentially very useful for ZDI mapping. 


\section{UV spectroscopic diagnostics of massive star magnetospheres}\label{sec:uv_spectroscopy}



   \begin{figure*}[t]
   \centering
   \includegraphics[width=0.95\textwidth,trim = 50 50 50 0]{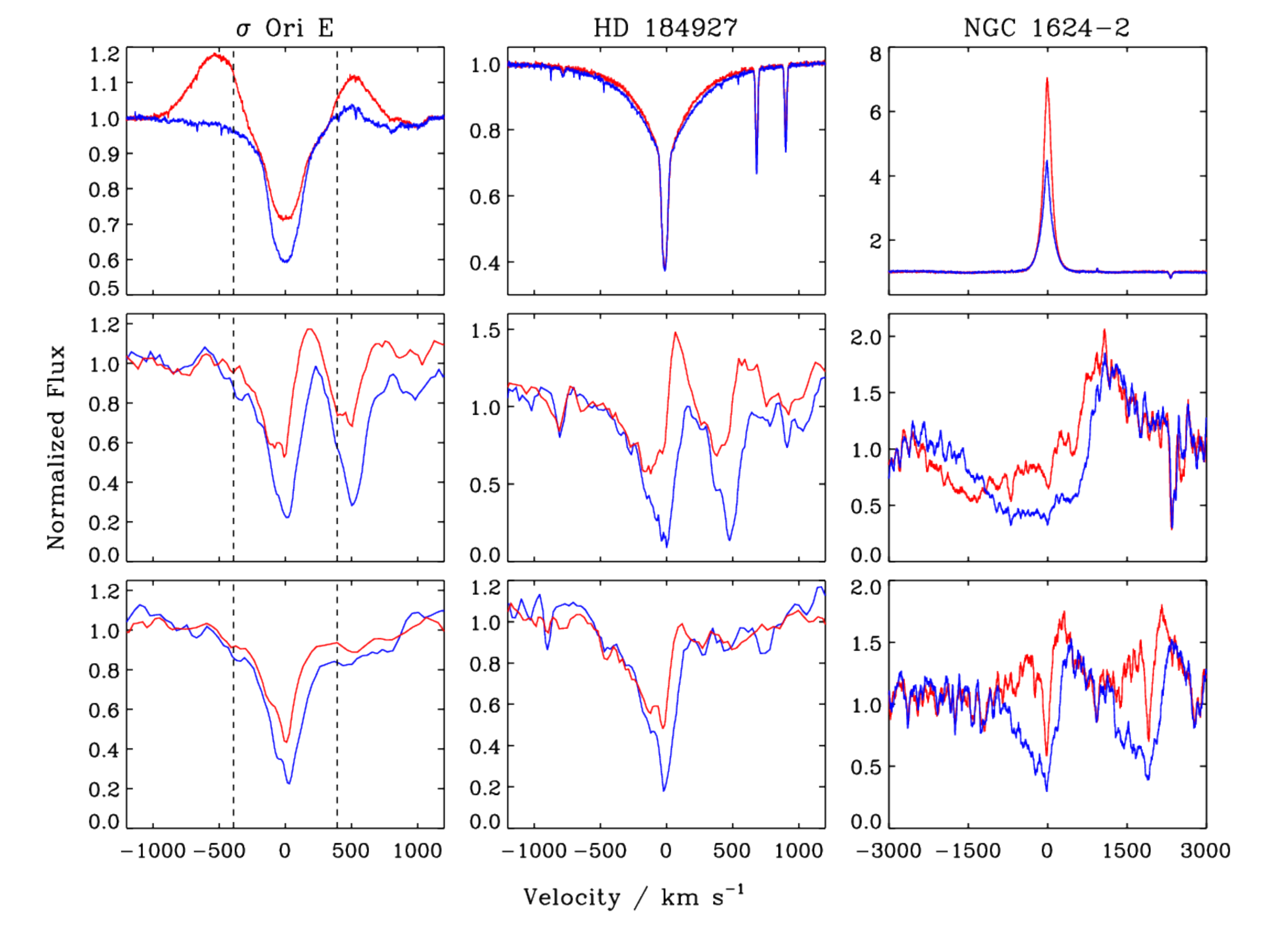}
      \caption{Comparison of H$\alpha$ (top), the C~{\sc iv} $\lambda\lambda$1548,1551 doublet (middle), and the Si~{\sc iv} $\lambda\lambda$1394,1403 doublet (bottom) for 3 representative stars (note that the red component of the Si~{\sc iv} is not visible for $\sigma$ Ori E and HD\,184927). High and low state are respectively indicated by red and blue. For $\sigma$ Ori E, dashed lines indicate $\pm R_{\rm K}$. All lines have been shifted to the stellar rest frame.}
         \label{polstar_uv_halpha}
   \end{figure*}

The interest of UV observations has been spectacularly demonstrated in the study of magnetic O stars \citep[e.g.][]{smi05,gru09,mar12,mar13,naz15,2019MNRAS.483.2814D,dav21}. As UV line profiles trace the density and velocity structure of the wind along the line-of-sight, line profile variations were expected for magnetic stars as the stellar rotation modifies our angle-of-view on the non-spherical magnetosphere. The available observations revealed enhanced absorptions at low-velocity when the slow confined winds appear in front of the star (equator-on view). In contrast, high-velocity absorptions were recorded for optically-thick lines when the fast polar wind entered into the line-of-sight (pole-on view). These changes were of course magnified for extremely magnetic stars (such as NGC 1624-2) or totally damped for very weakly magnetic stars ($\zeta$ Ori A). They could also be qualitatively reproduced by dedicated models \citep{mar13,naz15,erb21}. 

These first attempts at UV probing also opened up new research avenues. Indeed, spectra at only two phases were usually available (equator-on/edge-on). When improving the phase coverage, clear departures from a smooth transition are seen \citep{dav21}. Furthermore, when the magnetosphere geometry allows the two magnetic poles to enter our view, the associated line profiles do not appear identical \citep{naz15}. All this hints at more complex magnetospheric structures than the generally assumed dipolarity. Moreover, the magnetic star $\theta^1$ Ori C, despite being a prototype, appears as a clear exception to the general phenomenology \citep{smi05} and the reasons forthat remain unknown. By observing a large sample of stars with a dense phase coverage, Polstar will reveal the exact magnetic topology and shed a new light on the magnetospheric structure.

UV spectroscopy has also been crucial to detecting and understanding the magnetospheres of B-type stars \citep[e.g.][]{2001A&A...372..208S,2003A&A...406.1019N,2003A&A...411..565N,2006MNRAS.370..629D,2008A&A...483..857S,2011MNRAS.412L..45P,2013A&A...555A..46H,2021MNRAS.506.2296E}. Indeed, UV spectra have been crucial to detecting magnetic fields in B stars lacking the surface chemical peculiarities characterizing the universally magnetic Bp class \citep[e.g.][]{2003A&A...406.1019N,2006MNRAS.370..629D,2011MNRAS.412L..45P,2017MNRAS.471.2286S}. As surface chemical abundance patches are unable to form when the stellar wind strips the surface faster than patches can accumulate, this enabled detection of magnetic fields in stars earlier than B2. 

As with magnetic O-type stars, magnetic B-type stars exhibit periodic variations modulated on the rotational timescale. The UV lines of magnetic B stars are characterized by a `high state' with a red-shifted emission feature and a blue-shifted absorption, similar to a the classical P Cygni profile originating in a spherically expanding wind, but appearing at much cooler effective temperatures than those at which P Cygni profiles are generally seen. At `low state', the red-shifted emission disappears and the blue-shifted emission deepens. The high state corresponds to maxima of \bz, the line-of-sight magnetic field averaged over the visible stellar hemisphere, i.e.\ high state is seen when the magnetic pole is closest to the line of sight. Conversely, low state corresponds to \bz~nulls, i.e.\ when the magnetic equator bisects the stellar disk. This is generally interpreted as a result of a plasma torus in the magnetic equatorial plane, which either produces emission when projected off the limb of the star, or absorption when eclipsing the star, together with an absorption component produced by the escaping wind itself. 

Magnetic B-type stars frequently display variability in the N~{\sc v}~123.9, 124.3 nm doublet, which has an ionization temperature higher than the photospheric effective temperature of these stars. This is believed to be due to over-ionization from X-rays produced in colliding wind shocks \citep[e.g.][]{2001A&A...372..208S,2014MNRAS.441.3600U}, and is entirely absent in non-magnetic B-type stars.


\subsection{Comparing Visible and Ultraviolet Magnetospheric Diagnostics}\label{subsec:vis_uv_spectroscopy}

Fig.\ \ref{polstar_uv_halpha} compares the variable magnetospheric line diagnostics available from H$\alpha$ to two UV resonance lines, the C~{\sc iv} 154.8, 155.1 nm and Si~{\sc iv} 139.4, 140.3 nm doublets. 

In the case of the CM star $\sigma$ Ori E (B2\,Vp), H$\alpha$ emission variability is predominantly located outside of the Kepler corotation radius (indicated with vertical dashed lines), a consequence of the accumulation of cool, dense plasma at and beyond $R_{\rm K}$ which, at high state, is projected off of the stellar limb. By contrast, neither of the UV doublets demonstrate any variable emission outside of $\pm R_{\rm K}$. Instead, there is an emission feature shifted to the red of the line centre. At low state, the high-velocity emission disappears in H$\alpha$, replaced by enhanced absorption in the line core; this is due to the CM eclipsing the star. Similar enhanced absorption, also due to eclipsing, is detectable in the UV during low state. Note that the amplitude of variation is largest in the C~{\sc iv} doublet, however as the two components are very close to one another their emission overlaps, whereas the Si~{\sc iv} doublet has weaker emission, but the larger separation of its components enables the contribution of each line to be more effectively isolated. 

There is no H$\alpha$ emission in the case of HD\,184927 (B2\,Vp): while this star has a similarly strong magnetic field to that of $\sigma$ Ori E, it has a much longer rotational period and, therefore, does not possess a large CM. Its UV variability is remarkably similar to that of $\sigma$ Ori E. Together with the absence of UV variations outside the Kepler radius in $\sigma$ Ori E's UV lines, this suggests that, amongst B-type stars, the UV is primarily probing the dynamical part of the magnetosphere, i.e.\ the region closest to the star. 

The O-type star NGC\,1624-2 has the strongest magnetic field of any star at the top of the main sequence. Its long rotational period means that its magnetosphere is purely dynamical. Its H$\alpha$ line shows extremely strong emission at high state, due to its large Alfv\'en radius and the powerful wind that easily fills the DM. H$\alpha$ emission is only slightly weaker during low state. The emission is confined within about $\pm 200$~\kms of line centre. The velocity broadening is expected to be primarily thermal and turbulent, and the low velocity of the material again reflects the fact that H$\alpha$ is probing the dense, cool plasma around the magnetic equator. 

By contrast, NGC\,1624-2's UV profiles demonstrate both strong emission and absorption. The C~{\sc iv} doublet provides a sensitive probe of the unconfined wind, as revealed by its P Cygni-like profile, whereas Si~{\sc iv} is apparently more sensitive to the magnetically confined plasma. 

Fig.\ \ref{polstar_uv_halpha} demonstrates two important advantages of the UV: 1) unlike H$\alpha$, UV emission is detectable for essentially all magnetic stars; 2) even when H$\alpha$ is available, UV probes magnetospheric regions that are not reachable by other means. 





\subsection{Comparing X-ray and Ultraviolet Magnetospheric Diagnostics}\label{subsec:xray_uv}

Because of the (dipolar) magnetic fields, part of the stellar winds are channeled towards the magnetic equator where they collide. This shock is able to generate X-ray emission in addition to that naturally generated in massive star winds \citep{1997A&A...323..121B,2021arXiv211005302D}. For example, in O-stars, the wind line driving generates a soft X-ray emission with a luminosity scaling with the bolometric one ($L_{\rm X}/L_{\rm BOL}\sim 10^{-7}$, \citealt{1996A&AS..118..481B,1997A&A...322..878F}) while the magnetic O-stars show a clear enhancement, (with $\log (L_{\rm X}/L_{\rm BOL})\sim-6.2$, \citealt{2014ApJS..215...10N,2011MNRAS.416.1456O}). For most B-stars, the X-ray luminosity and its ratio with respect to bolometric luminosity are smaller. However, as for O-stars, they agree well with predictions of the X-ray Analytic Dynamical Magnetospre or XADM model \citealt{2014ApJS..215...10N}. A few discrepancies however remain and require data at other wavelengths, especially in the UV, to understand the peculiarities of the magnetospheres of these targets (e.g. the complex magnetic geometry of $\tau$\,Sco, or the potential impact of fast rotation) and the exact temperature stratification (to understand notably why X-rays are not as hard as predicted see e.g. HD191612, \citealt{2016ApJ...831..138N}, or to understand notably the diversity in X-ray hardness \citep{2014ApJS..215...10N}. 

In a few cases (e.g. $\theta^1$\,Ori\,C, \citealt{2005ApJ...628..986G}
; HD191612, \citealt{2010A&A...520A..59N}
; CPD --28 2561, \citealt{naz15}
; NGC1624-2, \citealt{2015MNRAS.453.3288P}
), sufficient X-ray data were collected to detect variability of the X-ray flux on the rotational period. The extremely magnetic NGC1624-2 showed a clear increase of absorption when the confined winds were seen magnetic equator-on, demonstrating that the X-ray emitting source lies in the confined winds and that their cool component can be dense enough to absorb X-rays.  Other objects rather show an occultation effect as part of the X-ray emission source is hidden by the stellar body as the confined winds are seen edge-on. The large amount of variability here suggests that the X-ray emitting region is not symmetric, as in a ring for example \citep{2016AdSpR..58..680U} 
but only additional information, thanks to UV polarimetry will enable determination of the exact magnetospheric geometry. 

Extremely few magnetic stars could be studied at high-resolution in X-rays, because of the low sensitivity of such facilities (although this may change when Athena becomes operative). The narrow and symmetric X-ray lines agree with predictions of magnetohydrodynamic (MHD) models \citep{2003A&A...398..203M,2005ApJ...628..986G,2007MNRAS.375..145N,2008AJ....135.1946N, 2012ApJ...746..142N,2009A&A...495..217F} 
but a detailed, quantitative assessment of the plasma kinematics can only be done at other wavelengths. Polstar here represents a unique opportunity.

\subsection{Numerical Modeling of Hot Star Magnetospheres}\label{subsec:numerical_models}

\begin{figure*}[t]
\vfill
\includegraphics[width=0.33\textwidth]{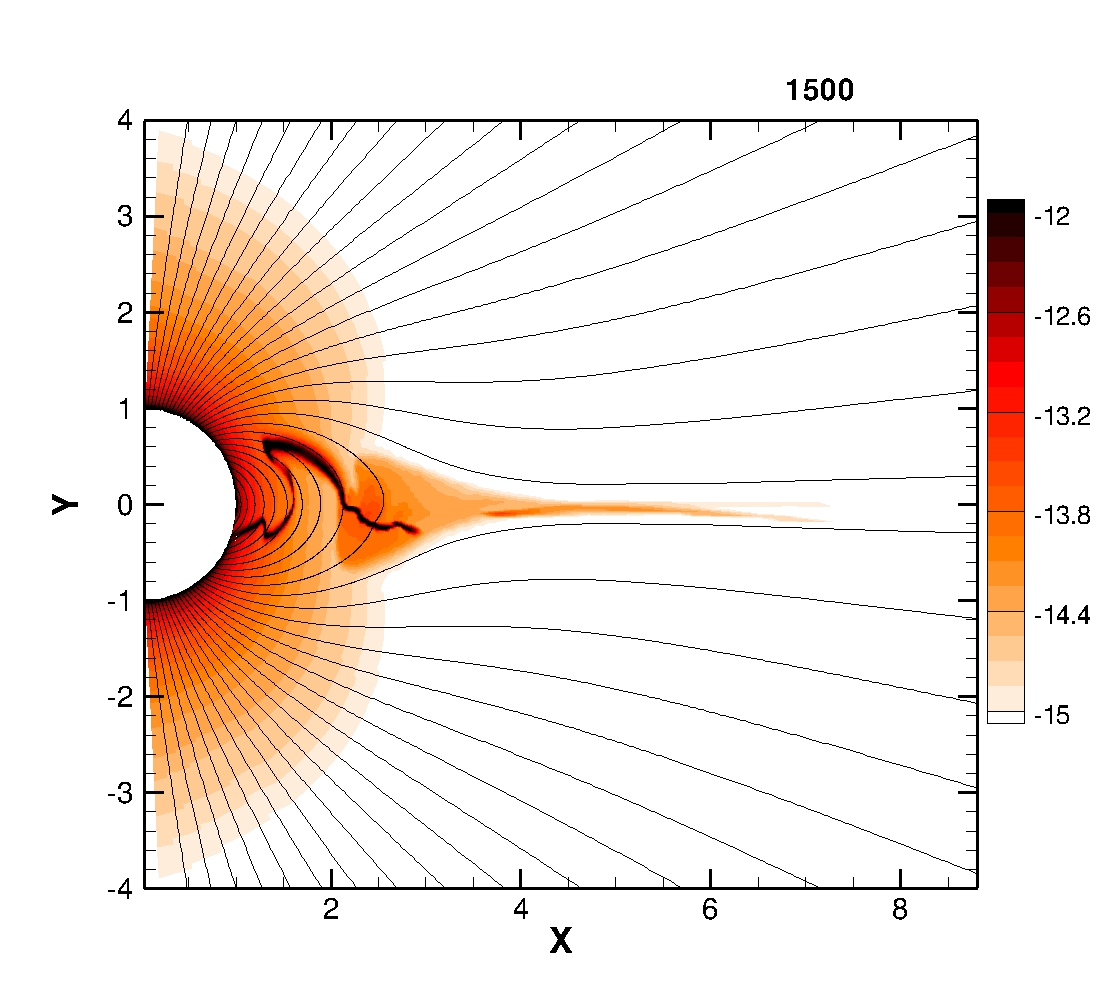}
\includegraphics[width=0.33\textwidth]{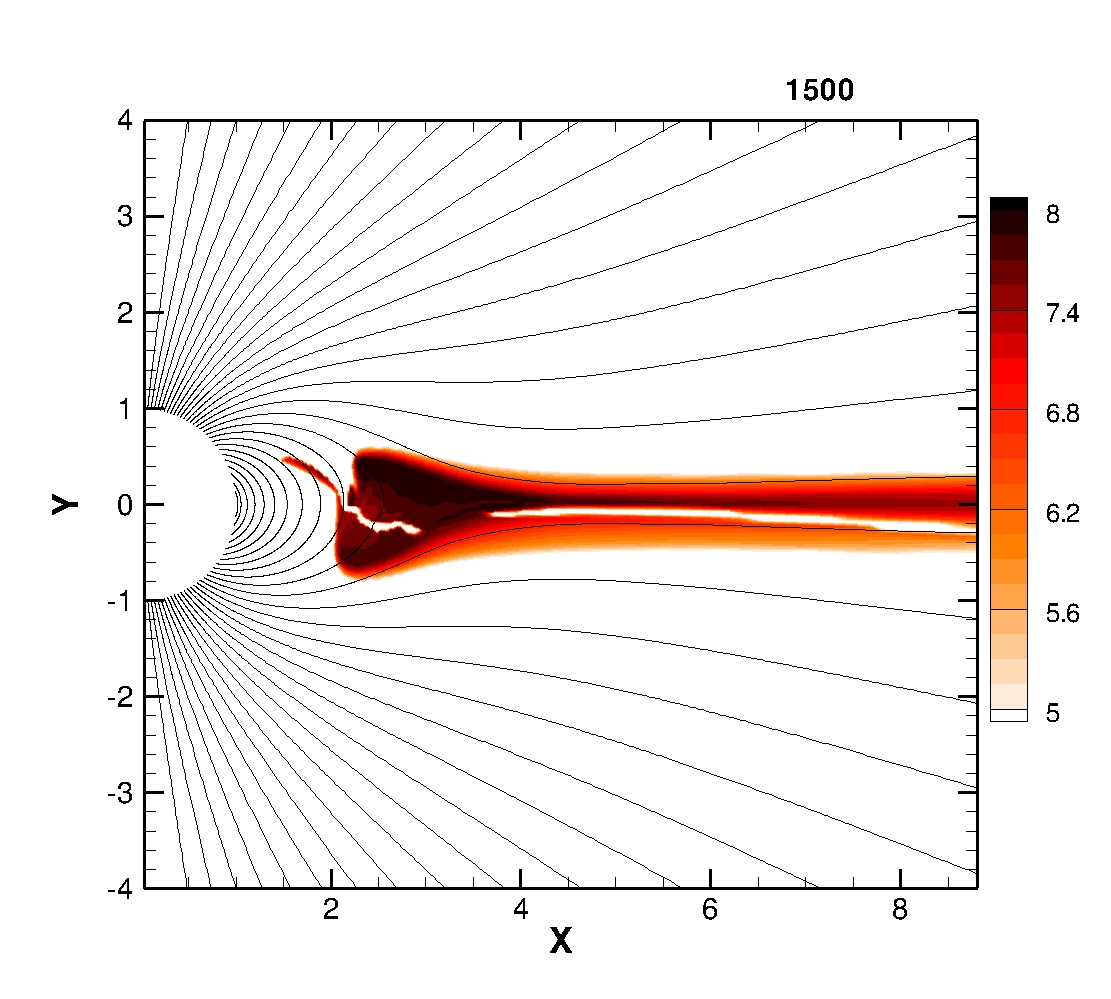}
\includegraphics[width=0.33\textwidth]{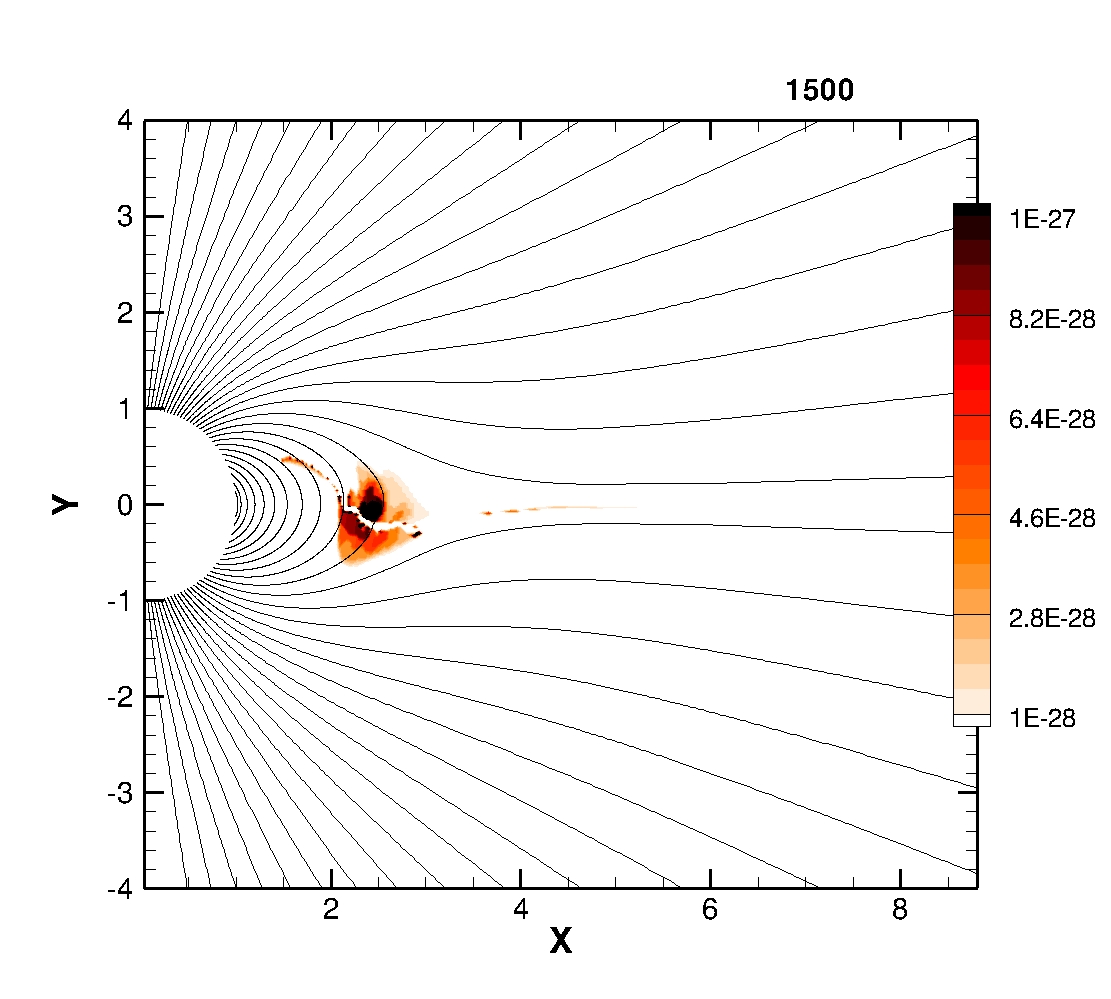}
\caption{
For a 2D MHD simulation of a magnetized wind with confinement parameter $\eta_\ast = 100$,
 color plots of log density (left) and log temperature (middle) in cgs units for arbitrary snapshot many dynamical times after initialization.
 Note that magnetic loops extending above $R_A/R_\ast \approx 100^{1/4} \approx 3.2$ are drawn open by the wind, while those with an apex below $R_A$ remain closed.
The right panel plots associated X-ray emission from the magnetically confined wind shock (MCWS) near the apex of closed loops. 
}
\label{fig:rhotxem}
\end{figure*}

\begin{figure*}[t]
	\includegraphics[scale=0.60]{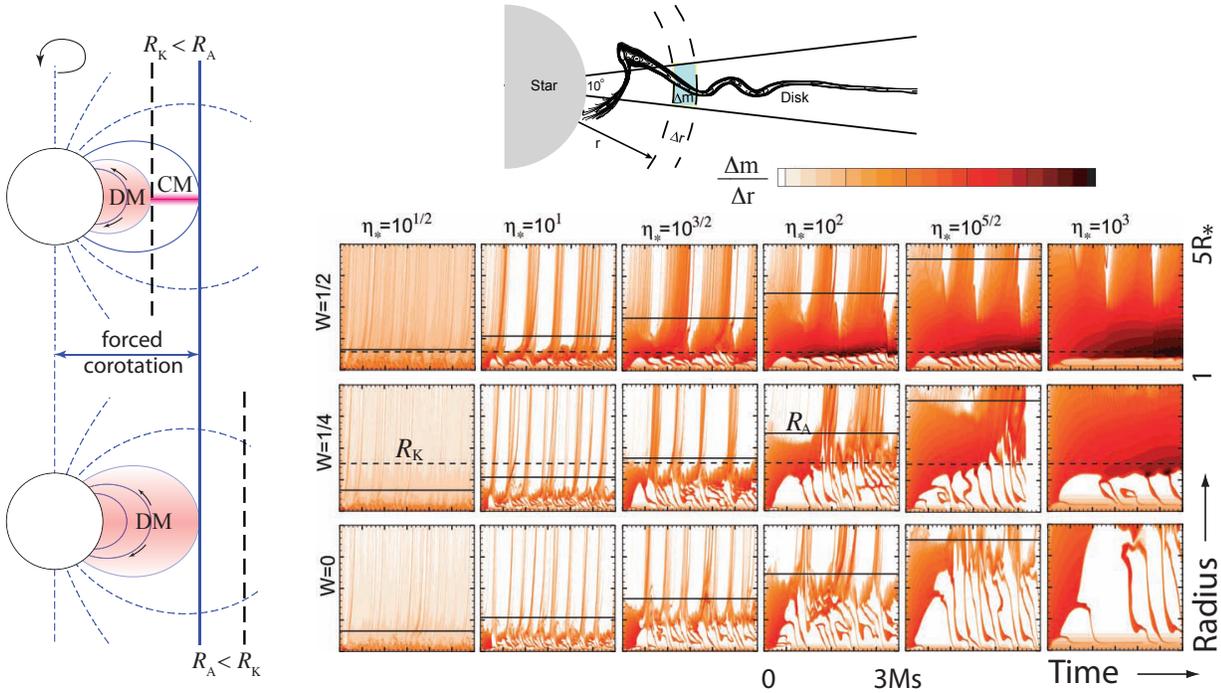}
\vspace{-0.7in}
\caption{
{\em Left:}
Sketch of the regimes for a dynamical vs.\ centrifugal magnetosphere (DM vs. CM).
{\em Upper right:}
 Contour plot for density at an arbitrary snapshot of an isothermal 2D MHD simulation, overlaid with illustration to define the radial mass distribution, $\Delta m/\Delta r$ near the equator.  
{\em Lower right:}
Color plots for  log of $\Delta m/\Delta r$, plotted versus radius (1-5 $R_\ast$) and time (0-3~Msec), for a mosaic of 2-D MHD models with a wide range of magnetic confinement parameters $\eta_\ast$, and 3 orbital rotation fractions $W$. 
The horizontal solid lines indicate the Alfv\'en radius $R_{Alfven}$ (solid) and Kepler radius $R_{Kepler}$ (dashed).
\label{fig:dmdr} 
}
\end{figure*}

\begin{figure*}[t]
	\begin{center}
	\includegraphics[scale=.6]{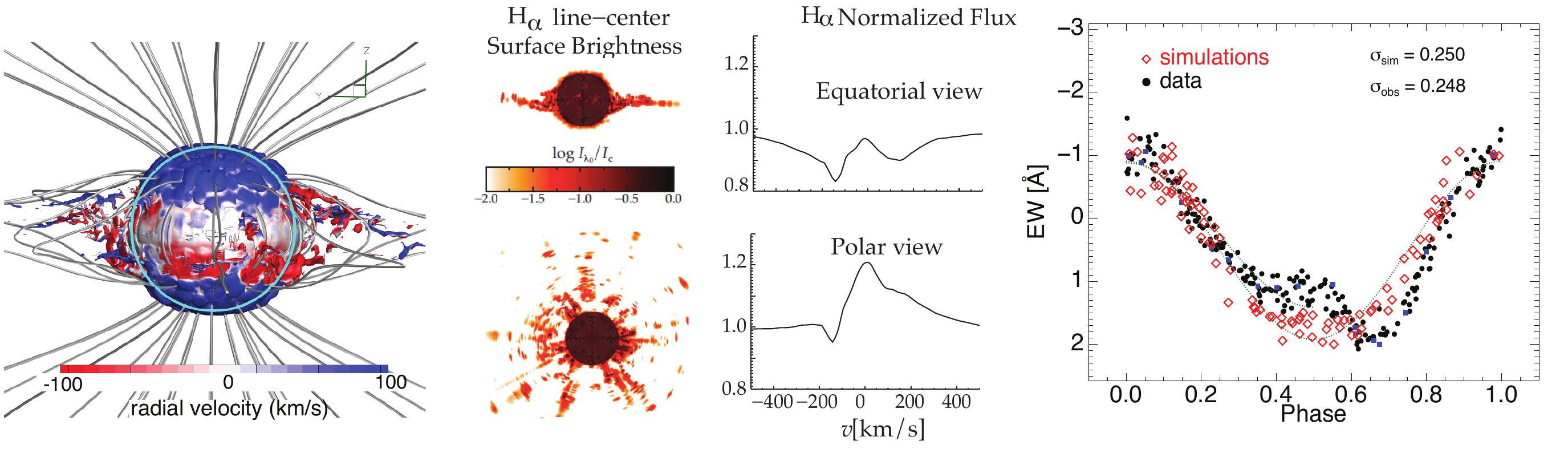}
	\caption{\small{
3D MHD model of the dynamical magnetosphere for the young, slowly rotating (15.4-day period) O7V star $\theta^1$~Ori~C
The left panel shows a snapshot of wind structure drawn as isodensity surface,
colored to show radial component of velocity.
The middle panels shows the predicted equatorial and polar views of H$\alpha$ line-center surface brightness, along with corresponding line-flux profiles.
The right panel compares the observed rotational modulation of the H$\alpha$ equivalent width (black) with 3D model predictions (red) assuming a pure-dipole surface field tilted by $\beta = 45^\circ$ to the rotation axis, as viewed from the inferred observer inclination of $i = 45^\circ$.
 }
 }
 \label{fig:3DT1OC}
\end{center}
\end{figure*}

Some characteristics of UV observations can be predicted by numerical models of hot star magnetospheres. In the past two decades, extensive observational surveys with theoretical analyses provided a strong basis to  develop a quite successful general paradigm for characterizing the properties of hot star magnetospheres in terms of their rotation (setting the Kepler co-rotation radius $R_K$) and level of wind magnetic confinement (setting the Alfv\'en radius $R_A$; 
(see figures \ref{fig:rhotxem} and  \ref{fig:dmdr}). 


As with X-rays, MHD simulations have been used advantageously to reproduce the overall variability phenomenology of UV resonance lines \citep{mar13, naz15}. As a sample of a fully self-consistent 3D MHD simulation of a hot star magnetosphere, figure \ref{fig:3DT1OC} \citep{udDoula2013} shows how wind material trapped in closed loops over the magnetic equator in $\theta^1$~Ori~C (left panel) leads to circumstellar emission that is strongest during rotational phases corresponding to pole-on views (middle panel). For a pure dipole with the inferred magnetic tilt  $\beta=45^\circ$, an observer with  the inferred inclination $i=45^\circ$ has  perspectives that vary from magnetic pole to equator, leading in the 3D model to the rotational phase variations in H$\alpha$ equivalent width shown in the right panel (shaded circles). Such models can also help us synthesize UV line profiles that Polstar can observe.

Additionally, dynamic flows formed in the DM as material is launched from the surface and then falls back in complex patterns, occurring on timescales of tens of ks \citep{udDoula2008}, could lead to small-scale stochastic variability in several diagnostics. This was tentatively detected for $\theta^1$~Ori~C \citep{udDoula2013}, in which short-timescale variability is found in the equivalent width measurements of H$\alpha$, on top of larger, rotationally-modulated variations. However, such short-term variability has yet to be detected in the UV as no magnetic star has been observed with sufficiently closely-spaced or high-SNR spectroscopic time-series. To best disentangle the various scales of variability that might arise as a consequence of dynamic flows in the DMs of magnetic massive stars, targets with very dense DMs and very slow rotation \citep[such as HD 108;][]{2010MNRAS.407.1423M,2017MNRAS.468.3985S} appear most promising.

\subsection{Analytical Models of Highly Magnetized Hot Star Winds}\label{subsec:analytic_models}

\begin{figure}[t]
\centering
\includegraphics[width=0.5\textwidth]{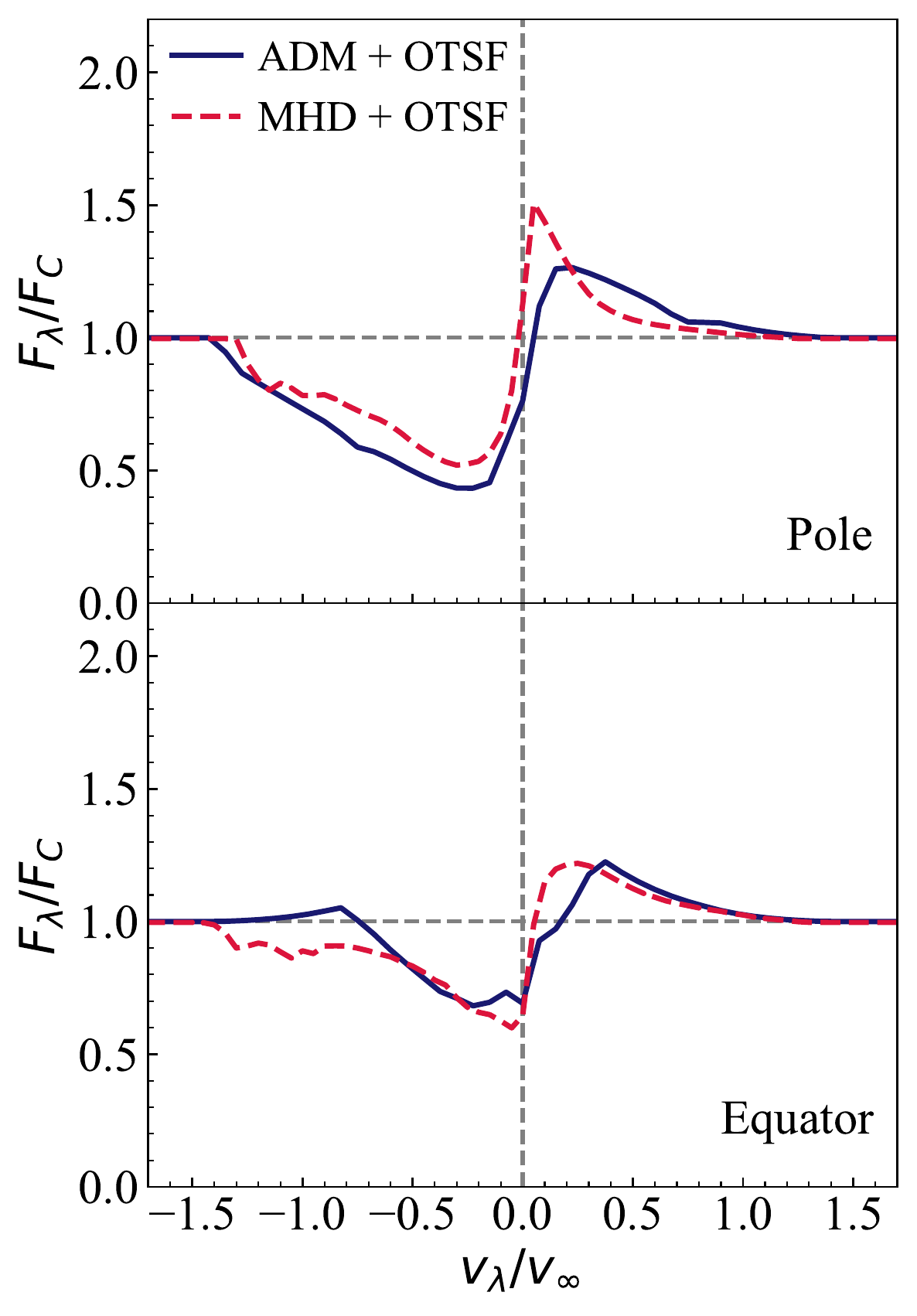}
\caption{
Synthetic UV line profiles calculated with an optically thin source function (OTSF) using the ADM formalism (blue solid line) and a snapshot of a 3D MHD magnetosphere (generated for $\theta^1$~Ori~C, \citealt{udDoula2013}). Both the ADM and MHD models have been coupled with the same radiative transfer technique. Both sets of models were calculated for a magnetic pole-on and equator-on view, using the same line strength parameter, and $\theta^1$~Ori~C characteristics ($R_{\rm A} = 2.3~R_*, \; v_\infty = 3200$~\kms). Overall the results from the ADM model agree well with the predictions from MHD. Reproduced from \citet{erb21}.}
\label{fig:mhd_uvadm_compare}
\end{figure}


Ultimately, proper self-consistent treatment of the global mass budget, including complex cycles of upflow and infall in DMs, and eventual {\em centrifugal breakout} (CBO) of trapped plasma in CMs, requires self-consistent MHD simulations that account for the competition between field and material flow. However, in the limit of arbitrarily strong fields (effectively with $R_A \rightarrow \infty$),  analyses based on an idealization of purely rigid fields have led to both {\em analytic dynamical magnetosphere} (ADM) and {\em rigidly rotating magnetosphere} (RRM) models that have shown great promise for analyzing broad observational trends in multiple diagnostics without the computational complexity and expense of full MHD simulations. 

\citet{2005ApJ...630L..81T} have shown that the RRM model can successfully reproduce the periodic modulations observed in the light curve, H$\alpha$ emission-line profile, and longitudinal field strength of a rapidly rotating, strongly magnetized star with a large CM \citep{2005ApJ...630L..81T,2008MNRAS.389..559T,2013ApJ...769...33T}. In this model, plasma accumulates at minima within the gravitocentrifugal potential of a rotating star, which define a warped accumulation surface approximately in the magnetic equatorial plane, with the densest regions at the intersections of this plane with the rotational equatorial plane, and an inner edge approximately defined by the Kepler corotation radius $R_{\rm K}$ (with a gap between $R_{\rm K}$ and the star). The model can be extended from a simple tilted dipole to any arbitrary magnetic geometry, e.g.\ as determined via ZDI, a method which \cite{2015MNRAS.451.2015O} used to reproduce the spectroscopic and photometric properties of the prototypical magnetosphere star $\sigma$ Ori E (see also Fig.\ \ref{sigorie_pol_to_model}). The RRM model has also been used to predict the broadband polarimetric observables of $\sigma$ Ori E, although in this case there is severe tension between the best fit achieved by polarimetry and photometry \citep[][see also Sect.\ \ref{sec:continuum_linpol}]{2013ApJ...766L...9C}.

The Analytic Dynamical Magnetosphere (ADM) model \citep{Owocki2016} was developed to provide an approximate, static view of the magnetospheric structure of a slowly-rotating massive star. Compared to time-averaged MHD simulations, this analytic prescription provides a satisfactory description of magnetospheres, and can be used to synthesize various diagnostics (e.g. H$\alpha$ and optical photometric measurements; \citealt{Owocki2016,munoz20}), as well as the strong, wind-sensitive lines found in the ultraviolet \citep{Hennicker2018,erb21}. Figure \ref{fig:mhd_uvadm_compare} shows that when coupled with appropriate radiative transfer techniques, UV line profiles synthesized using the ADM formalism generally compare well with those produced using an MHD magnetosphere.

The ADM model relies on a few simplifying assumptions: rotation is not taken into account, the dynamic flows formed by alternating episodes of wind launching and infall back onto the stellar surface are approximated by using an unphysical superposition of an upflow and a downflow component, and the magnetic field is idealized, originally taken to be a pure dipole. The latter assumption can be relaxed to take into account various values of $R_\textrm{A}$ (e.g. \citealt{erb21}), and even an arbitrary magnetic geometry \citep{2017IAUS..329..369F}.

Given its success in reproducing a range of observational diagnostics, the ADM model represents a computationally inexpensive alternative to MHD simulations, especially given the wide parameter space probed in UV studies of magnetospheres
. \citet{erb21} produced a grid of synthetic UV line profiles using the ADM formalism coupled with a simplified radiative transfer technique (using their \textit{UV-ADM} code). They presented the first large-scale parameter study of the many factors (e.g. magnetosphere size, line strength, observer's viewing angle) that affect UV wind line formation, with parameters chosen to correspond with particular well-known magnetic massive stars (e.g. HD 191612, NGC 1624-2). 
The application of the ADM model by \citet{erb21} has also provided evidence for spectral features in the ultraviolet that appear to be unique to magnetic massive stars, including the presence of red-shifted absorption and a strong desaturation of the high-velocity absorption trough, leading to a spectrum which appears to be of a later type than would be assumed from other observations (e.g. optical). Such signatures could help establish ultraviolet spectroscopy as a means of indirectly detecting magnetic fields in massive stars, similarly to other rotationally-modulated variations across the electromagnetic spectrum \citep[e.g. optical photometric variations, H$\alpha$ emission, or gyrosynchrotron radio emission][]{2019MNRAS.487..304D,2020MNRAS.499.5379S,2021MNRAS.507.1979L}. Of particular interest to Polstar, the red-shiftd absorption dip is diagnostic of infalling plasma. As can be seen in the lower panel of Fig.\ \ref{fig:mhd_uvadm_compare}, this is a subtle feature requiring high $S/N$ and high spectral resoluton to detect.




\section{Circumstellar magnetometry in the UV}\label{sec:circumstellar_magnetometry}

\citet{erb21}'s UV line synthesis technique has recently been extended to model magnetospheric polarization using UV wind lines, following the method outlined by \citet{Gayley2010} and \citet{Gayley2017}. Although often challenging to detect, Zeeman splitting is present in the spectral lines of magnetic massive stars \citep[e.g.][]{2009ARA&A..47..333D}. The split line components are circularly polarized, which is then detected and measured using Stokes~$V$ $= I_{L} - I_{R}$ profiles. 

The UV-ADM code (Sect. \ref{subsec:analytic_models}) can be modified to produce synthetic Stokes~$V$~profiles by updating the calculation for the line profiles following \citet{Gayley2017} \citep[see also][]{Gayley2010,Gayley2015,Kochukhov2015} for calculating the antiderivative of the Stokes~$V$~polarization profile. The UV-ADM code is used to obtain the field-weighted intensity, and the Stokes~$V$~profile can then be obtained from the derivative with respect to wavelength. A grid of synthetic UV intensity (Stokes~$I$) and Stokes~$V$ profiles is then produced.

\begin{table}[t]
\begin{center}
\caption{Adopted stellar parameters for synthetic Stokes $V$ profiles
\label{tab:modelparams_stokesv_uvadm}}
\begin{tabular}{lcc}
\hline \hline
& \textbf{B-type Star} & \textbf{O-type Star} \\
$\log(\dot{M})$ & -10 $M_{\odot}$~yr$^{-1}$ & -7 $M_{\odot}$~yr$^{-1}$ \\
$v_{\infty}$ & 1200~km~s$^{-1}$ & 2700~km~s$^{-1}$ \\
$R_{\ast}$ & 4~$R_{\odot}$ & 10~$R_{\odot}$ \\
\hline
\end{tabular}
\end{center}
\end{table}

\begin{figure}[t]
\centering
\includegraphics[width=0.6\textwidth]{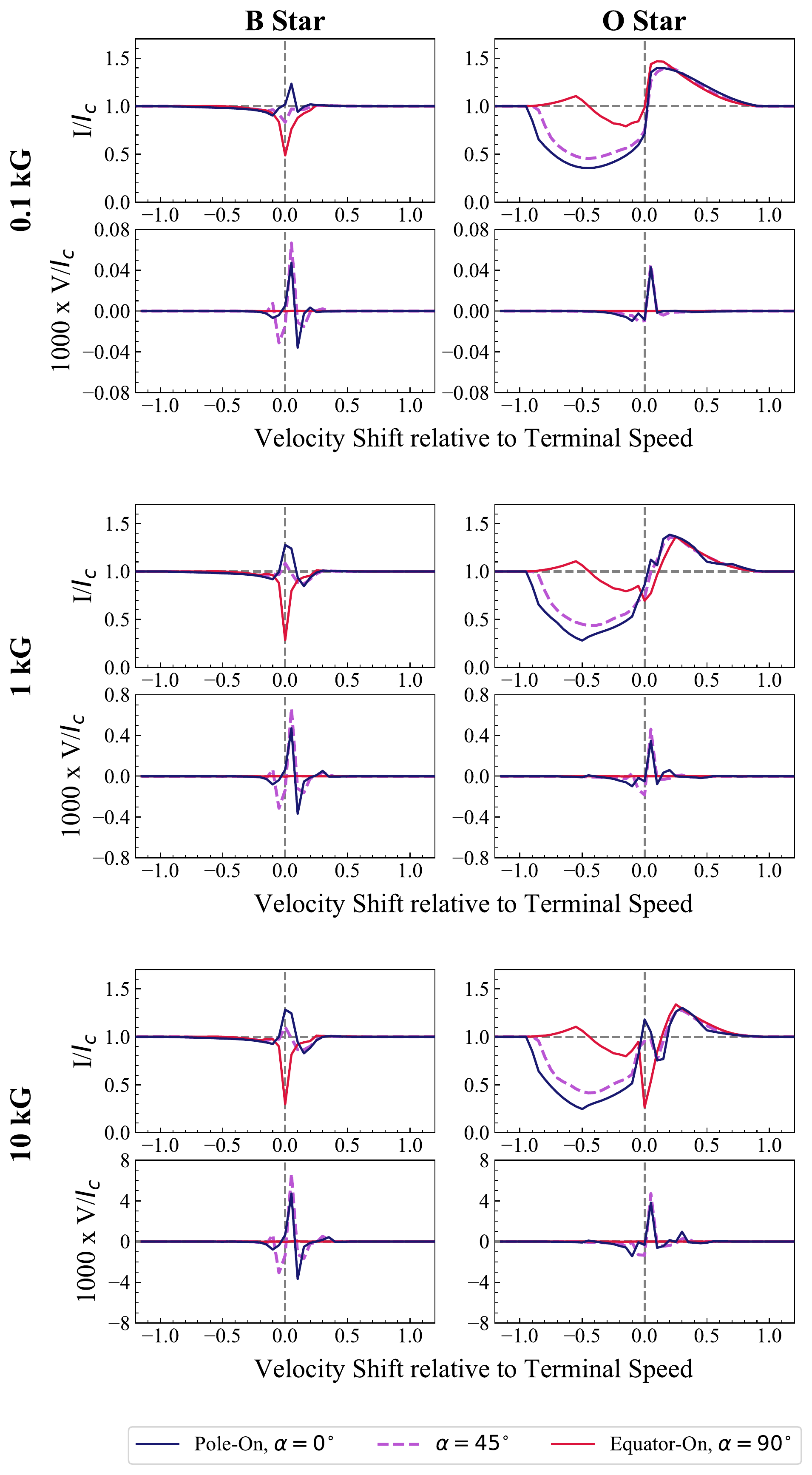}
\caption{Synthetic intensity (top row) and Stokes~$V$ (bottom row) profiles created using the UV-ADM code for representative O and B-type stars with surface fields of 0.1 kG, 1 kG, and 10 kG. Profiles are provided for magneteospheric viewing angles of $\alpha$ = 0$^{\circ}$ (pole-on; blue solid lines), $\alpha$ = 45$^{\circ}$ (purple dashed lines), and $\alpha$ = 90$^{\circ}$ (equator-on; red solid lines). The Stokes~$V$~profiles use the right component of the wind-sensitive C~\textsc{iv}~doublet as a representative spectral line. 
}
\label{fig:stokesv_uvadm}
\end{figure}

In Figure \ref{fig:stokesv_uvadm}, we show 36 synthetic Stokes $I$ profiles synthesized for the representative O- and B-type stars listed in Table \ref{tab:modelparams_stokesv_uvadm}. For each set of stellar parameters, we have calculated models for surface field strengths of 0.1, 1, and 10 kG, for viewing angles\footnote{The viewing angle $\alpha$ is defined to be the angle between the line-of-sight to the observer and the north magnetic pole \citep{erb21}.} $\alpha$ of 0$^{\circ}$ (pole-on; blue solid lines), 45$^{\circ}$ (purple dashed lines), and 90$^{\circ}$ (equator-on; red solid lines). The profiles were synthesized for line strengths and Land\'e factors corresponding to those of the right component of the wind-sensitive C~\textsc{iv}~154.8, 155.1 nm doublet.

The Stokes~$V$ signature from the equator-on view of each sample magnetosphere is effectively null, as required by the top-bottom symmetry with reversed line-of-sight magnetic field. Furthermore, the amplitude of the Zeeman signature in the pole-on view is extremely small for a 0.1~kG surface field, as the weak-field Zeeman effect is linear with magnetic field strength. However, for a 10~kG field, these results suggest the amplitude of the Stokes~$V$ signature is reasonably attainable at an expected $V/I$ sensitivity level of about 0.01\%.

An assumption in the model is that the Zeeman shift does not affect the total photon scattering, only the wavelengths at which scattering occurs, in the opposite circular polarization signatures. Hence, the total area under the Stokes $V$ profile
should be zero, consistent with its representation as a derivative of a function that begins and ends at zero at its anchors in the continuum on opposite sides of the line profile. This feature can help distinguish a true signal from noise, as anything that produces a net area under the $V$ curve would be noise under these conditions. However, it should be noted that Stokes $V$ asymmetries originating with velocity gradients along the line of sight have been detected in solar magnetic fields \citep{1989A&A...221..338G}. Similar asymmetries, which may have the same origin, have been detected in the ultraweak fields of Am stars \citep{2011A&A...532L..13P,2016A&A...586A..97B,2018CoSka..48...53F}. Whether such asymmetrical Stokes $V$ signatures should be expected in hot star magnetospheres is not known. 

As for the true Stokes $V$ signal, two physical effects combine to produce it \citep{Gayley2015}. The first is due to the wavelength derivative of the Stokes $I$ profile itself, an effect that would be  present even if the line-of-sight B field were constant across the profile. The other is due to the gradient in the average line-of-sight B field across the profile, an effect that would be present even if the Stokes $I$ profile were constant (i.e., flat). Which of these dominates the signal at any wavelength can be established by the gradient in Stokes $I$: the first effect will only be dominant when the Stokes $I$ signal is strong, whereas deviations of the Stokes $V$ signal from the derivative of Stokes $I$ would clearly indicate that the second effect is active as well. The latter possibility shows that Stokes V is not purely a diagnostic of the strength of the field, it is a combined diagnostic of field strength and field structure, requiring forward modeling to interpret.


\section{Linear spectropolarimetry in the UV}\label{sec:lin_specpol}




The basic idea of linear spectropolarimetry is rather straightforward: electrons in an extended circumstellar medium scatter radiation from the stellar surface,  
giving rise to a certain linear polarization level. 
If the sky projected electron distribution is perfectly 
circular, the linear 
Stokes $Q$ and $U$ vectors cancel, and linear polarization 
remains absent.
If the geometry is not circular, but   
an asymmetry is involved, this results in some level of continuum linear polarization.  

\begin{figure*}[t]
    \centering
\includegraphics[width=0.45\textwidth]{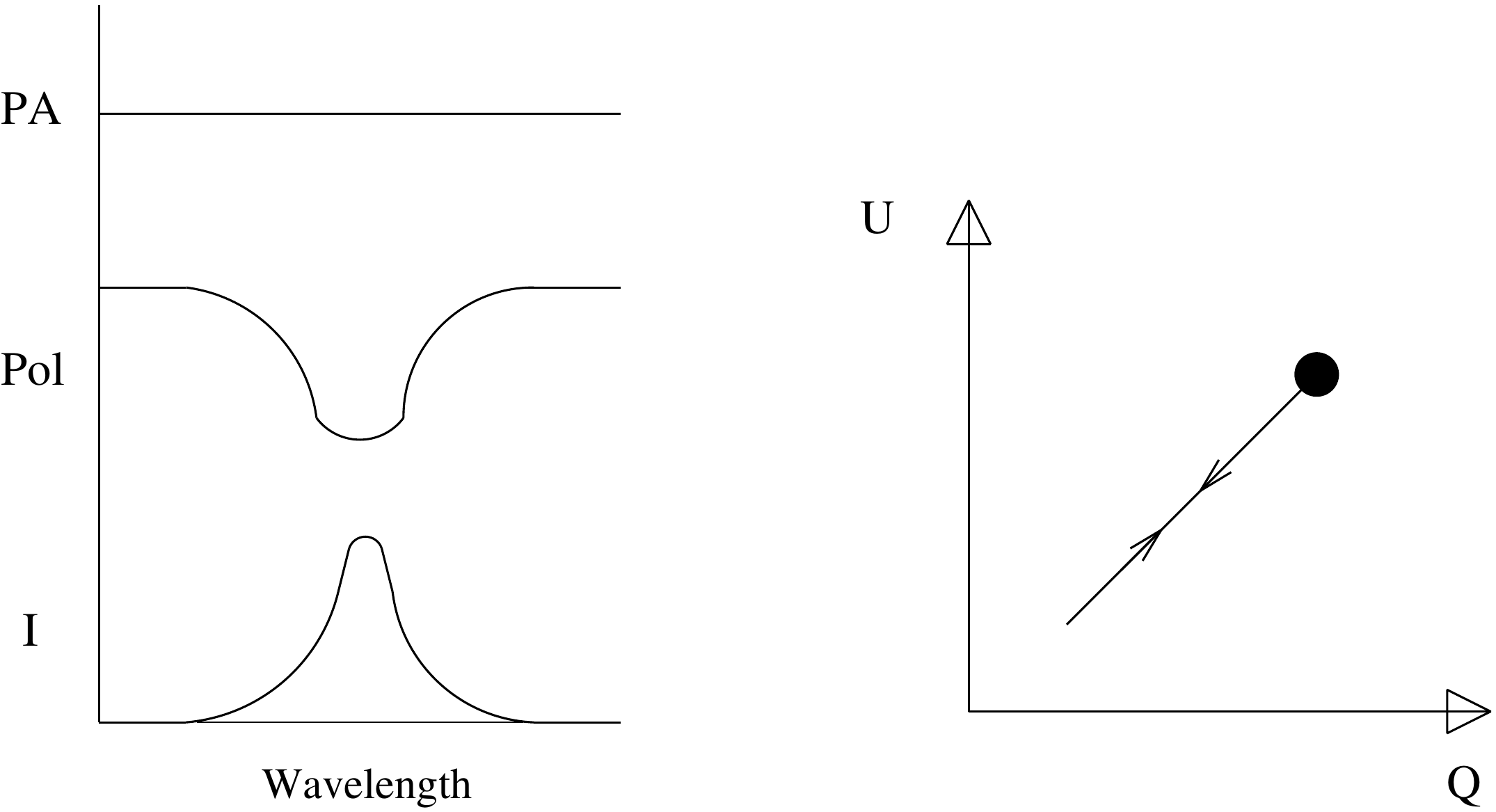}
\includegraphics[width=0.45\textwidth]{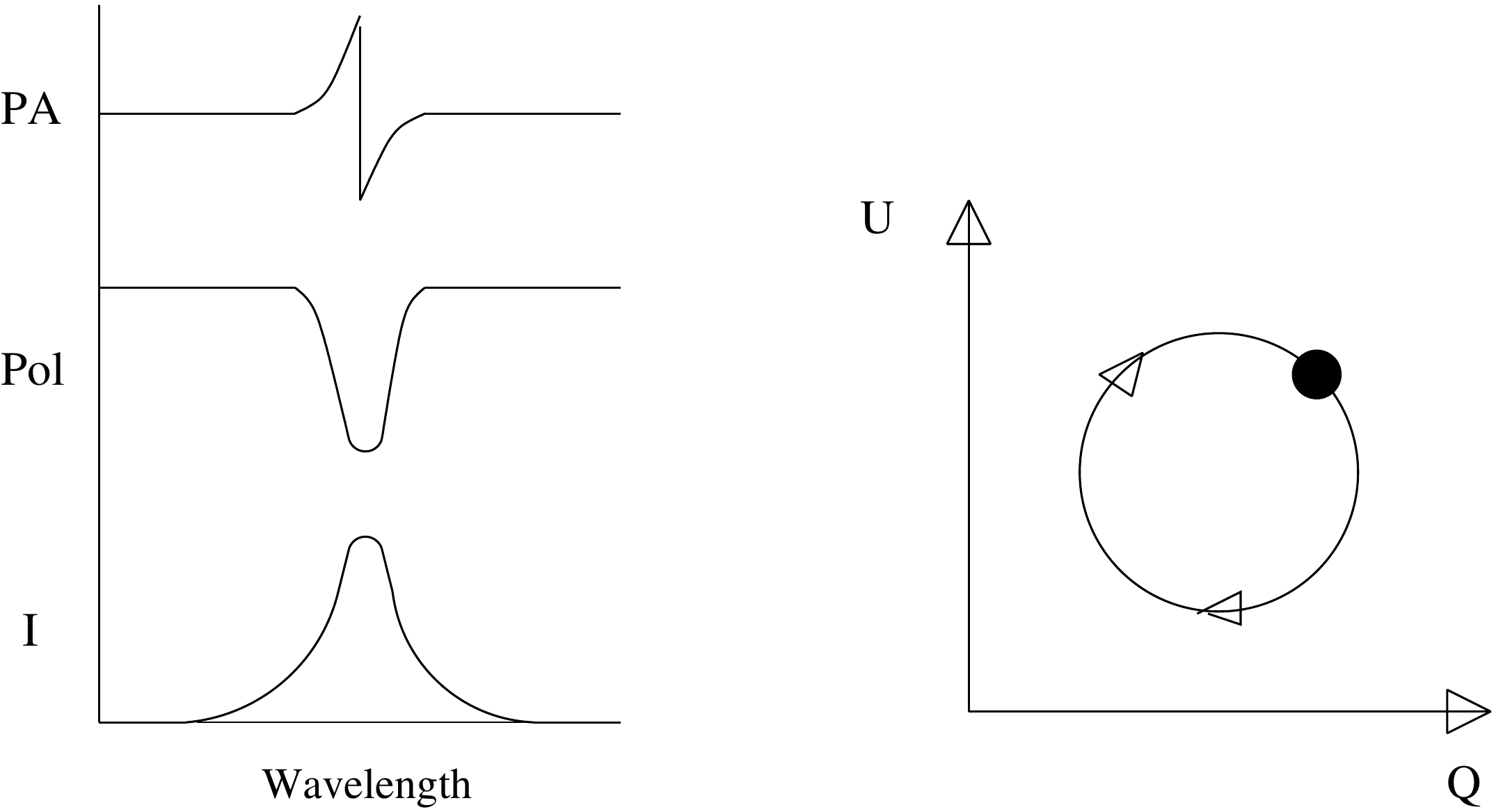}
    \caption{Cartoons representing 
line {\it depolarization} (left-hand side) and compact line emission
scattered off a rotating disk (right-hand sie) as triplots and $QU$ diagrams. Stokes~$I$
profiles are plotted in the lower triplot panels, \% Pol in the middle panels, and 
the position angles (PAs) are shown in the upper triplot panels. Line depolarization is as broad as the Stokes I emission, while {\it line} 
polarization is narrow by comparison. Depolarization translates 
into $QU$ space as a linear excursion, 
while {\it line} polarization PA flips are associated with $QU$ loops.}
    \label{fig:linepol}
\end{figure*}

One of the advantages of linear spectro-polarimetry over continuum polarimetry is 
that it is possible to perform differential measurements between a line and the 
continuum -- independent of interstellar and /or instrumental  
polarization.
One example may occur across an emission line. This ``line effect'' relies on the expectation 
that recombination lines arise over a much larger volume than the 
continuum, and becomes {\it de}polarized (see the left hand side of 
Fig.\,\ref{fig:linepol}. 
Depolarization immediately indicates the presence or absence of 
asphericity. 

The bulk of linear spectro-polarimetry studies involved these depolarization line effects, but in some situations there is 
evidence for intrinsic {\it line} polarization, such as predicted by \cite{1993A&A...271..492W} and 
found observationally in pre-main sequence T Tauri and Herbig Ae/Be stars \citep{2002MNRAS.337..356V,2005A&A...430..213V}.
In such cases the line photons are assumed to originate from a more compact source, e.g.
as a result of magnetospheric phenomena, and these photons are scattered off a rotating 
disk, leading to a flip in the position angle (PA), resulting in a rounded loop (rather than a linear excursion) 
in the $QU$ diagram (sketched on the right hand side of Fig,\,\ref{fig:linepol}.

\begin{figure*}[t]
    \centering
\includegraphics[width=0.45\textwidth]{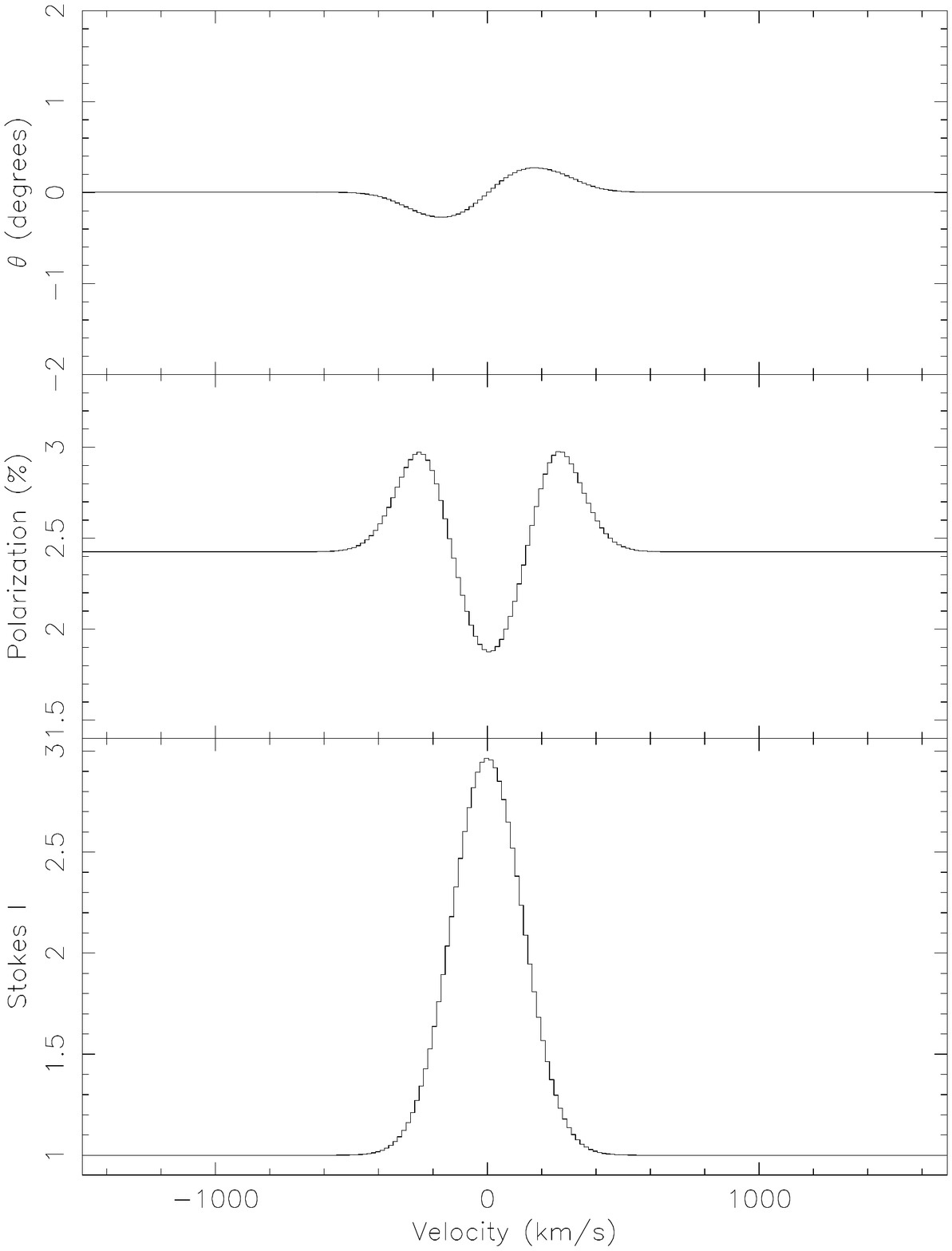}
\includegraphics[width=0.45\textwidth]{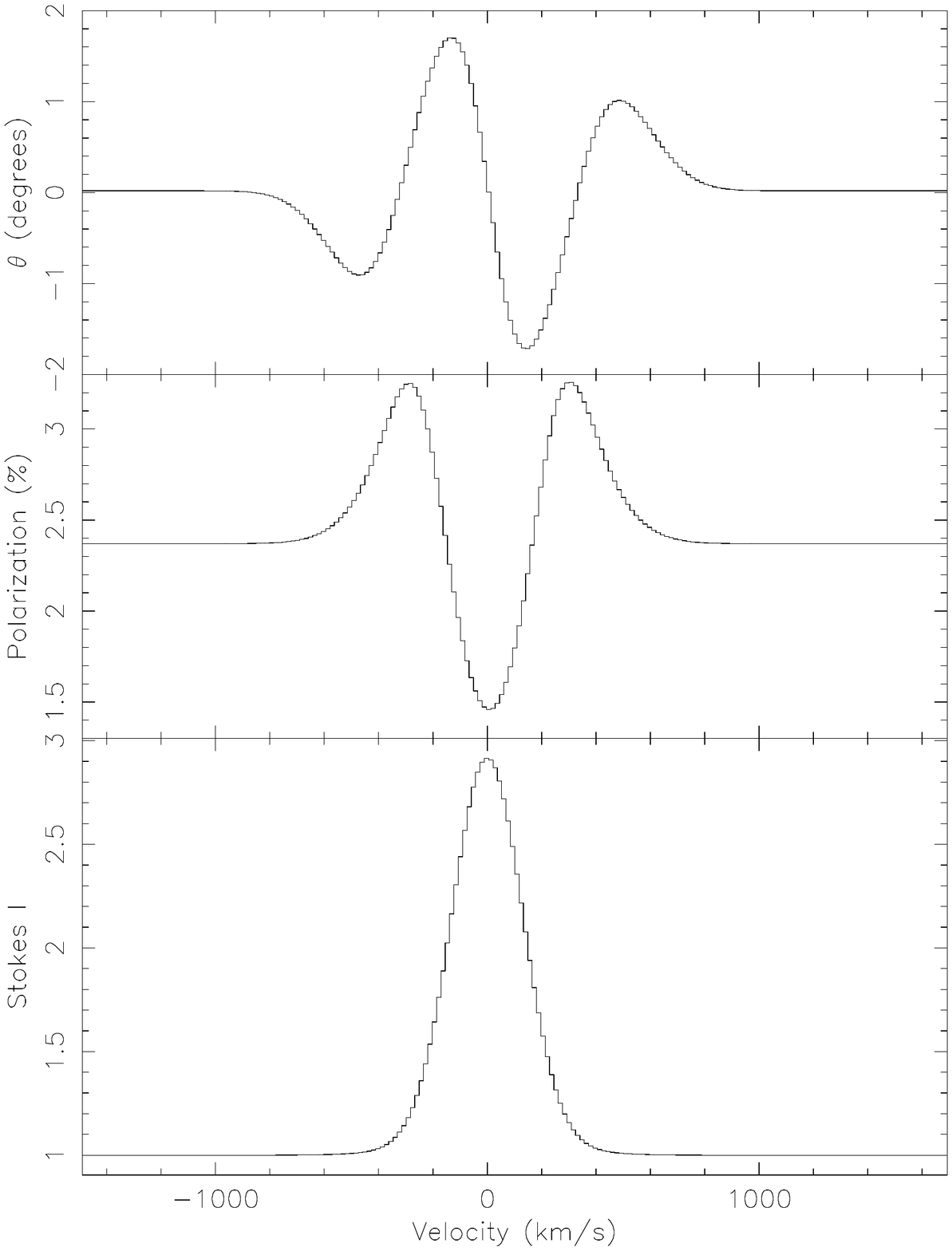}
    \caption{Monte Carlo line polarimetry for the case of a disk with an inner hole (left hand side) 
and without an inner hole -- a result of the finite size of the 
star (right hand side). From \cite{2005A&A...430..213V}.}
    \label{fig:linepol-MC}
\end{figure*}

Motivated by the high incidence of $QU$ loops in T Tauri and Herbig Ae stars, \cite{2005MNRAS.359.1049V}
developed Monte Carlo polarization models of scattering off rotating disks -- with and without inner holes. 
Figure \ref{fig:linepol-MC} shows a pronounced difference between scattering off a disk that hits the stellar surface 
(right-hand side), and one with a sizeable inner hole (left-hand side). 
The single PA flips on the left are similar to those predicted analytically \citep{1993A&A...271..492W}, but 
the double PA flips on the right -- associated with undisrupted disks -- have only been found numerically. They are thought to be unique to the appropriate geometric treatment of a finite-sized star that interacts with the velocity structure of the disk. 

The numerical models demonstrate the potential of instrinsic line polarimetry to determine not only disk inclination, but also the physical sizes of the inner regions associated with magnetospheres. 
Linear line polarimetry is so far the only method 
capable of doing this on such small spatial scales, within just a few stellar radii from the stellar surface.

Several studies linearly polarized H$\alpha$ emission have been conducted for AGB, T Tauri, and Herbig Ae/Be stars by \cite{2005MNRAS.359.1049V}, \cite{2007ApJ...667L..89H,2009ApJS..180..138H}, and \cite{2016MNRAS.461.3089A,2017MNRAS.472..854A}. These have found line polarization amplitudes of order 0.1-1\%, detectable at an unbinned S/N of $\sim200-1000$. This is comparable to the S/N required for circumstellar magnetometry, therefore the same data that will enable the measurement of circumstellar magnetic fields should also enable the detection of circumstellar line polarization effects. 

\section{Continuum linear polarimetry in the UV}\label{sec:continuum_linpol}


The presence of an obliquely rotating magnetosphere is largely responsible for the periodic modulation of the observable quantities of magnetic massive stars. As the star rotates, its own magnetosphere periodically occults the source star’s light. For hot stars with winds primarily dominated by the electron scattering opacity, the bulk of their photometric and polarimetric variability can be estimated under the single-electron scattering approximation. In this case, the photometric variability is determined by the column density, while the polarimetric variability is characterised by the general shape of the magnetosphere. 

Broadband linear polarimetry of stellar magnetospheres is observationally challenging: the typical levels of polarization are quite low, of order $10^{-4}-10^{-3}$ \citep[e.g.][, and Munoz et al. (accepted)]{munoz20,2013ApJ...766L...9C}. As a consequence, accurate monitoring of magnetospheric polarization on the relevant rotational timescales (ranging from days to years) requires understanding and eliminating competing instrumental effects. 

\begin{figure*}[t]
\centering
\includegraphics[width=7.5cm]{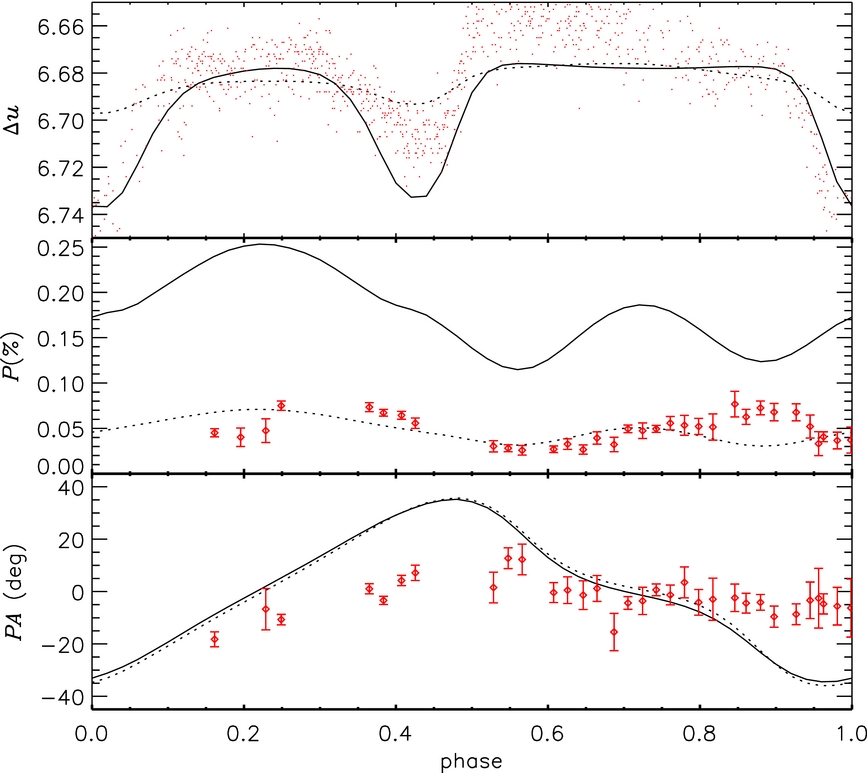}\hspace{0.5cm}
\begin{minipage}[b]{0.4\linewidth}
\centering
\includegraphics[width=7.5cm]{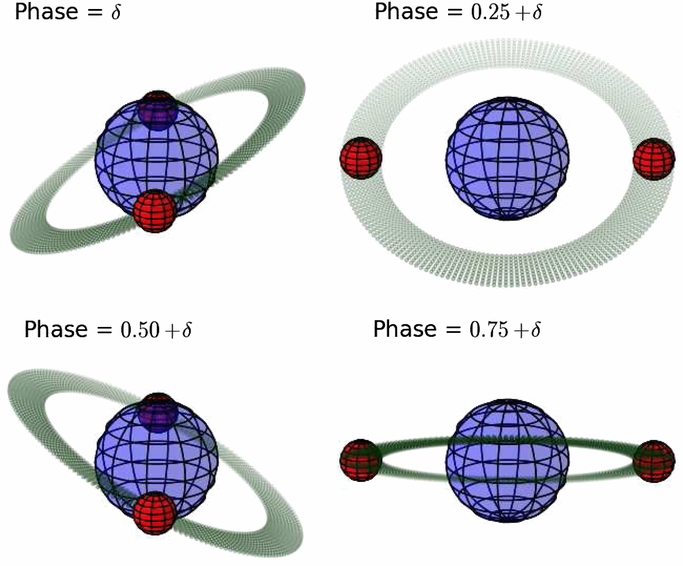}\vspace{0.4cm}
\end{minipage}
    \caption{{\em Left:\ } Modeling of the intrinsic polarization of $\sigma$ Ori E using the RRM model (observations are in red). The only free parameter is the maximum number density in the magnetosphere, which was set to $10^{12}$ cm$^{-3}$ (solid lines) to reproduce the depth of the eclipses and $2.5\times 10^{11}$ cm$^{-3}$ (dotted lines) to reproduce the amplitude of the linear polarization.  {\em Right:\ } Geometric conception of the "dumbbell + disk" model to scale. From \citet{2013ApJ...766L...9C}.}
    \label{magpol1}
\end{figure*}

Even in the optical, few studies of magnetospheric linear polarization have been carried out. The first such studies investigated the Centrifugal Magnetosphere of the archetypical He-strong Bp star $\sigma$~Ori E by \citet{1977ApJ...218..770K} and \cite{2013ApJ...766L...9C}. As shown in Fig.~\ref{magpol1}, Carciofi et al. obtained dense polarimetric coverage of the star's 1.19-day rotational cycle, measuring both the $Q$ and $U$ Stokes parameters with a typical precision of $10^{-4}$ (0.01\%, about an order of magnitude better than in the pioneering study by \citealt{1977ApJ...218..770K}). They attempted to reproduce the observed polarimetric variations by feeding the density distribution for $\sigma$~Ori E computed using the Rigidly Rotating Magnetosphere (RRM) model into a radiative transfer code. They were unable to find a model capable of simultaneously fitting both the photometry and the polarimetry, noting that a higher density model (solid line in Fig.~\ref{magpol1}) that matched the depth of the photometric eclipses predicted a polarization amplitude much larger than observed, while a lower-density model (dashed line in Fig.~\ref{magpol1}) that reproduced the amplitude of the polarization failed to reproduce the photometric amplitude. They were able to resolve these descripancies using an ad hoc ``dumbbell + disc" model with a density distribution motivated by the RRM predictions. This study serves as an excellent illustration of the power of polarization to test theoretical models of magnetospheric density and geometry.

\begin{figure*}[t]
    \centering
    \includegraphics[width=8.5cm]{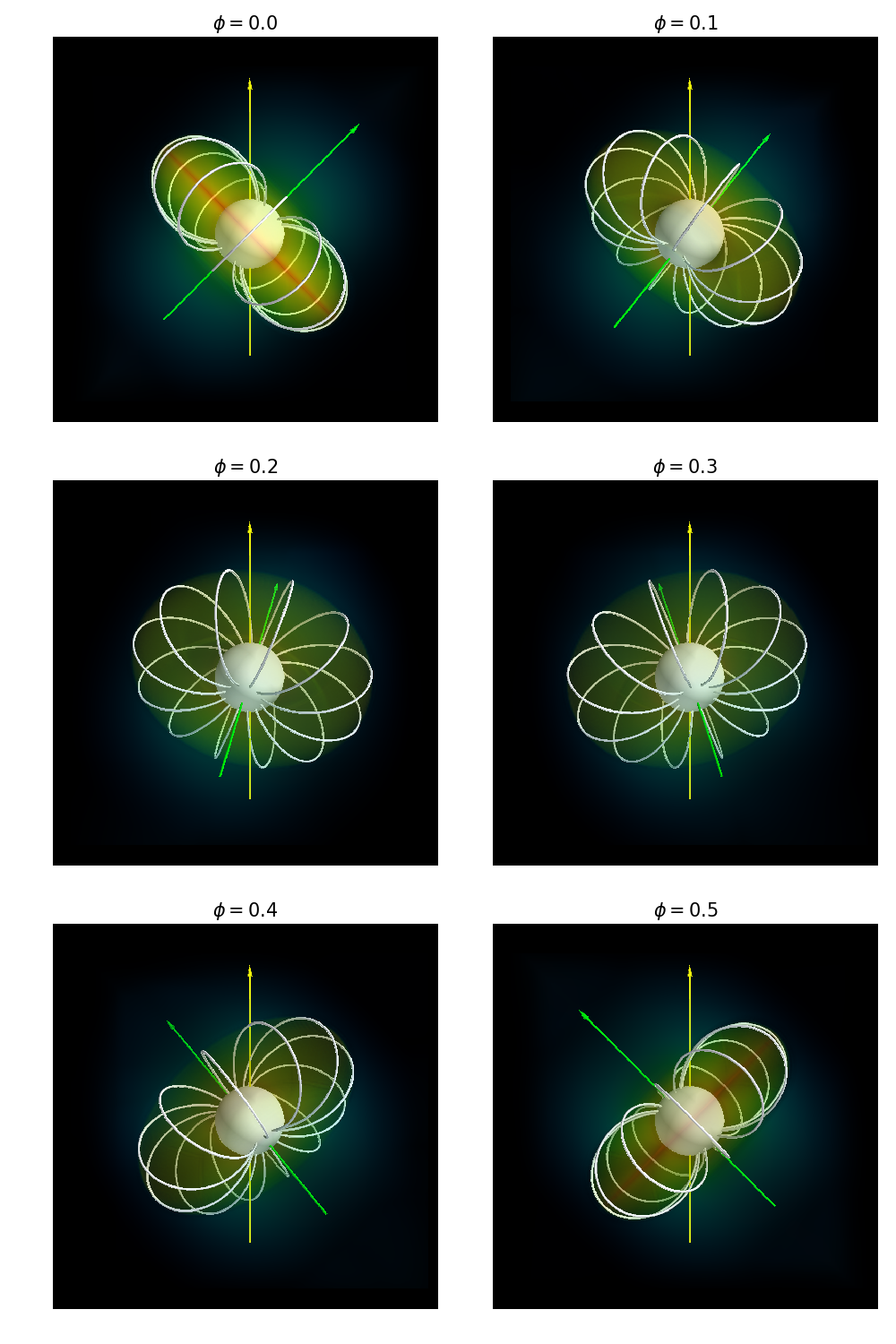}
\begin{minipage}[b]{0.4\linewidth}
\centering
\includegraphics[width=7.5cm]{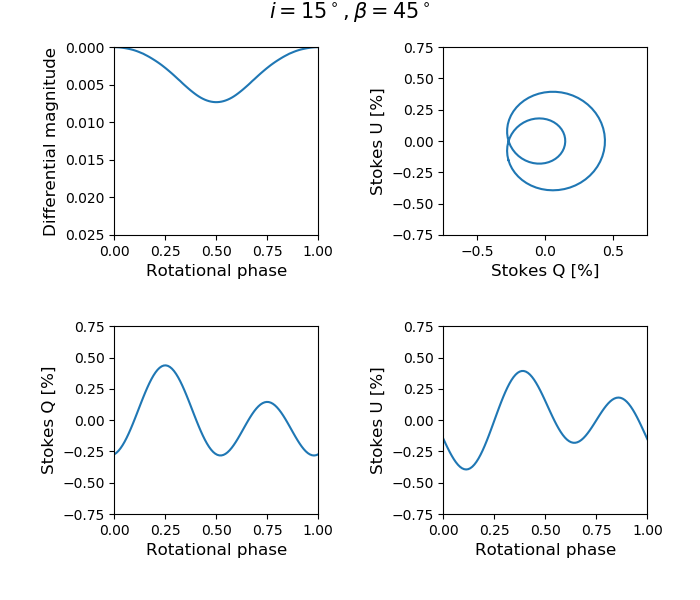}
\includegraphics[width=7.5cm]{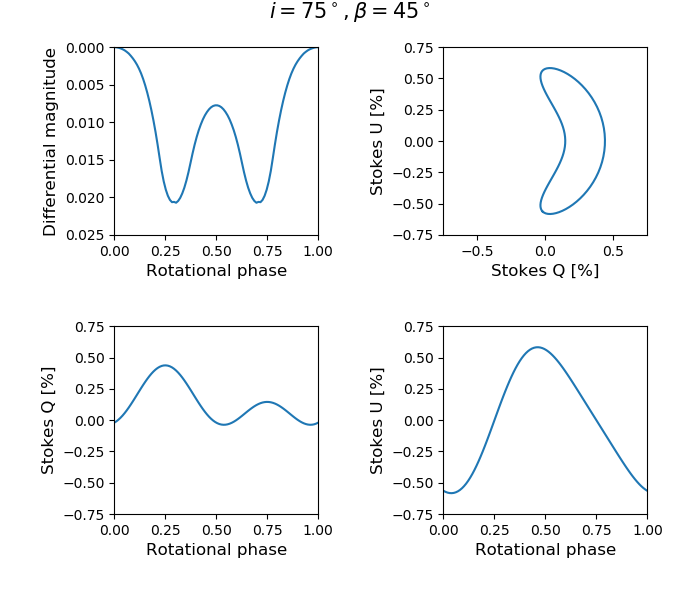}
\end{minipage}
    \caption{{\em Left:\ } Illustration of the density field of the Analytic Dynamical Magnetosphere (ADM) model at 6 rotational phases. {\em Right:\ } Photometric and polarimetric phase variations predicted for ADM models having two different geometries ($i=15\degr$, $\beta=45\degr$ (top), $i=75\degr$, $\beta=45\degr$ (bottom)). From \citet{munoz20}.}
    \label{magpol2}
\end{figure*}

More recently, \citep[][, and Munoz et al. (accepted)]{munoz20} have developed a capability to compute magnetospheric polarization in the framework of the Analytic Dynamical Magnetosphere (ADM) model (Fig.~\ref{magpol2}). Under the assumption of single electron scattering, they examined the behaviour of Stokes $Q$ and $U$ with changing magnetospheric (field strength and geometry) and stellar (mass, radius, and wind) properties. They demonstrated that linear polarization is uniquely able to disentangle the angular parameters $i$ and $\beta$ describing the magnetic geometry. They applied their model to the magnetic O star HD\,191612, obtaining a self-consistent solution to the magnetic and stellar parameters capable of simultaneously reproducing the polarimetric measurements and Hipparcos photometry.

Today, high precision broadband polarization measurements are available for only a handful of magnetic stars. The potential of Polstar to collect similar data for a significant fraction of the population of known magnetic objects is extremely exciting, as it will accurate determination of their magnetic and wind parameters, and provide a broad parameteric basis for testing of the assumptions underpinning magnetospheric models. From the models and observations that have so far been published, polarization amplitudes of the order of 0.01\% to 0.1\% are expected \citep{2013ApJ...766L...9C,munoz20}.

\section{Hanle Effect}\label{sec:hanle}

\begin{figure}
\centering
\includegraphics[width=0.8\textwidth]{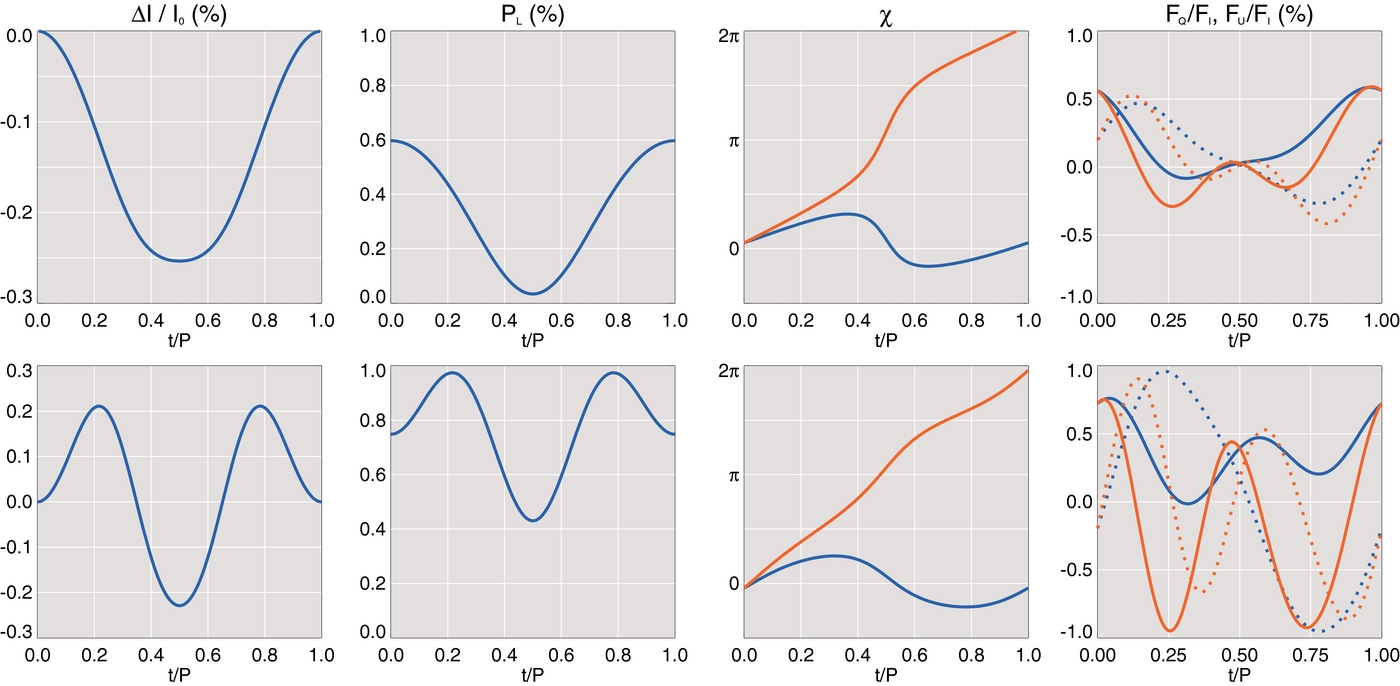}
\caption{Figure from \cite{2012ApJ...760....7M}. Cyclical variation of lines from an
unresolved oblique magnetic rotator with a dipole field showing
intensity fluctuation normalized to the intensity at $t = 0$ (1st
col), linear polarization (2nd col), polarization position angle
(3rd col); and $q$ (solid) and $u$ (dotted) variations (4th col).
Upper panels are when the combination of viewing inclination, field
obliquity, and rotation never prevents the dipole axis from ever
being viewed pole-on, with two examples (blue and red).  Lower
panels are when the dipole axis fluctuates about a pole-on view,
again with two examples (blue and red).}
\label{hanle1}
\end{figure}
 
\begin{figure}
\centering
\includegraphics[width=0.8\textwidth]{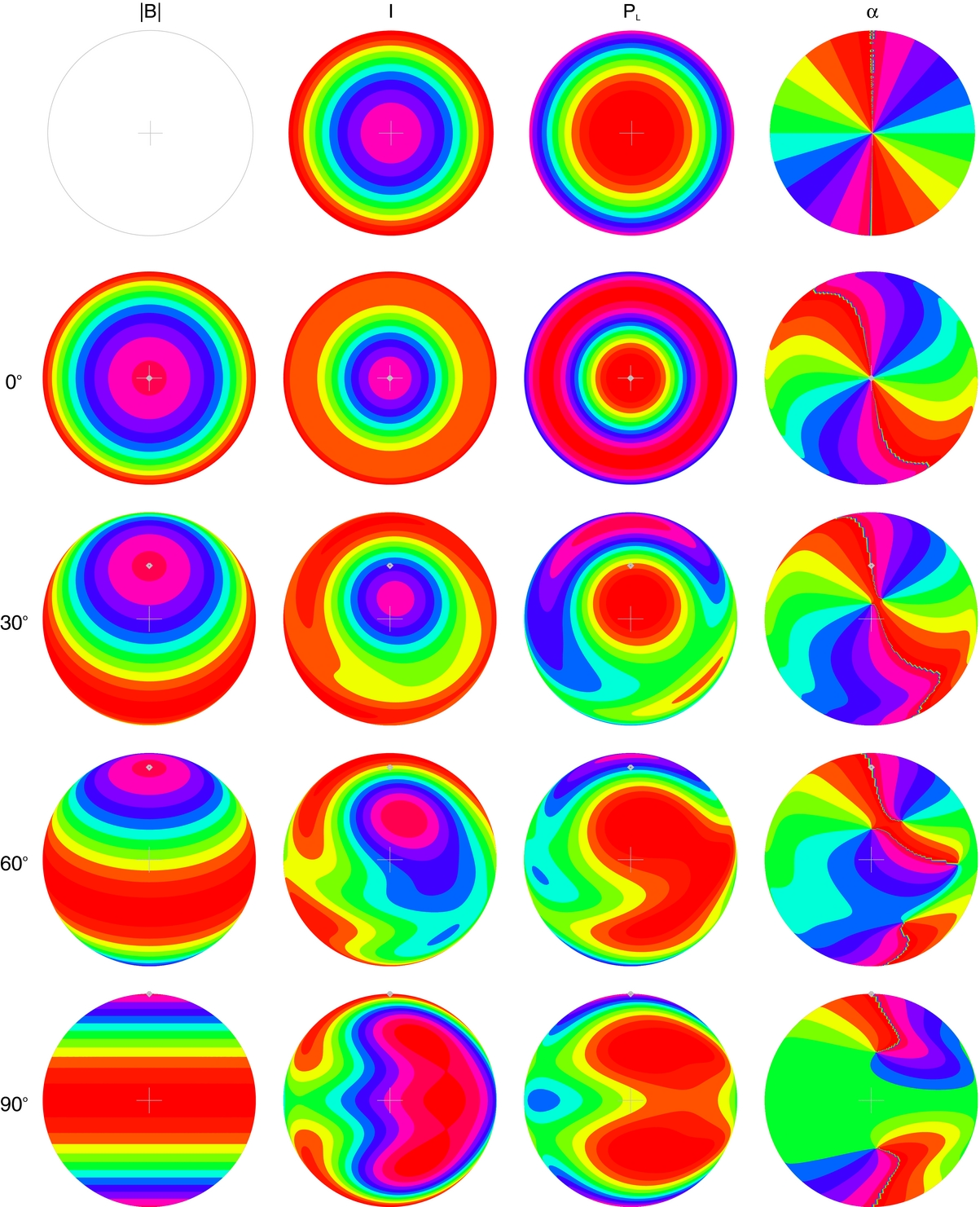}
\caption{Figure from \cite{2012ApJ...760....7M}. False color maps for the model surface distribution of the
field (1st col), line intensity (2nd col), linear polarization (3rd col),
and polarization position angle (4th col).  The top row is the case
of no magnetic field.  The next 4 rows are for viewing inclinations
as labeled, from top view ($i=0^\circ$) to edge-on view ($i=90^\circ$).
The surface field is at the Hanle field strength for these examples.}
\label{hanle2}
\end{figure}
 
\begin{figure}
\centering
\includegraphics[width=0.8\textwidth]{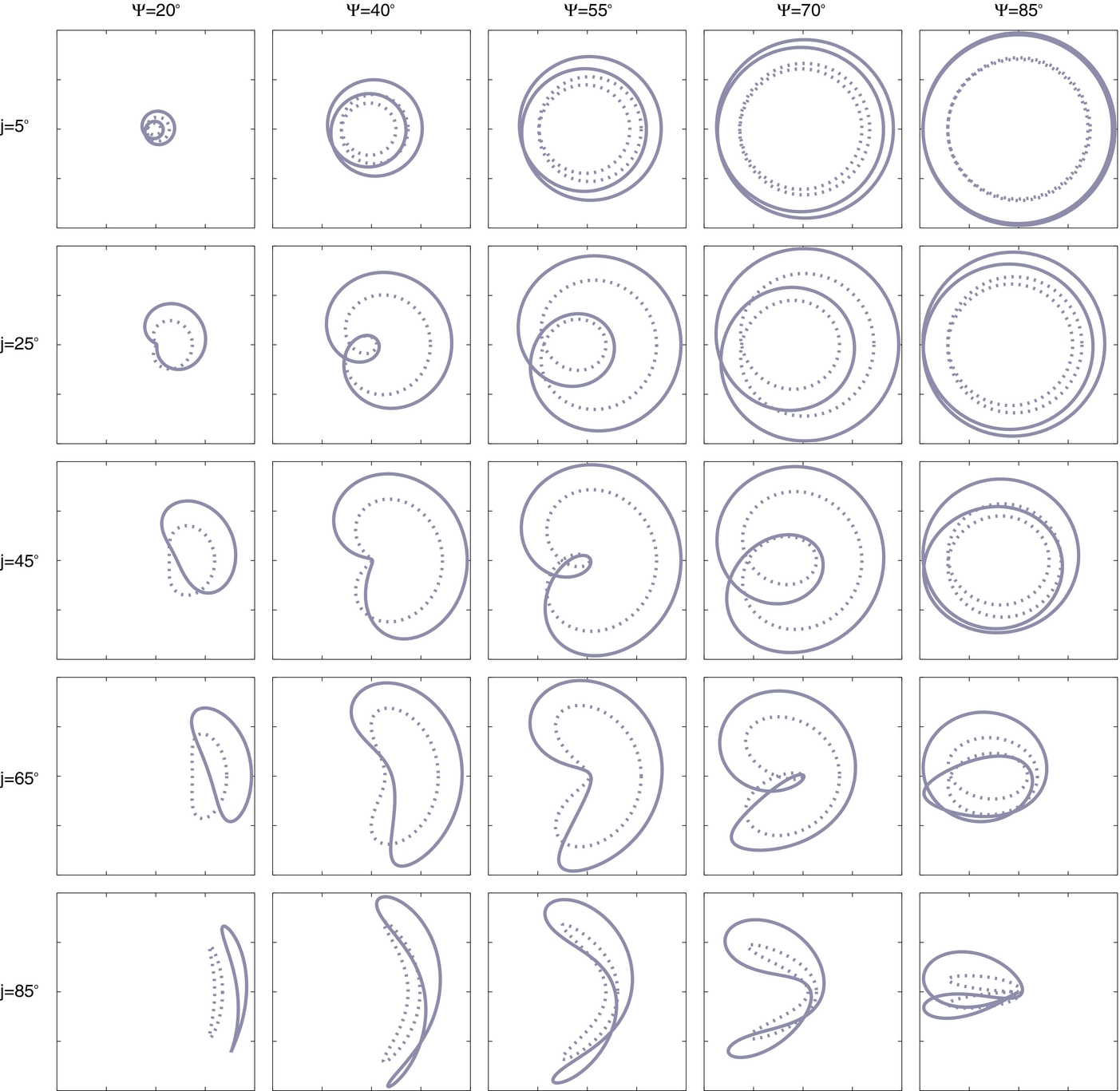}
\caption{Figure from \cite{2012ApJ...760....7M}.  Q--U loops for the total line polarization from
a photospheric line with the Hanle effect.  Solid curve is at
the Hanle field strength level; dotted is for at the saturated field
level (see text).  Using the notation of the authors, row are for different viewing inclinations ($j$)
as labeled; columns are for different dipole field obliquities ($\Psi$)
as labeled.}
\label{hanle3}
\end{figure}
 
\begin{figure}
\centering
\includegraphics[width=0.8\textwidth]{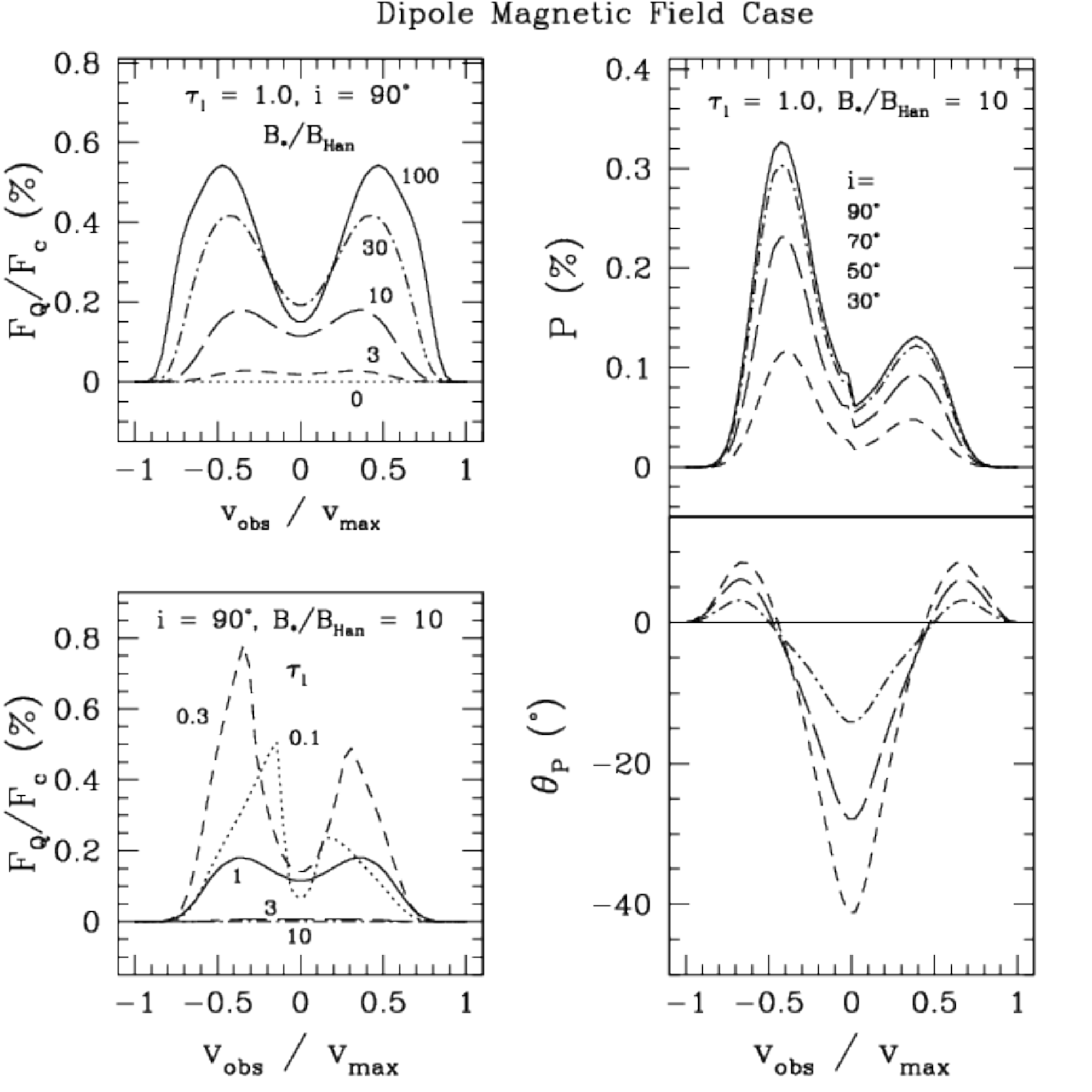}
\caption{Figure from \cite{2004ApJ...609.1018I}. Example wind line polarizations for a spherical wind
with a dipole magnetic field.  Here $v_{\rm max}$ is the wind terminal
speed, and profiles are displayed in normalized velocity shift.
Upper left:  Edge-on for different surface field strengths 
$B_\ast$ relative to
the Hanle level $B_{\rm Han}$.  Lower left:  Surface field as
labeled, also edge-on, now for different line optical depth.
Upper and lower right:  Polarized line profiles with corresponding
position angle $\theta_P$ across the line.}
\label{hanle4}
\end{figure}
 
\begin{figure}
\centering
\includegraphics[width=0.8\textwidth]{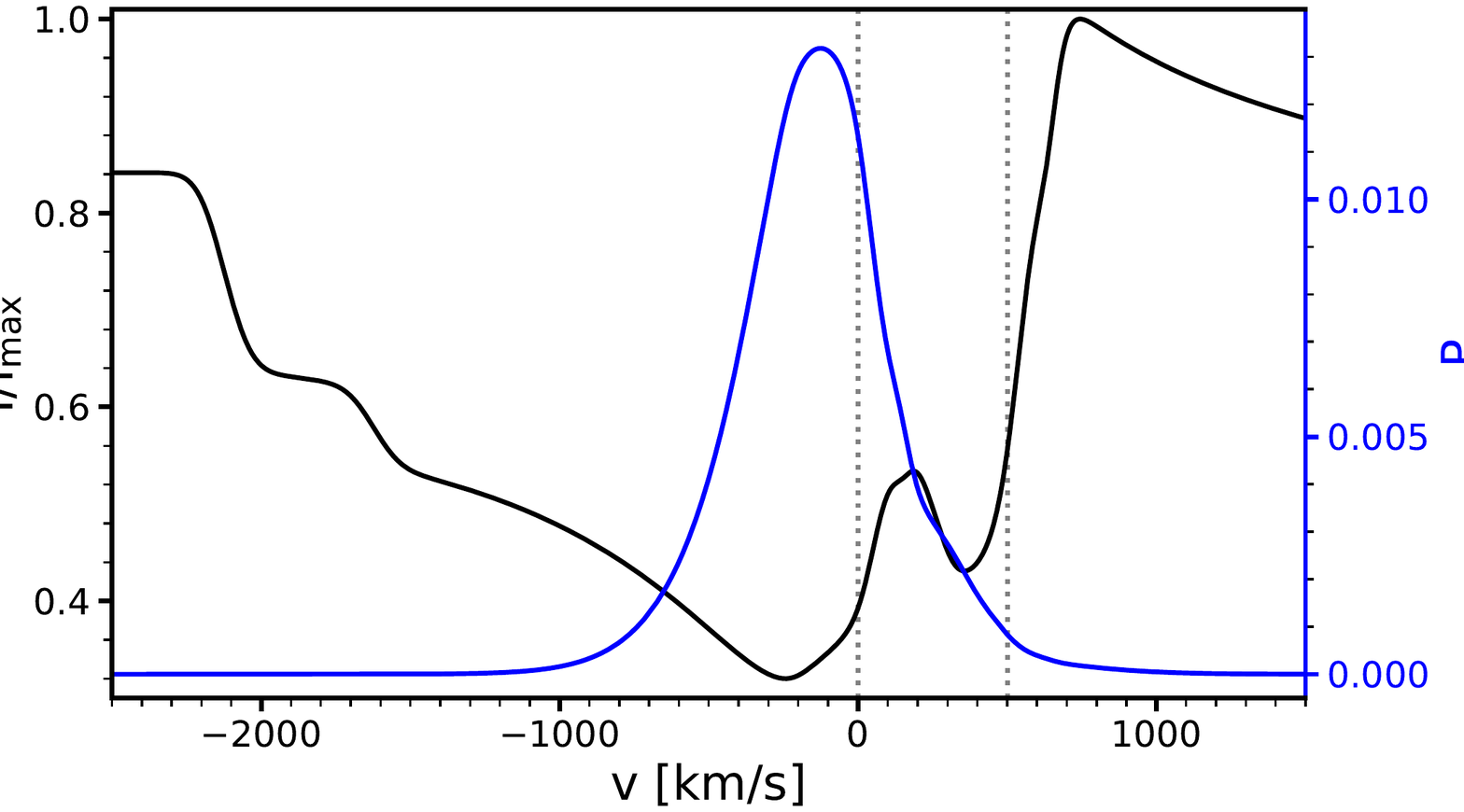}
\caption{
An illustration of the Hanle effect in the CIV doublet (154.8nm and 155.0 nm) using
HanleCLE showing the doublet absorption lines (black) and
polarization from a dipole field (blue) for a spherical wind.  Dotted vertical
lines indicate locations of the doublet component rest wavelengths (here
in relative velocities with zero for 154.8nm component).
As described in text, only the shortward component of this
Li-like doublet produces polarization.}
\label{hanle5}
\end{figure}

Polstar offers, for the first time, the opportunity to routinely 
measure magnetic fields in a star other than the Sun using the 
Hanle effect.
Whereas the better known Zeeman effect is an incoherent
process, where light polarization is manifested purely because of the 
energy splitting of the sublevels due to the action of a magnetic field, 
in the Hanle effect the main source of polarization is radiation 
anisotropy, i.e., scattering polarization \citep{1994ASSL..189.....S,2004ASSL..307.....L}.
In fact, the Hanle effect 
is the modification of scattering polarization by the action of a 
(weak) magnetic field, in a regime where the sublevels are only 
marginally separated -- at the level of the natural broadening
\citep{2002ApJ...568.1056C}.
In this regime, the sublevels interfere quantum-mechanically with
each other, leading to phase coherence effects in the polarization,
mostly, a depolarization and a rotation of the scattered linear 
polarization \citep{2008pps..book..247C}. 

Because this effect peaks when the Larmor frequency $\omega_B$ of the 
field is comparable to the Einstein A-coefficient of the transition, the 
Hanle effect is characterized by a ``critical'' field value. For a
two-level atom $(J_0,J)$, this is determined by the following condition,
\begin{equation} \label{eq:Hanle_formula}
g_J\,\omega_B \sim A_{JJ_0}\;,\qquad \omega_B=(\mu_0/\hbar)\,B\;,
\end{equation}
where $g_J$ is the Land\'e factor of the upper level $J$, and $\mu_0$ is
Bohr's magneton.

The Hanle effect has become a standard diagnostic tool of quiet-Sun magnetism \citep{1994A&A...285..655F,1998A&A...337..565B,2004Natur.430..326T,2011ApJ...743...12M,
2016ApJ...830L..24D,2018ApJ...863..164D}, and in solar structures such as prominences and filaments, where radiation processes are dominated by scattering \citep{1977A&A....60...79L,1994SoPh..154..231B,2002Natur.415..403T,2003ApJ...598L..67C,2014A&A...566A..46O,2015ApJ...802....3M,2021A&A...647A..60B}. 
The Hanle effect in the FUV (e.g., \ion{H}{1} Lyman $\alpha$, \ion{O}{6} 103.2\,nm doublet) also has a particular interest for the plasma and magnetic diagnostics of the solar corona and wind \citep{1982SoPh...78..157B,1991OptEn..30.1161F,2002A&A...396.1019R,2011A&A...529A..12K,2019ApJ...883...55Z,2021ApJ...912..141Z}.

Applications to stars other than the Sun have been little addressed.
\cite{2011A&A...527A.120L}, \cite{2011A&A...530A..82I}, \cite{2012A&A...539A.122B}, and \cite{2012ApJ...760....7M} have
explored the influence of the Hanle effect in spectropolarimetry
of the photospheric lines of unresolved stellar atmospheres.  In
particular, \cite{2012ApJ...760....7M} calculated the total line polarization in Stokes
$Q$ and $U$ for stars with either dipole or quadrupole magnetic
fields.  Figure~\ref{hanle1} shows an example for the variable
intensity and polarization in the line-integrated
flux as functions of rotational phase.  An interesting property of
the Hanle effect is a net change to the polarization for the line
as a whole.  Indeed, the models of Figure~\ref{hanle1} assume a
spherically symmetric star for which no net line polarization would
result in the absence of magnetism. 

Those same authors produced example polarization maps
across the projected star, as shown in
Figure~\ref{hanle2}.  For distant stars that are not
spatially resolved, these snapshots at a particular rotational
phase produce a single datum in the lightcurves
of Figure~\ref{hanle1}.  Rotation of the star generally leads
to a changing map leading to observable lightcurve variations.

Another common tool of analysis for variable linear polarization
are $Q$-$U$ diagrams.  Figure~\ref{hanle3} shows such $Q$-$U$ ``loops''
in the case of a dipole magnetic field for a range of viewing
inclinations (listed along the left side; the authors use $j$ for inclination)
and dipole obliquities (listed along the top; the authors use $\Psi$ for obliquity). Note that the two
curves are for two different field strengths.  Solid is at the so-called
``Hanle field'' or ``critical field'', $B_{\rm Han}$, conveniently defined roughly
at when the Larmor frequency equals the A-value of the transition
(modulo the Land\'e factor; c.f., eq.[\ref{eq:Hanle_formula}]).
Dotted curves are for the so-called ``saturated limit''.  This is
when  $\omega_B \gg A_{JJ_0}$.  In this limit the Hanle diagnostic
becomes insensitive to field strength, aside from the fact that
$B \gg B_{\rm Han}$, but remains sensitive to orientation of the 
vector magnetic field.

The Hanle effect has also been explored in the context of circumstellar
media.  \cite{1997ApJ...486..550I} and \cite{1999ApJ...520..335I}
used simplifying assumptions to explore the
potential of the Hanle effect for tracing the magnetic field in
stellar winds.  Using a last scattering approximation combined
with the concept of the Sobolev optical depth, \cite{2004ApJ...609.1018I}
considered somewhat more realistic magnetized wind models.
An example in Figure~\ref{hanle4} shows a spherically symmetric
wind with a dipole field, showing the polarization across wind lines.
Similar to \cite{2012ApJ...760....7M}
for the case of photospheric lines, spherical symmetry
of the wind was adopted to highlight how the Hanle effect can produce
net line polarizations.

Of chief importance for observations with {\em Polstar} is that the
Hanle effect applies to resonance line scattering, and many resonance
lines of massive stars are located in the UV waveband.  To first
order, resonance line scattering produces linearly polarized light
according to dipole scattering.  In this respect resonance line
scattering is qualitatively similar to electron scattering, but
with a relative efficiency determined by the particular atomic
transition \citep{1960ratr.book.....C}.  Indeed, some resonance
lines scatter isotropically, producing no polarized light from
scattering, and therefore they cannot subject to the Hanle effect.
Even without magnetism or the Hanle effect, resonance line scattering
can contribute to line polarization of circumstellar media and the
shaping of wind emission profiles; examples of such effects have
been explored by
\cite{1998A&A...332..686I,1998A&A...337..819I,2000A&A...363.1106I}.

Using the Hanle-RT code of \cite{2016ApJ...830L..24D}, Figure~\ref{hanle5} shows an example of
a detailed radiative transfer calculation for the Hanle effect in
the \ion{C}{4} 155\,nm doublet.  The two doublet components are indicated
by the vertical lines for their rest wavelengths (i.e., in relative velocity shift).
The calculation is for an O~star at 30,000~K with wind terminal speed 2450 km/s
and a typical $\beta=1$ wind velocity law.
For illustration purposes the wind is treated as spherically symmetric to
highlight the Hanle effect.  A dipole field inclined by $30^\circ$ toward
the observer is assumed, with polar field strength of 90~G.

This figure highlights ways in which the
Hanle-effect diagnostic enabled by {\em Polstar} will be used to achieve 
the science objective to map the magnetospheres of massive stars and
how they limit the escape of mass from wind driving.  There are 3
main considerations for use of the Hanle effect as a magnetic
diagnostic of massive stars: (1) the Hanle effect is typically
senstive to weaker magnetic fields when compared to the Zeeman
effect, (2) specifically for the commonly strong Li-like resonance
doublets for massive stars, one of the line components is insensitive
to the Hanle effect (unpolarizable upper level), and (3) the Hanle 
effect has different responses for different lines, and so a multi-line 
approach is beneficial and generally employed (REFS about the
differential Hanle effect, Stenflo). \emph{The large presence of
resonance lines in the UV particularly favors this aspect of the Hanle
diagnostic of stellar environments.}

Regarding item \#1, the Zeeman effect is certainly sensitive to
any field strength. However, its detectability depends in large part 
on how the Zeeman splitting compares to the line broadening,
whether thermal, rotational, or from wind velocity distribution. In
particular, indicating with $\Delta\omega$ the line width in frequency
units, the longitudinal Zeeman effect (circular polarization Stokes $V$)
is first order in $v=\omega_B/\Delta\omega$, and the transverse effect 
(linear polarization Stokes $Q$ and $U$) is second order in $v$. A
direct consequence of this is that the Zeeman effect suffers from low
poor spectral resolution during observations, as $v\to 0$ in that case.

Because in the Zeeman effect the polarization is generated by
the magnetic field, and the direction of polarization is directly
related to the magnetic field direction along the line-of-sight, the polarization 
of unresolved targets tends to cancel out in Zeeman diagnostics, if the 
magnetic field vector takes all possible directions within
a spatial resolution element. By contrast, the Hanle effect is not 
affected by such limitation, as scattering polarization can be
produced also in a magnetic free environment. The Hanle effect 
modifies such a scattering polarization in a characteristic way
depending on the topology and strength of the magnetic field, but 
such that a complete cancellation is practically never attained. For
example, in the presence of an unresolved, macroscopically turbulent
field, for which the Zeeman effect polarization vanishes identically,
the Hanle effect would lead to a reduction of polarization to 20\% with
respect to the magnetic free case. 

For the UV resonance lines of interest, the field sensitivities to
the Hanle effect are roughly in the range of 1--100\,G (see
Tab.~\ref{hantable}).  For many of the massive stars observed by
{\em Polstar}, the Zeeman effect will still be employed to infer
the magnetism at the photosphere (see \S~XX), and in limited cases
the circumstellar field near the photosphere (see \S~XY). But these
sources will often have kG-level fields.  Thus the Hanle effect
will be employed for stars with lower surface fields (see
Tab.~\ref{hantargs}), or as a tool for mapping the magnetic fields
relatively far from the photosphere, in the wind acceleration zone
or in the wind-confining magnetospheric lobes.

\begin{table}
\label{hantable}
\begin{center}
\caption{Sample FUV Lines for Hanle
Effect} \begin{tabular}{cccccc}
\hline\hline Ion & $\lambda$ & $A$ & $g_u$ &
$g_{\rm eff}$ & $B_H$ \\
    &   (nm)    & $10^8$ s$^{-1}$ &   &   &  (G) \\ \hline
HI      &121.57        &6.26   &1.33 &1      &53.5 \\ 
HeII &164.03        &3.59   &1.33 &1      &30.7 \\ 
CII &133.45 &2.41   &0.8    &0.83 &34.3 \\ 
CIV     &154.82        &2.65 &1.33 &1.17 &22.6 \\ 
NV      &123.88        &3.40   &1.33 &1.17 &29.0 \\ 
MgII    &123.99        &0.0135 &1.33 &1.17 &0.115 \\ 
SiII    &126.04        &25.7   &0.8    &0.83 &365 \\
SiII    &180.80        &0.0254 &0.8    &0.83 &0.361 \\ 
SiIV &139.38        &8.80   &1.33 &1.17 &75.1 \\ 
SII     &125.38 &0.512  &1.73 &1.87 &3.36 \\
SII     &125.95        &0.510 &1.6    &1.3    &3.63 \\ \hline \end{tabular}
\end{center}
\end{table}

Regarding item \#2, the longer wavelength component of the doublet
(upper level $J=1/2$) scatters isotropically to produce no polarization,
and therefore no Hanle effect. The shorter wavelength component 
(upper level $J=3/2$) scatters with a 50\% dipole-like efficiency 
(i.e., half dipole-like and half isotropic).
This is seen in Figure~\ref{hanle5}, where in Stokes $I$ a wind line
is seen at each of the doublet components, but only one displays 
polarization.
The diagnostic value of this situation is that both of the doublet
components of \ion{C}{4} form in the same spatial zones (one having half
the optical depth of the other). Consequently, any polarization
at the wavelength of the non-polarized component must arise from
the source continuum polarization or from interstellar polarization,
or both.  Such a polarization serves as an internal
calibration that can be
applied to the other doublet component to infer the polarization
arising strictly from resonance scattering and the Hanle effct.

Regarding item \#3, it is difficult to know {\em a priori}
whether the resonance line polarization is influenced by the
Hanle effect, because of the need to know first the polarization 
the line would produce in the absence of a magnetic field. \emph{The
fact that the interpretation of the Hanle effect relies on the
modeling of such a zero-field polarization is arguably the main 
limitation of this diagnostic.} On the other hand, there
are multiple viable resonance lines (c.f., Tab.~\ref{hantable})
available for analysis, each with different Hanle sensitivites.
As \cite{1997ApJ...486..550I} have discussed, it then becomes possible to disentangle the 
Hanle effect from non-magnetic resonance scattering polarization 
through analysis of the differing responses of the different lines.  
In this way, the 3D magnetic field in the region of line formation 
can be reconstructed, or at least fit with models.

As described already, not only will {\em Polstar} provide exquisite
spectropolarimetric capability at UV wavelengths, but the mission
emphasizes the importance of time-series observations.  When direct
imaging is not an option for these distant stars, the only viable path
for mapping the asymmetric 3D distribution of magnetic, density,
and velocity structures about the stars is by seeing the system
from varying perspectives.  This naturally arises through stellar
rotation, but requires numerous visit to the same targets.  It is
this principle that will be employed for extracting a picture of
circumstellar structure from the spectropolarimetric data at different
rotational phases.  In the case of the Hanle effect, this makes use
of the high levels of line broadening, well beyond thermal broadening,
owing to rotation ($\sim 10^2$ km/s) and/or wind flow ($\sim 10^3$
km/s).

One aspect of defining the fractions of escaping and confined
winds for magnetic massive stars is mapping the circumstellar
magnetic field.  The Hanle effect serves to complement other
methodologies that will be adopted in the analysis of
time-series spectropolarimetric data provided by {\em Polstar},
which include variable P-Cygni line behavior (\S XX) and
the Zeeman effect (\S XY).  

{\em Polstar} will for the first time allow for routine measurement
and diagnostic use of the Hanle effect in other stars, representing
a fruitful collaboration between heliophysics and stellar astrophysics.
Consequently, use of the Hanle effect for massive star studies
involves initially a limited selection of targets, as given in
Table~\ref{hantargs}.

\begin{table}
\label{hantargs}
\begin{center}
\caption{Targets for Hanle Studies}
\begin{tabular}{ccccc}
\hline\hline Name &	$\eta_\ast$    & $v_\infty$ & $v_{\rm rot}$ & Rationale \\
     &          & (km/s)  & (km/s)  & \\ \hline
$\beta$ CMa        &13.5    &1885    &24      &good confinement \\
$\epsilon$  CMa        &14.5    &1720    &21      &good confinement \\
$\zeta$ Ori        &0.20    &2000    &127+    &weak confinement \\
$\tau$  Sco        &7.9     &2540    &8       &complex field \\
$\zeta$ Cas        &430     &1880    &58      &strong confinement \\
HD 64740        &2510    &---     &---     &strong confinement \\ \hline
\end{tabular}
\end{center}
\end{table}

Our selection of targets for the Hanle effect spans a range of
$\eta_\ast$ values.  The selection includes $\zeta$~Ori as an example
of $\eta_\ast \ll 1$ -- a powerful wind threaded by a magnetic field
that is drawn out with the wind to test the low $\eta_\ast$ limit
as described in \cite{udDoula2002}.  We have $\tau$~Sco
with modest $\eta_\ast$ but complex magnetic topology
\citep{2006MNRAS.370..629D}.  There are nearly twin early B-type
stars in $\beta$ CMa and $\epsilon$ CMa; not only are the stars very
similar, they also have very similar intermediate values of
$\eta_\ast$.  This means good confinement of the wind.  Then there
are two stars with extreme $\eta_\ast$ values in $\zeta$ Cas at
100's and HD64740 at 1000's.  These are cases where Hanle can be
tested for probing trapped wind plasma at large radius.  We will
use the Hanle effect to directly measure circumstellar magnetism
ranging from wind-dominated cases ($\eta_\ast\ll 1$) to wind-confined
scenarios and dynamical magnetospheres ($\eta_\ast \sim 10$) to
highly confined large and centrifugal magnetospheres ($\eta_\ast
\gg 1$).  Not only can we measure circumstellar winds, but we can
also test model predictions for the Hanle effect in photospheric
lines.

In order to achieve these science goals with {\em Polstar}, we
require spectral resolving powers of $R=30,000$ for photospheric
line studies owing to modest rotation speeds, and slighly less
resolution at $R\sim 10,000$ for wind lines.  Both $\beta$ CMa and
$\epsilon$ CMa are key targets as near twins and suitable for both
photospheric and circumstellar field mapping.  The former provides
proof-of-concept against known solutions of the surface magnetism
from the Zeeman effect in optical lines.  To verify our solutions,
it would be useful to obtain 2 complete rotations of $\beta$ CMa
and $\epsilon$ CMA.  In the wind-dominated case of $\zeta$~Ori, we
expect to obtain 4 rotation periods to understand how magnetism
influences discrete absorption components and structured wind flow
when the field is relatively weak compared to the wind.

\section{Sample description}\label{sec:sample}

   \begin{figure}[t]
   \centering
   \includegraphics[width=0.6\textwidth]{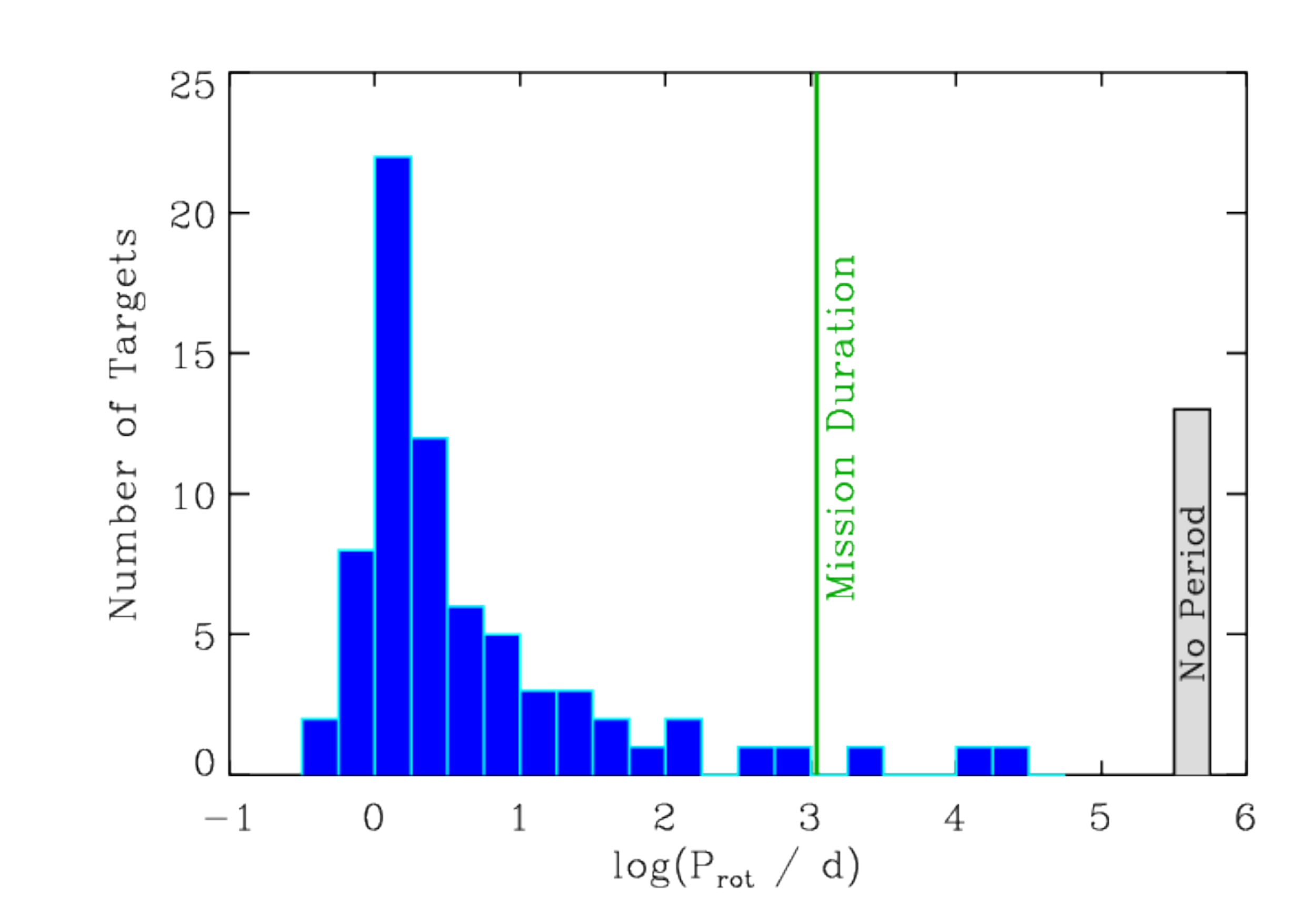}
      \caption{Distribution of rotational periods.}
         \label{prot_hist}
   \end{figure}

The initial target list consists of those OB stars for which magnetic fields have been detected, comprising 84 stars in total, with most of the list having been drawn from the O-type stars listed by \cite{petit2013} and the B-type stars examined by \cite{2018MNRAS.475.5144S,2019MNRAS.490..274S,2020MNRAS.499.5379S}. For the majority of these stars, rotational periods are known (see Fig.\ \ref{prot_hist}) and magnetic oblique rotator models are well characterized, with the 13 exceptions being either stars exhibiting no variability, or new discoveries for which sufficient followup data has not yet been obtained. Only 3 stars have periods longer than the 3-year mission duration; while rotational phase coverage cannot be completed for these targets, they are ideal for exploration of alternate science goals (e.g. examining intrinsic rather than rotationally modulated magnetospheric variability). 

   \begin{figure}[t]
   \centering
   \includegraphics[width=0.95\textwidth]{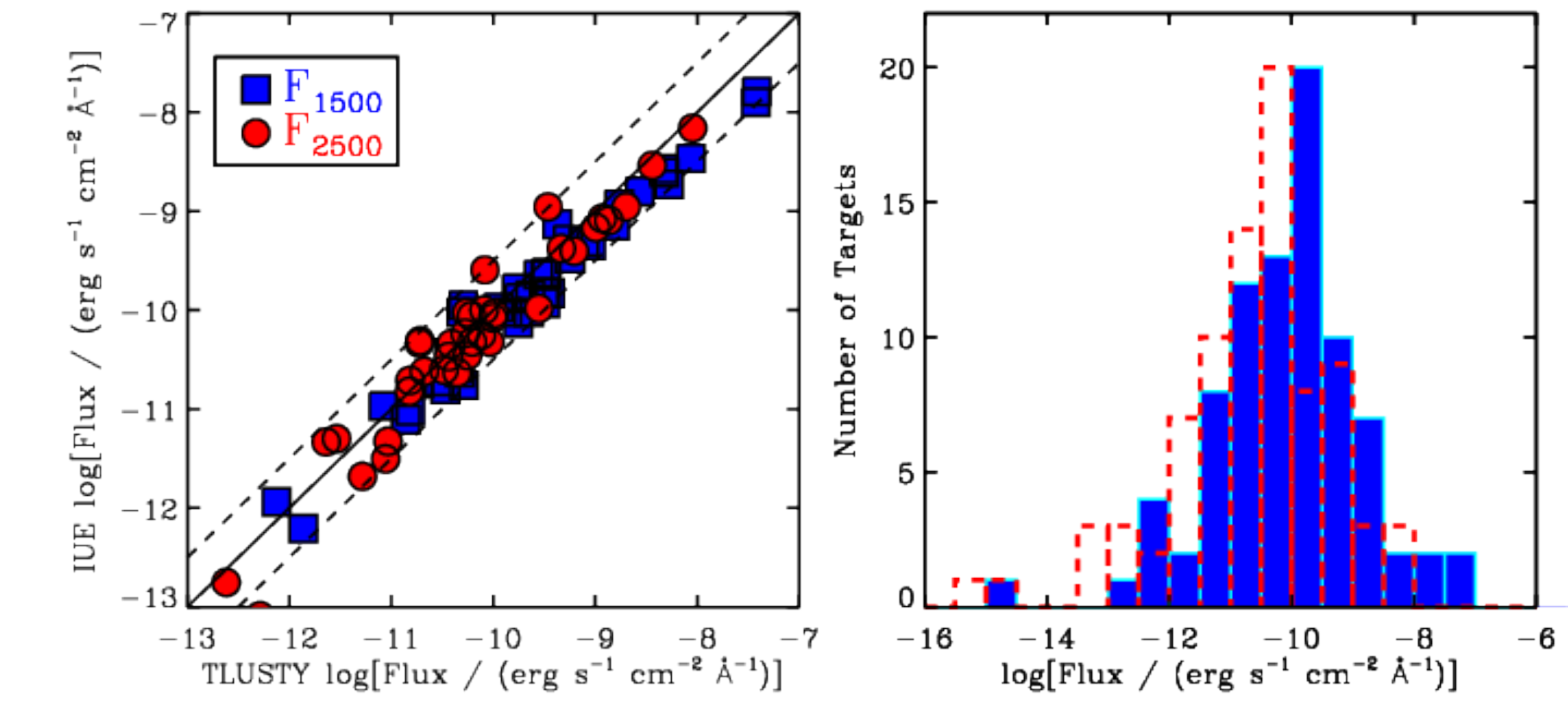}
      \caption{{\em Left}: fluxes measured at 1500 \AA~and 2500 \AA~as a function of fluxes predicted from TLUSTY spectra; the solid and dashed lines show $x=y$ and the approximate scatter. {\em Right}: histograms for 1500 and 2500 \AA~fluxes for the full sample, combining both IUE fluxes (where available) and TLUSTY fluxes (where not).}
         \label{f1500_f2500}
   \end{figure}

To evaluate the signal-to-noise ($S/N$) that can be achieved for a given target, IUE spectra were acquired from the IUE archive. Where possible low-resolution spectra were utilized, as these more accurately preserve the true flux level than the high-resolution IUE data; otherwise high-resolution data were used. For each spectrum, the mean flux was calculated at 1500 \AA~and 2500~\AA, as the approximate middle of the spectral ranges of Channels 1 and 2 respectively, with windows of $\pm 25$~\AA. When multiple spectra were available for a given target, the mean was calculated after discarding 3$\sigma$ outliers. 

Since IUE data are not available for all targets, synthetic spectra calculated using non-Local Thermodynamic Equilibrium (NLTE) TLUSTY models were also utilized in order to estimate the flux \citep{2003ApJS..146..417L,2007ApJS..169...83L}. These were adjusted according to the radius of the star, the star's {\em Gaia} parallax distance, and the reddening inferred for the star's position on the sky and heliocentric distance using the tomographic STILISM three-dimensional tomographic dust map \citep{2014A&A...561A..91L,2017A&A...606A..65C}. The left panel of Fig.\ \ref{f1500_f2500} compares the IUE and TLUSTY fluxes, demonstrating a generally good agreement. The right panel shows histograms of the fluxes. When IUE fluxes are available, these are used; when not, we use TLUSTY fluxes.

   \begin{figure}[t]
   \centering
   \includegraphics[width=0.6\textwidth]{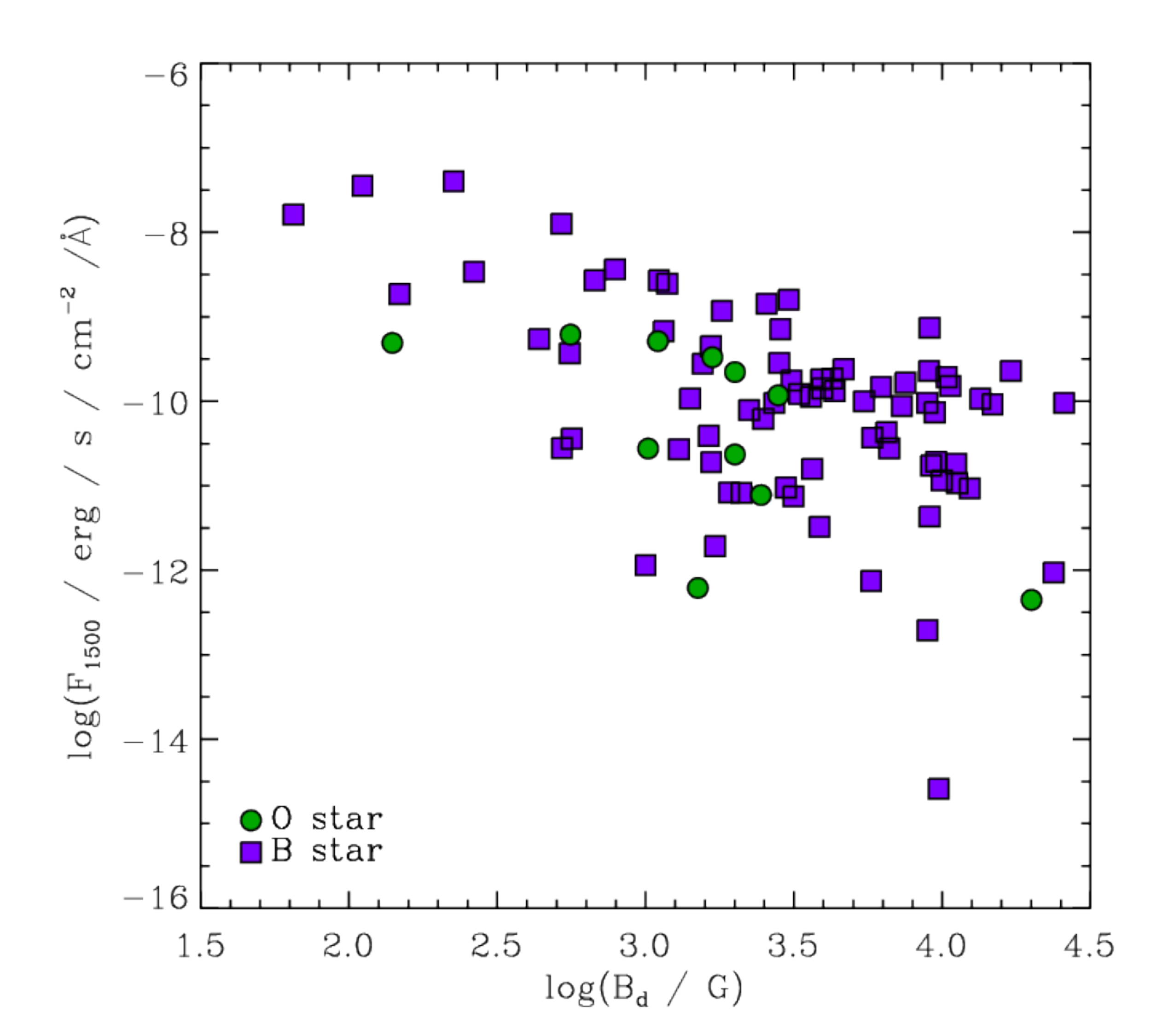}
      \caption{$F_{1500}$ as a function of $B_{\rm d}$.}
         \label{bd_f1500}
   \end{figure}

Fig.\ \ref{bd_f1500} shows an important property of the sample that guides the observation strategy: there is a close relationship between $F_{\rm 1500}$ and the measured values of the surface dipole strength $B_{\rm d}$, such that the weakest fields are found in stars with the highest flux, whereas stars with lower flux have systematically stronger magnetic fields. This is a straightforward consequence of observational bias. Weak fields are intrinsically difficult to detect and therefore have only been measured in very bright targets. Conversely, the absence of very bright stars with extremely strong ($\sim$10 kG) magnetic fields is a result of their rarity. Since a stronger magnetic field can be measured with a lower $S/N$, rather than aiming for a uniform $S/N$, we adopt a uniform 1-hour exposure time, which as will be demonstrated below results in the detectability of circumstellar magnetic fields in the majority of the sample. 
   
The $S/N$ that can be achieved for a given target in a 3600 sec spectropolarimetric sequence was calculated according to:

\begin{equation}\label{snr}
S/N = \sqrt{\frac{\dot{S}_{\rm p}^2\Delta t^2}{\dot{S}_{\rm p}\Delta t + 216 N_{\rm pix}(\Delta t / 3600 {\rm s}) + 175N_{\rm pix}}},
\end{equation}

\noindent where $\dot{S}_{\rm p}$ is the photon count rate, $N_{\rm pix}$ the number of pixels, and $\Delta t$ is the total exposure time for all 6 sub-exposures. The second and third terms in the denominator of Eqn.\ \ref{snr} originate from the dark count rate and the read noise. $N_{\rm pix} = 2 \times 2$ for Channel 1 and $2.5 \times 2.5$ for Channel 2. The photon count rate was estimated using 

\begin{equation}\label{spdot}
\dot{S}_{\rm p} = \frac{f_\lambda}{g_\lambda},
\end{equation}

\noindent where $f_\lambda$ is the flux at a given wavelength in units of ${\rm erg~s^{-1}~cm^{-2}}$~\AA$^{-1}$ and $g_\lambda$ is a factor with units of ${\rm erg~cm^{-2}}$~\AA$^{-1}$~given by

\begin{equation}
g_\lambda = \frac{\lambda}{R(\lambda)}\frac{\lambda}{hc}A_{\rm eff}(\lambda),
\end{equation}

\noindent where $R$ is spectral resolution, $h$ and $c$ are Planck's constant and the speed and light, and $A_{\rm eff}$ is the wavelength-dependent effective area. 

   \begin{figure}[t]
   \centering
   \includegraphics[width=0.6\textwidth]{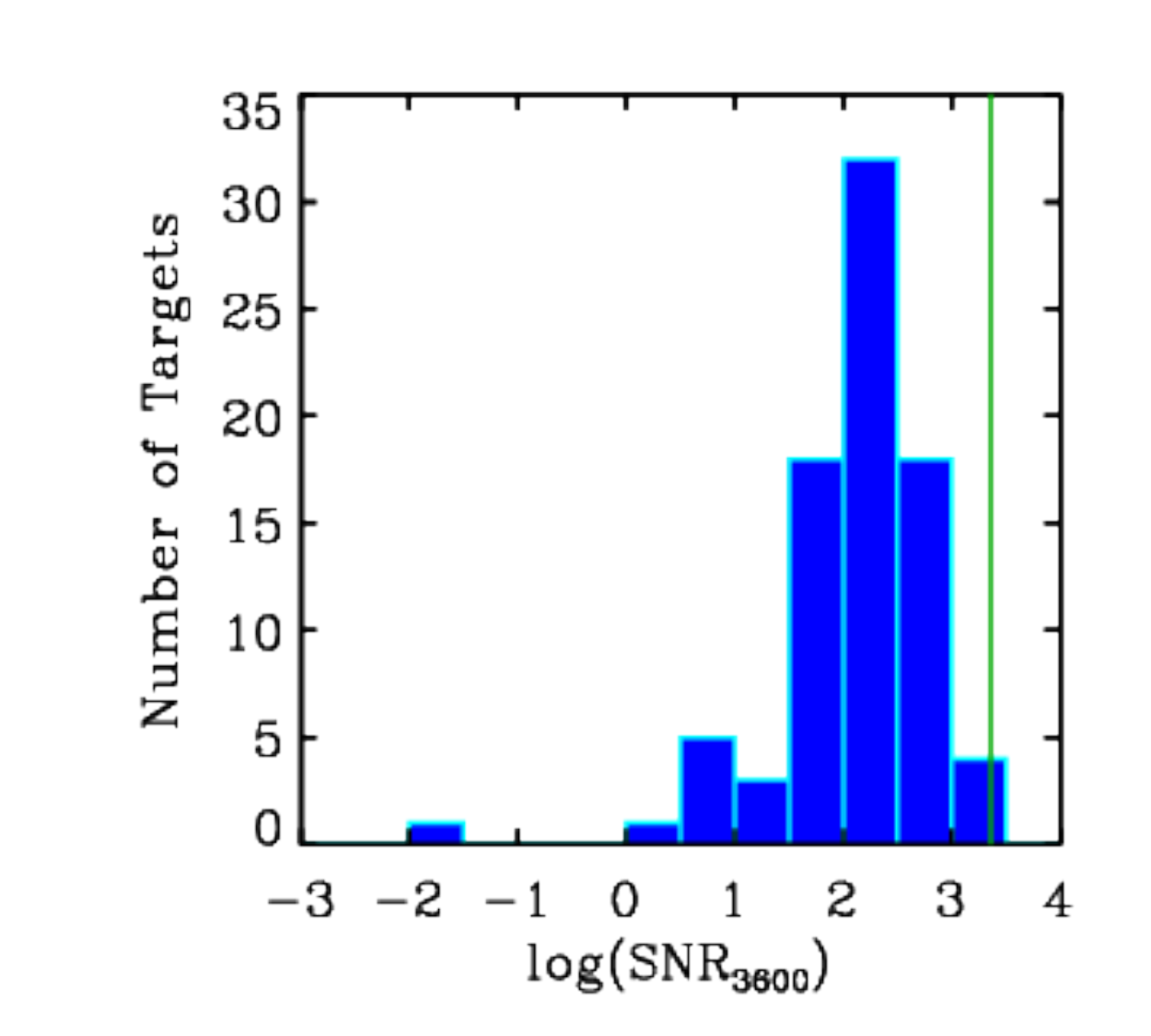}
      \caption{$S/N$ for a 3600 s Channel 1 spectropolarimetric sequence. The green line shows the saturation $S/N$.}
         \label{snr3600}
   \end{figure}

The $S/N$ that can be achieved using a 3600 s spectropolarimetric sequence in Channel 1 is shown in Fig.\ \ref{snr3600}. The median $S/N$ is 192. As explained below, we expect a $S/N$ of 100 to be the approximate lower bound for surface magnetometry, while a $S/N$ of 10 is sufficient for spectroscopy alone (this being a typical value for IUE spectra). Only 7 stars are below the spectroscopic threshold. If 10 3600 s observations are obtained for each of the 77 targets for which at least a $S/N$ of 10 can be achieved, the total observing time necessary to complete Channel 1 coverage is 770 hours. 

   \begin{figure}[t]
   \centering
   \includegraphics[width=0.95\textwidth]{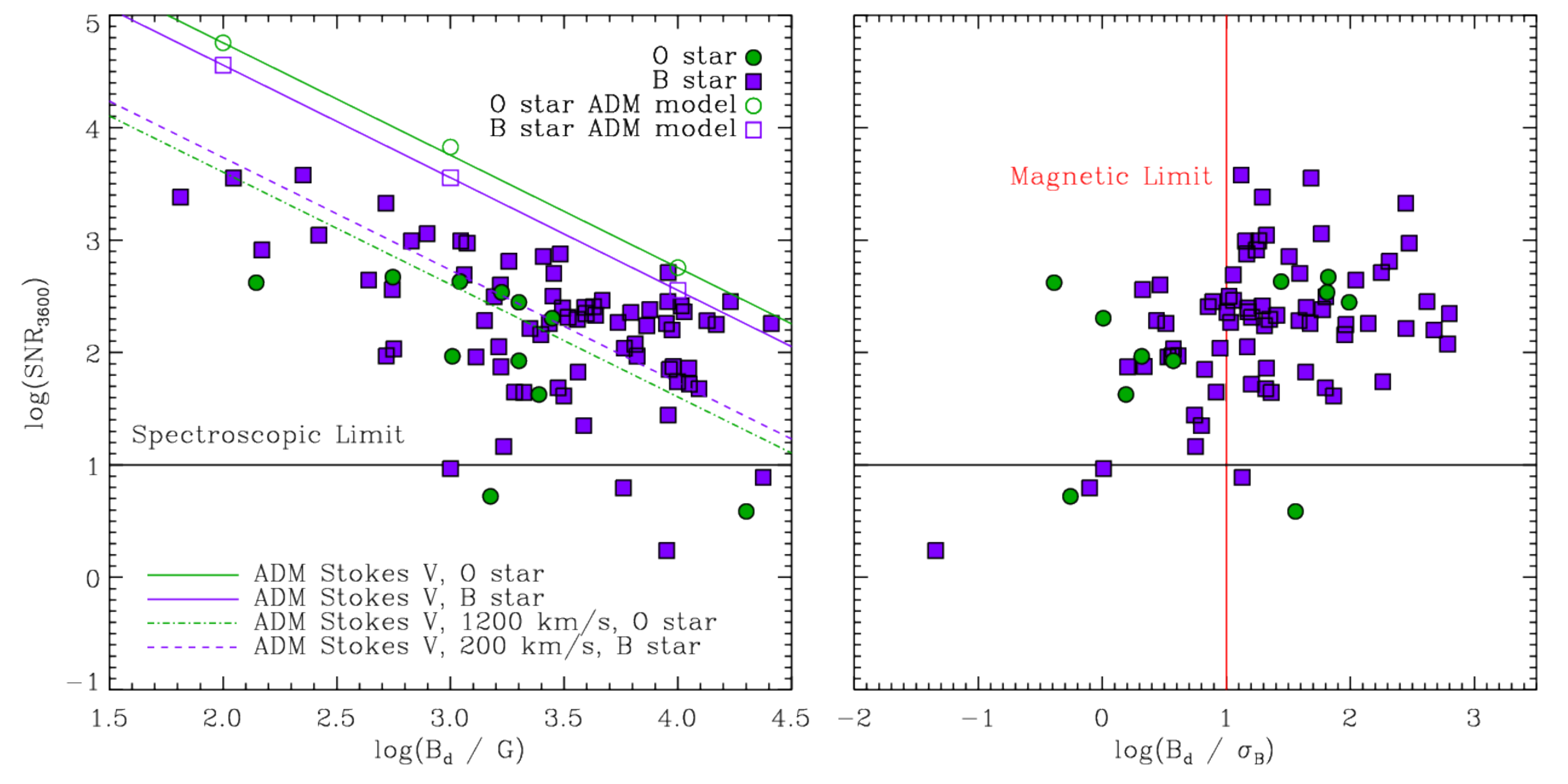}
      \caption{{\em Left}: $S/N$ for a 3600 s Channel 1 spectropolarimetric sequence as a function of $B_{\rm d}$. Diagonal solid lines indicate the $S/N$ required to detect a circumstellar magnetic field, as determined from ADM models for O and B-type stars. Diagonal dashed and dot-dashed lines indicate the $S/N$ that can be achieved using bins of 200 \kms~ and 1200 \kms, resulting in 5 wavelength bins across a resonance line for a B and an O-star, respectively. The solid horizontal line indicates the lower bound for useful spectroscopy. {\em Right}: $S/N$ for a 3600 s Channel 1 spectropolarimetric sequence as a function of the ratio of $B_{\rm d}$ to the \bz~error bar, where the error bar is inferred from UV LSD models and \vsini. In order for the magnetic field to be measureable, the error bar must be about 10\% of $B_{\rm d}$.}
         \label{bd_snr_circumstellar_detection}
   \end{figure}

The ability of these measurements to detect circumstellar magnetic fields, as inferred from ADM Stokes $V$ models (Sect. \ref{sec:circumstellar_magnetometry}), is demonstrated in Fig.\ \ref{bd_snr_circumstellar_detection}. Since B-type stars are expected to yield a larger Stokes $V$ amplitude in resonance lines for a given field strength than O-type stars, their circumstellar magnetic fields can be detected at a slightly lower $S/N$. Three of the most strongly magnetic B-type stars can be detected without wavelength binning. For the remainder of the sample, some degree of binning is necessary. The amount by which a given line can be binned is proportional to its line width. Taking the C~{\sc iv} doublets in Fig.\ \ref{polstar_uv_halpha}, B stars are expected to span about 1000~\kms, while O stars should span about 6000 \kms. In order to have a minimum of 5 measurements across the line, a B-type star can therefore adopt a maximum bin size of 200 \kms, while an O-star can be binned to a maximum of 1200 \kms. Such a strategy can detect circumstellar magnetic fields in 4 O-type stars and 32 B-type stars, or almost half the full sample. If each line in a resonance doublet can be co-added in order increase the $S/N$ still further, in a process similar to LSD, the number of detectable B-stars increases to 44. Note that many of the B stars are detectable without applying this maximal degree of wavelength binning. 

The right panel of Fig.\ \ref{bd_snr_circumstellar_detection} demonstrates the quality of the photospheric magnetometry that can be expected, showing the 1500 \AA~$S/N$ as a function of the ratio of $B_{\rm d}$ to the the error bar $\sigma_B$ in \bz. Since the maximum value of \bz~is approximately $B_{\rm d}/3.5$, a ratio of at least $\sigma_B / B_{\rm d} = 0.1$ is required for the field to be securely detected and modelled. In this case, $\sigma_B$ was determined using the \vsini~and $S/N$-dependent relationships given by \cite{2016MNRAS.456....2W}, with an LSD $S/N$ gain inferred from the UV modelling in Sect.\ \ref{sec:lsd_magnetometry}. This figure demonstrates that the observations will be able to achieve the more challenging goal of surface magnetometry, as compared to the less challenging goal of spectroscopy, for 56 of the sample stars.

   \begin{figure}[t]
   \centering
   \includegraphics[width=0.6\textwidth]{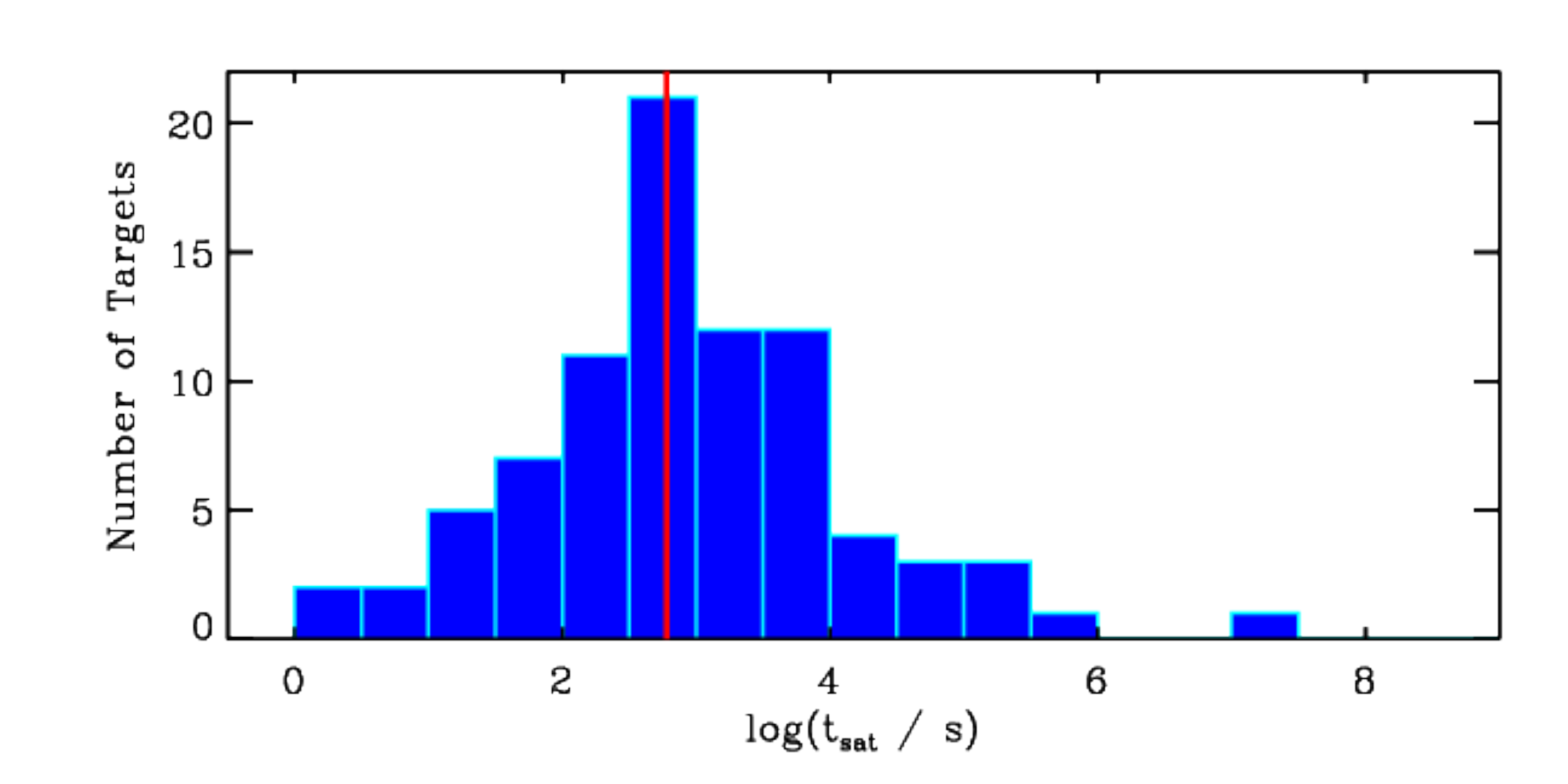}
      \caption{Saturation time for Channel 2 observations. The red line shows the maximum observation 600 s length.}
         \label{tsaturation}
   \end{figure}

Channel 2 has a much larger effective area and a much lower spectral resolution than Channel 1, and therefore more easily reaches the saturation $S/N$ of 2872. Fig.\ \ref{tsaturation} shows the Channel 2 saturation times estimated using 2500~\AA~fluxes. The median saturation time is 775 s. Two stars reach saturation time in less than the the minimum sub-exposure time of 2 s, and therefore cannot be observed in Channel 2. There are 38 observable stars which reach saturation time in less than 600 s. If the Channel 2 dataset is limited to these stars, and 30 observations are obtained for each target in order to obtain the necessary dense coverage of the $QU$ plane, completing this component of the observing program will require 76 hours. 

As discussed in Sect. \ref{sec:continuum_linpol}, the expected level of continuum polarization, as inferred from observations using visible data, ranges from on the order of 0.01\% to 0.1\%. Taking the lower bound, this implies that a $S/N$ of at least 10,000 is necessary to obtain a precision sufficient to obtain a 5$\sigma$ measurement of the weakest expected signals. This can be easily achieved by wavelength binning: while Channel 2 has a low spectral resolution, the required high $S/N$ is easily achievable by binning around 10 wavelength elements. 

   \begin{figure}[t]
   \centering
   \includegraphics[width=0.6\textwidth]{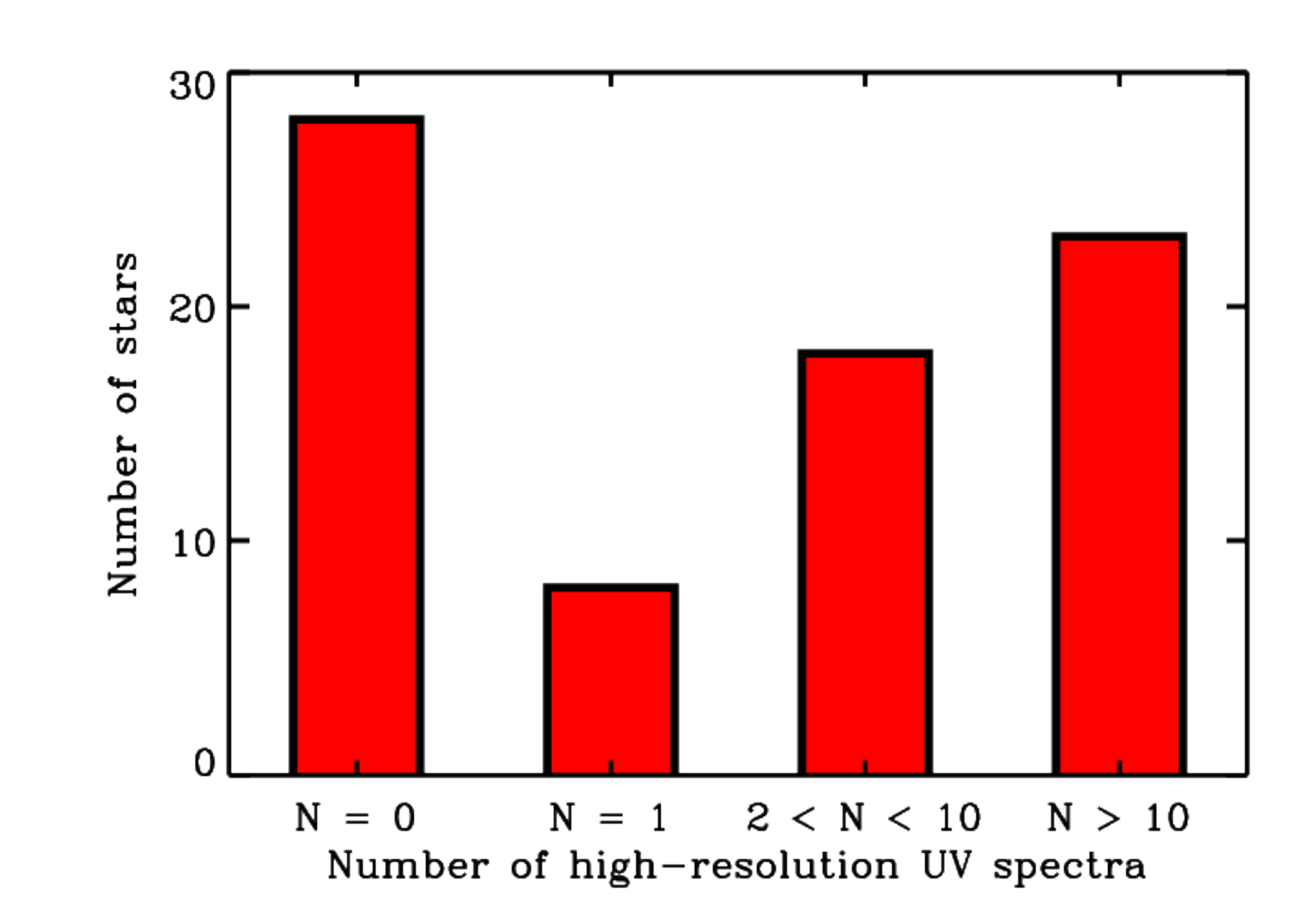}
      \caption{Histogram summary of the available high-resolution UV spectroscopy. The sample is about evenly divided into stars with no observations, stars with a large number of observations, and stars with only a few observations.}
         \label{nobs_hist}
   \end{figure}

Fig.\ \ref{nobs_hist} shows a histogram of the number of high-resolution UV observations available for the target stars. There are no data available for 28 stars. For 8 stars, only a single snapshot is available, therefore variability cannot be assessed. Between 2 and 10 observations are available for 18 stars; in these cases, some limited evaluation of variability is possible, but coverage of the rotational period is poor. For 23 stars more than 10 observations are available; in these cases detailed studies have generally been possible \citep[e.g.][]{2003A&A...406.1019N,2003A&A...411..565N,2013A&A...555A..46H}. A key program goal is to obtain sufficient observations to characterize the ultraviolet variability across the full rotational period of the entire sample. A particular point of interest is the red-shifted absorption dip identified by ADM models \citep[Sect.\ \ref{subsec:analytic_models}][]{erb21}, which is diagnostic of the infalling plasma. Detecting this feature will require both a high spectral resolution (as it has an expected velocity width of about 100 \kms), and high $S/N$. 

   \begin{figure*}[t]
   \centering
   \includegraphics[width=0.95\textwidth, trim=100 0 100 0]{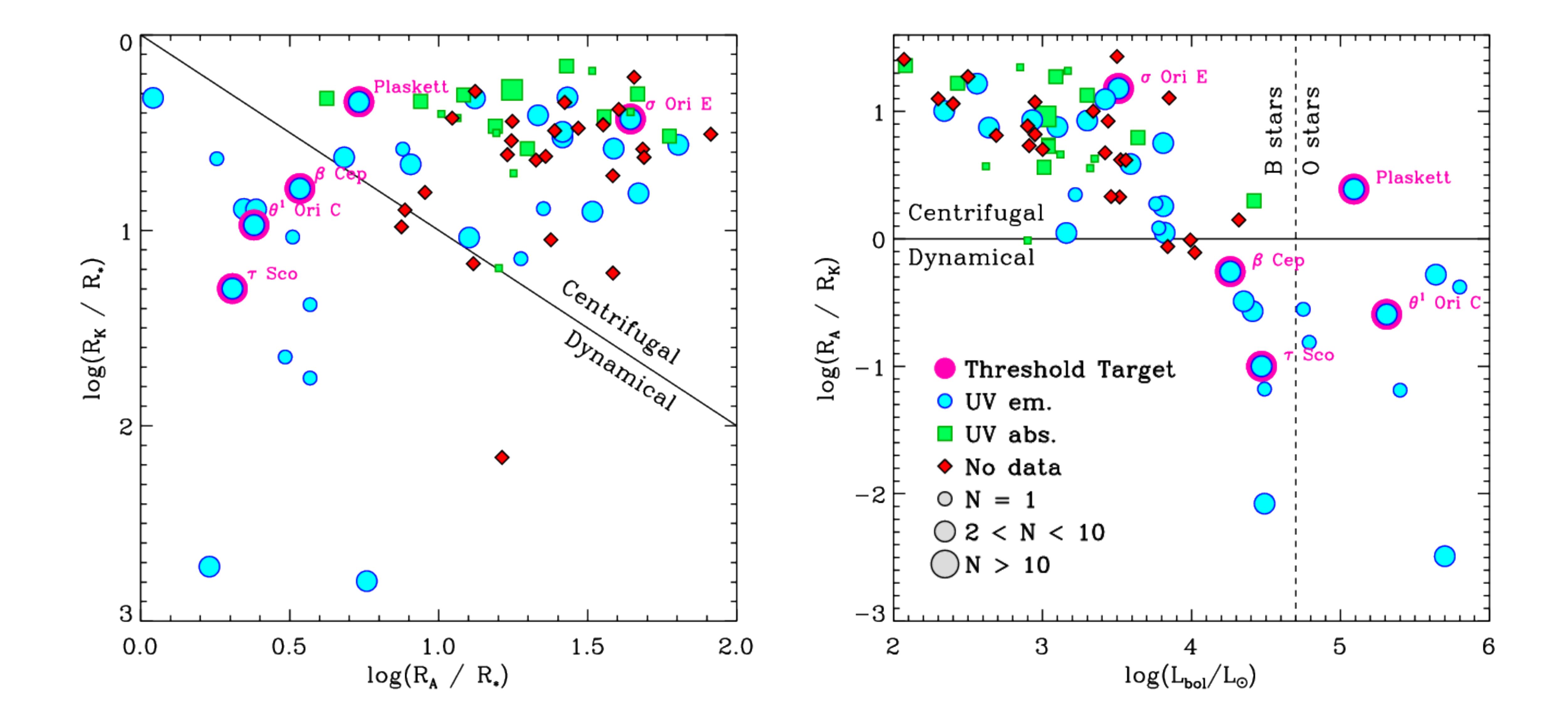}
      \caption{{\em Left}: survey sample on the rotation-magnetic confinement diagram. As indicated by the slanted line, stars with Alfv\'en radii $R_{\rm A}$ greater than Kepler corotation radii $R_{\rm K}$ have centrifugal magnetospheres, while those with $R_{\rm A} < R_{\rm K}$ have dynamical magnetospheres. Symbol size is proportional to the number of available high-resolution UV spectra. Colour indicates whether magnetospheric emission is detected, not detected, or if no data are available. Threshold targets are highlighted in magenta and labelled. {\em Right}: as left, in the bolometric luminosity-$\log{R_{\rm A}}/R_{\rm K}$ plane.}
         \label{polstar_ipod}
   \end{figure*}

The left panel of Fig. \ref{polstar_ipod} shows the full sample on the rotation-magnetic confinement diagram, the `fundamental plane' of stellar magnetospheres \citep{petit2013}. This diagram shows the Kepler corotation radius $R_{\rm K}$ as a function of the Alfv\'en radius $R_{\rm A}$. As a star rotates more rapidly, $R_{\rm K}$ withdraws towards the stellar surface, becoming identical with the equatorial radius of the star at critical rotation. The Alfv\'en radius is a measure of the extent of magnetic confinement in the magnetic equatorial plane, and increases with increasing surface magnetic field and declining wind strength. As a star rotates more rapidly, the Kepler radius moves closer to the stellar surface. The right panel of Fig. \ref{polstar_ipod} shows the sample stars on the $\log{R_{\rm A}/R_{\rm K}}-\log{L_{\rm bol}}$ diagram. The ratio $\log{R_{\rm A}/R_{\rm K}}$ serves as a dimensionless proxy to the size of the CM. As can be seen in these diagrams, the majority of the O-type stars have dynamical magnetospheres, as they have powerful winds leading to small $R_{\rm A}$ and rapid spindown timescales, meaning slow rotation and therefore large $R_{\rm K}$. Conversely, the majority of B-type stars have centrifugal magnetospheres.

As indicated in Fig.\ \ref{polstar_ipod}, while at least 1 UV observation is available for all magnetic O-type stars, a considerable fraction of magnetic B-type stars have not been observed in the UV. Variable UV emission are seen at essentially all points in the two diagrams, raising the obvious question of why some stars that have been observed do not show obvious magnetospheric signatures (coded in the diagram as UV absorption). As suggested by the symbol size (proportional to the number of spectra), this may simply be the result of a small number of observations, which make it difficult or impossible to evaluate variability in the wind-sensitive doublets. 

\subsection{Threshold Targets}

A subset of the sample were selected as {\em threshold targets}, i.e.\ high-priority targets. The first criterion regarding these targets is that they be bright enough for Polstar to obtain a high signal-to-noise ratio ($S/N$). Beyond this, it is important to sample parameter space whilst also observing the most interesting targets. As can be seen in Fig.\ \ref{polstar_ipod}, the critical targets were chosen in such a fashion as to cover the various key parts of parameter space: B stars with large CMs, B stars with DMs, O stars with CMs, and O stars with DMs. The threshold targets are:

\noindent {\bf $\theta^1$ Ori C}: HD\,37022 is the most massive O-type star in the Orion Nebula Cluster and was the first O-type star in which a magnetic field was discovered \citep{2002MNRAS.333...55D}. It exhibits phase-locked ultraviolet variability consistent with a magnetospheric origin \citep{1996A&A...312..539S}, although the observed variability is contrary to expectations, being anti-correlated in phase \citep{2008cihw.conf..125U}. This points to an important discrepancy between models and observations, which may be resolved by the structural and magnetic information obtained via circumstellar polarimetry.

\noindent {\bf $\sigma$ Ori E}: HD\,37479 was the first star in which a magnetosphere was detected \citep{1978ApJ...224L...5L} and is the prototype of the $\sigma$ Ori E variable class (i.e.\ stars with H$\alpha$ emission and photometric eclipses from a CM). It has by far the strongest magnetospheric emission of any CM star \citep{2020MNRAS.499.5379S}. It has a large IUE dataset \citep[e.g.][]{2001A&A...372..208S}, and as the benchmark CM star has also been extensively studied at radio and X-ray wavelengths \citep[e.g.][]{2000A&A...363..585R,leto2012}. Furthermore, its surface magnetic field has been mapped via Zeeman Doppler Imaging, and a magnetospheric model extrapolated from this map used to reproduce its H$\alpha$ emission and photometric eclipses \citep{2015MNRAS.451.2015O}. It is also the only CM star with published broadband polarimetry \citep{2013ApJ...766L...9C}. Adding Polstar spectropolarimetry to the large multiwavelength datasets and extensive modeling already performed for this star will be a key step in calibrating the next generation of magnetospheric models that will be enabled by Polstar.

\noindent {\bf Plaskett's Star}: HD\,47129 is a magnetic colliding wind binary consisting of two O-type stars, one of which is the only O-type star with a CM \citep{2013MNRAS.428.1686G,2021arXiv211106251G}. The star's rapid rotation is believed to be a result of recent binary interactions. Polstar data will enable evaluation of the effects of rapid rotation on the circumstellar environment in an O-type star's magnetosphere, together with examination of the effects of a strong magnetic field on the colliding wind shock. 

\noindent {\bf $\tau$ Sco}: HD\,149438 is the hottest B-type star with a detected magnetic field, and in sharp contrast to the usual dipolar morphology has an extraordinarily complex surface field structure \citep{2006MNRAS.370..629D,2016A&A...586A..30K}. Its distinctive ultraviolet resonance line profiles were the crucial clue leading to the detection of its magnetic field \citep{2006MNRAS.370..629D}; similar ultraviolet signatures have successfully enabled the identification of other magnetic stars in the same mass range \citep{2011MNRAS.412L..45P}. It has been suggested to be a blue straggler and a possible binary merger product, with its complex surface field being the remnant of a merger-powered dynamo \citep{2019Natur.574..211S}, although its properties may also be consistent with single-star evolutionary models incorporating magnetic fields \citep{2021MNRAS.504.2474K}. As the surface magnetic field of this bright star is relatively weak, it is an excellent target for utilization of the Hanle effect. 

\noindent {\bf $\beta$ Cep}: HD\,205021 is a magnetic $\beta$ Cep pulsator. It features by far the largest IUE time series of any magnetic star, and exhibits clear and strong rotational modulation in all wind-sensitive resonance lines; indeed, it was precisely this modulation that led to the detection of its magnetic field \citep{2013A&A...555A..46H}, with comparable ultraviolet signatures in other $\beta$ Cep stars leading to the detection of further magnetic stars in this class \citep{2008A&A...483..857S,2017MNRAS.471.2286S}. 

\section{Enabled Science}\label{sec:enabled}

   \begin{figure*}[t]
   \centering
   \includegraphics[width=0.95\textwidth]{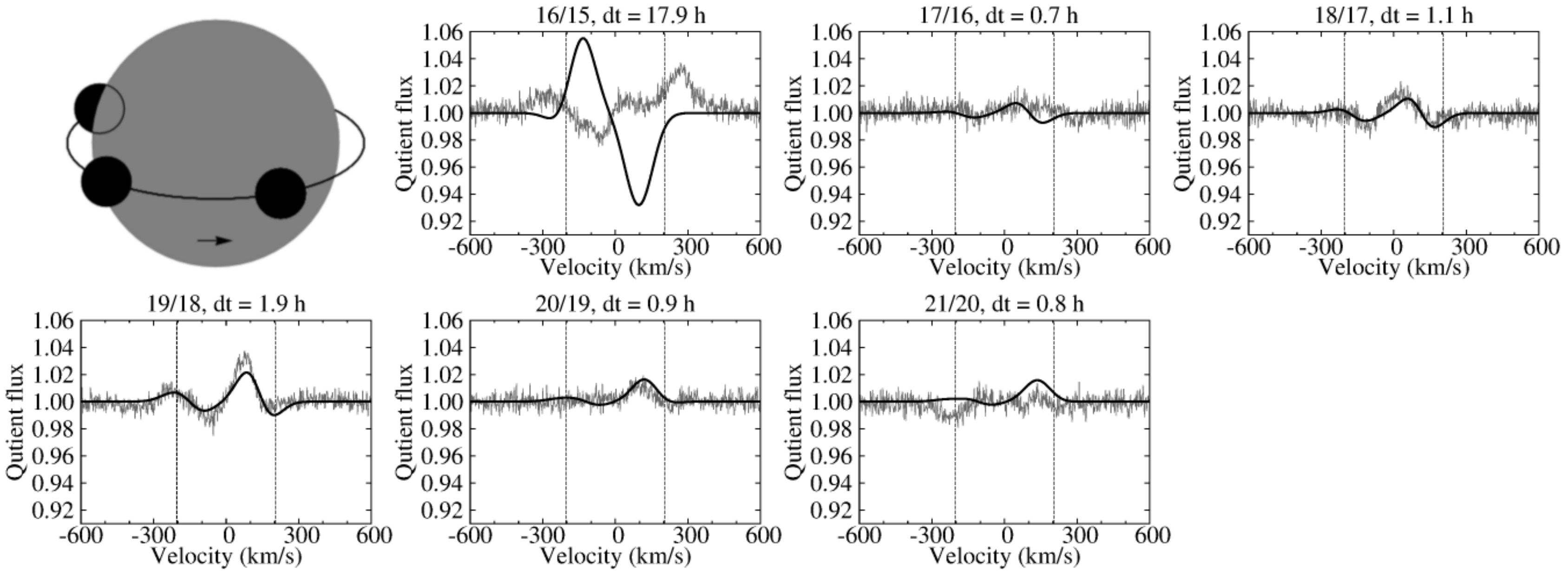}
      \caption{Model fits (black thick lines) for subsequent quotient spectra of He II $\lambda$4686 line for the star $\lambda$ Cep (gray thin lines) of the 2013 dataset, spectra 15--21. The top label gives the sequence numbers of the two spectra of the quotient, followed by their time interval in hours. The geometry which is depicted at the beginning of the series, is used for all figures in the sequence and shown for the epoch of the first spectrum in the series. The star rotates but the blobs keep their same relative position. The first and last figure of the series contains intentionally failed fits, to signify the extend over which the fitted configuration, carried around by the rotation, survives. From \cite{sudnik2016}.}
   \label{stel_prom}
   \end{figure*}

 \begin{figure}[t]
  \begin{minipage}{0.48\textwidth}
  \centering
  \includegraphics[width=1\textwidth]{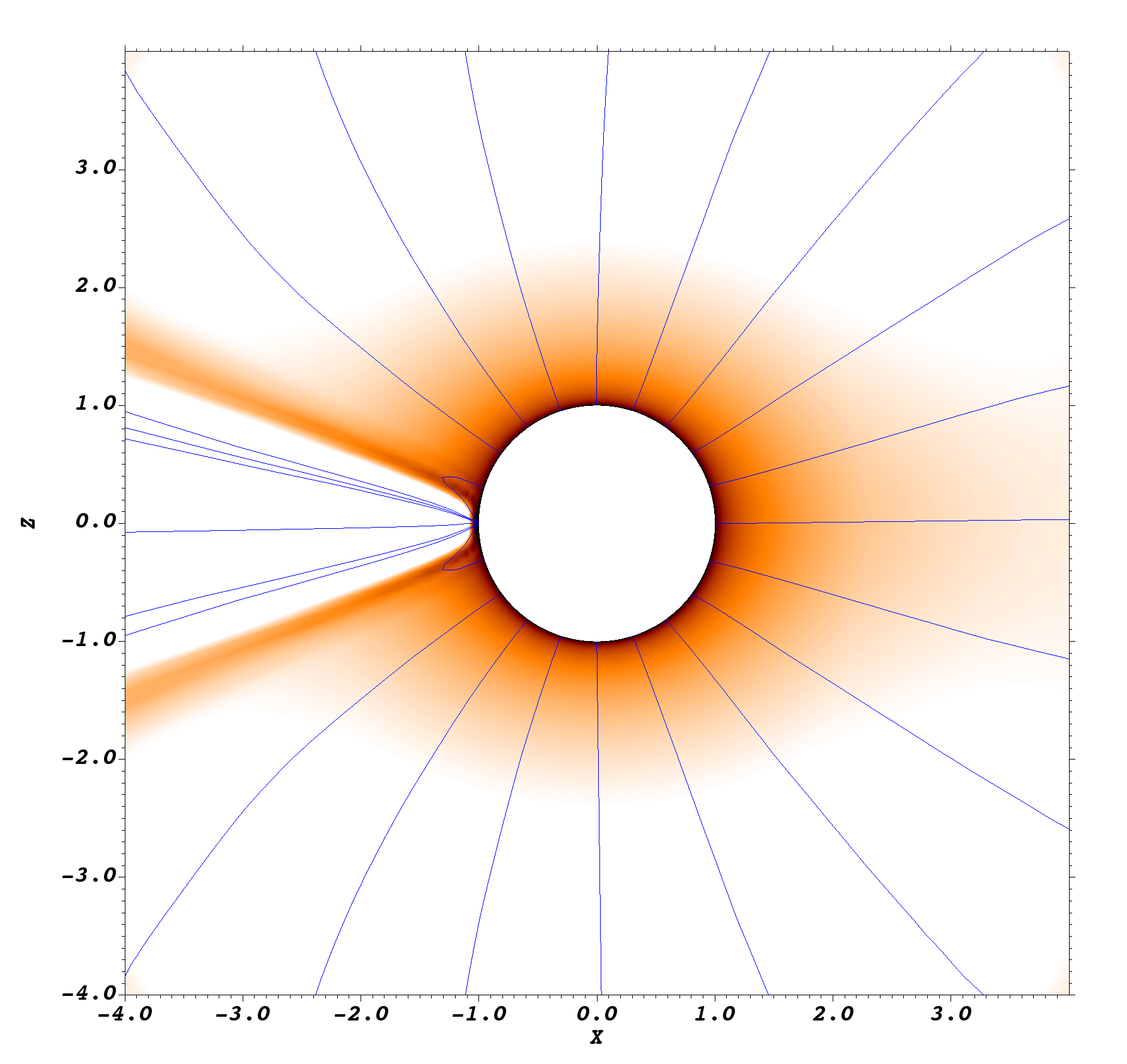}
  \end{minipage}\hfill
  \begin{minipage}{0.48\textwidth}
  \centering
  \includegraphics[width=1\textwidth]{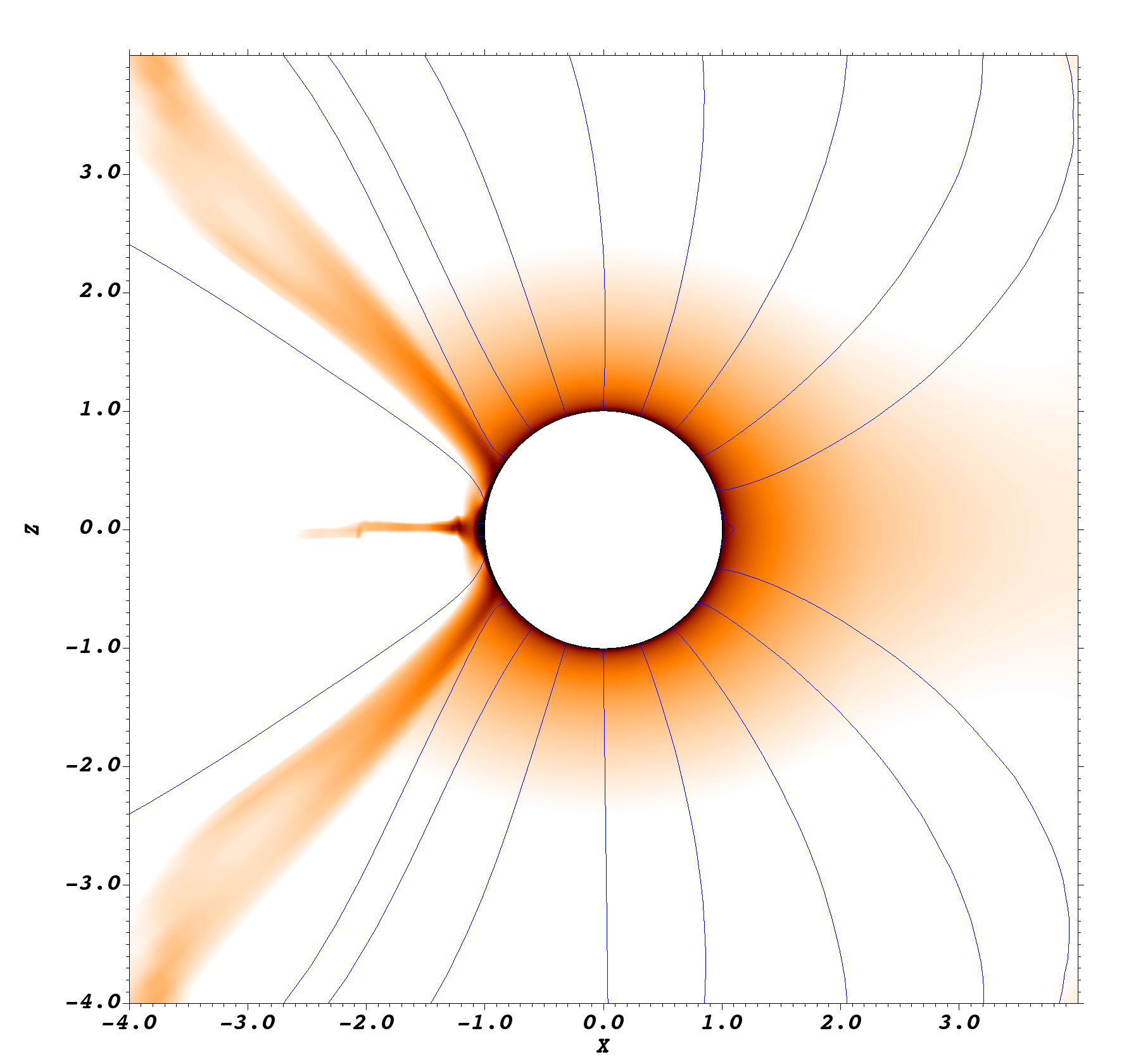}
  \end{minipage}
  \caption{Logarithm of density of ZX-plane of 3D MHD simulations of a sample O-star wind with a small scale off-center magnetic dipole field to mimic small scale strong magnetic field that can perturb the wind at the base. The angle between rotation ($W=0.5$ half critical) and magnetic axes $\beta=0^{\circ}$ ({\it left panel}) and $\beta=90^{\circ}$ ({\it right panel}). The color code represents the logarithm of density in a range $10^{-15}$ to $10^{-11}$\,g/cm$^3$. Blue lines represent time-evolved magnetic field lines which are strong only over limited area on the stellar surface and fall of steeply with distance away from the star.}
  \label{tiny_dipole}
\end{figure}






Massive OB stars show variability both spectroscopically and photometrically in many wavelength bands including the UV, optical and X-ray. Regular variability could be related to stellar pulsations, rotation modulation, and/or the presence of globally organized magnetic fields, while widespread stochastic variability on hourly to daily timescales is poorly understood. Stochastic variability could be caused by the presence of bright spots (possibly of magnetic origin), clumps, internal gravity waves \citep{bowman2019, bowman2020}, or subsurface convection \citep{cantiello2021}. 
A well-known example of irregular variability in UV wind-line profiles are discrete absorption components (DACs). The formation of these wind structures are ascribed to corotating interacting regions (CIRs, \citealt{mullan1984, mullan1986}), which arise above bright spots at the stellar surface \citep{cranmer1996}. The formation of bright surface spots can be caused by magnetic fields produced by dynamo action in the subsurface convection layer and brought up to the surface by buoyancy \citep{cantiello2011}. The lifetime of these fields is determined by the relatively short turnover time of the subsurface layer. The expected strength of these fields imply the local magnetic confinement parameter to be around unity \citep{udDoula2002} i.e.\ the fields are able affect the wind dynamically. 

The small-scale magnetic fields, that appear at the surface as bright spots, can be a driver of many surface related phenomena observed in OB stars. Short-lived bright spots give rise to low-amplitude ($\simeq 10$ mmag) photometric variability, as observed by space based high-precision photometry (MOST, BRITE, TESS) in the light curves of many OB stars. \cite{ramiaramanantsoa2014} simulated the MOST light curve of the O7.5 III(n)((f)) star $\xi$ Per with several corotating bright spots, presumably of magnetic origin, with random starting times and lifetimes up to several rotations varying from one spot to another. Analysing simultaneous photometry from the BRITE constellation and ground-based spectroscopy for the O4I(n)fp star $\zeta$ Pup, \cite{ramiaramanantsoa2018} established an empirical link between wind structures and photospheric activity, implying a photospheric origin. Two types of variability were found: one single periodic non-sinusoidal component with 1.78~d period, superimposed on a stochastic component. The periodic component is consistent with rotational modulation arising from evolving bright spots that were mapped at the surface. The signal was also found in the He II 468.6~nm wind emission line, showing signatures of corotating interaction regions that turn out to be driven by the bright photospheric spots observed by BRITE. The spots are suggested to originate from small-scale magnetic fields generated through dynamo action within a subsurface convection zone \citep[e.g.][]{cantiello2011,2020ApJ...900..113J}. The stochastic component observed in He II $\lambda$4686 line correlated with the amplitudes of stochastic light variations. Stochastic He~{\sc ii} variability was attributed to wind clumps, while stochastic photometric variability was proposed to be the photospheric drivers of the clumps.

\cite{daviduraz2017} investigated whether bright spots compatible with recent observation could be responsible for DACs observed in the UV line profiles of the O7.5 III(n)((f)) star $\xi$ Per. The authors successfully reproduced the behaviour of DACs with 4 equally spaced equatorial spots with sizes between $10^{\circ}$ and $20^{\circ}$. In this scenario spots appearing and disappearing randomly over time, with varying strengths and sizes, might explain the cyclical nature of DACs. 
Semi-analytic analysis for the spot size and amplitude needed to produce an overloaded wind that form the DAC was done by \cite{owocki2018}.

Based on spectroscopic analysis of He II $\lambda$4686 line for the 6I(n)fp star $\lambda$ Cep it was proposed that cyclical variations are caused by the presence of multiple, transient, short-lived, corotating magnetic loops, so-called stellar prominences, likely associated with bright surface spots \citep{sudnik2016}. The prominences are represented as corotating spherical blobs of emitting gas attached to the surface of the star. Depending on the location of the blob it contributes to the line profile as an emission or absorption feature. 
An example of the model fit of subsequent quotient spectra i.e., the following spectrum divided by the previous one, is shown in Fig.~\ref{stel_prom}. The proposed model applied to the He II 468.6~nm line can be fitted with 2\,to\,5 equatorial blobs with lifetimes between $\sim$1 and 24 h. The action of the subsurface convection zone would be the most likely driving mechanism that generates short-lived magnetic bright spots as the source of prominences.

Local small scale magnetic fields tied to stellar prominences can be viewed as an off-center small (tiny) dipole with its center located close to the stellar surface. Such small scale fields can cause variability at the stellar surface that can propagate outwards affecting what is observed in these winds. As a proof-of-concept, we have carried out some preliminary 3D MHD simulations of hot star winds wherein we assumed that the location of the strong local field coincides with bright spots using the approach of \cite{owocki2018}.  Fig.~\ref{tiny_dipole} shows the logarithmic density of two such sample models for a typical hot star viewed along  ZX-plane. The left panel shows a case where the small scale dipole field is aligned with rotation axis, and the right panel has an inclination angle of 90$^{\circ}$ with respect to the rotational axis. We assumed rotation of 0.5 critical, and a field strength of order kG at the stellar surface that falls off steeply with distance from the photosphere. As seen in the Fig.~\ref{tiny_dipole}, despite being localized, this field can substantially influence the wind dynamics globally.

While such small-scale fields cannot be detected with existing ground-based facilities, ultraviolet spectropolarimetry may provide an avenue for direct detection, for instance via the Hanle effect (if they are relatively weak), or even via the Zeeman effect (if the spots are large, the field strong, with an advantage being gained by the greater brightness of the spots in the UV). Linear spectropolarimetry will further provide important geometrical data that may help distinguish between CIR models with and without magnetic fields.

\section{Synergies with other science objectives}\label{sec:other_objectives}

\noindent {\bf Hot star winds:} \cite{2021arXiv211111633G} describe the utility of Polstar for calibrating the mass-loss rates of hot stars via their radiative winds in light of wind structures such as clumping and the corotating interaction regions (CIRs) thought to underlie discrete absorption components (DACs). The calibrated mass-loss rates obtained via this project for single stars stars without large-scale magnetic fields will inform the surface mass-flux for magnetospheric models described here. This will enable it to be determined if strong surface magnetic fields directly modify the surface mass-flux via radiative winds. In addition, CIRs are thought to be driven by bright spots, which are believed to be associated with small-scale magnetic fields. Magnetospheric models developed from the study of magnetic hot stars may prove to be an important component of modelling CIRs. Weak magnetic fields in the launch regions of CIRs may also be detectable via the Hanle effect.

\noindent {\bf The origin of rapidly rotating B-type stars:} \cite{2021arXiv211107926J} will use Polstar to search for hot subdwarf companions around apparently single classical Be stars and Bn stars, for which binary interactions are a leading origin scenario for the spin-up of Be stars to near-critical rotation. Binary mergers are a leading scenario for the origin of fossil magnetic fields, presenting an intriguing dichotomy in that no magnetic field has ever been detected in a Be star, and indeed magnetic fields should destroy their Keplerian disks. Techniques developed from binary searches for subdwarf companions around Be/n stars can be applied to magnetic hot stars in a similar fashion, enabling sensitive comparison of their respective binary fractions; at this point, it is already known that most, albeit not all, magnetic stars are apparently single, whereas Be stars when in binary systems are typically paired with post-main sequence companions). \citeauthor{2021arXiv211107926J} also describe the use of limb polarization to determine the critical rotation fraction of rapid rotators. While most magnetic stars are relatively slowly rotating, there are a handful of relatively rapidly rotating objects for which similar techniques will provide important constraints. Since the surface mass flux is sensitive to the effective temperature, gravity darkening due to rapid rotation may need to be incorporated in magnetospheric models of rapidly rotating objects. Finally, while magnetic fields are neither detected nor expected in Be/n stars, the Hanle effect may enable detection of weak magnetic fields in the near-star environment if present, and sensitive upper limits if not. 

\noindent {\bf Mass transfer and loss in B-type interacting binaries:} \cite{2021arXiv211114047P} describe Polstar's application to interacting binary systems, which can be used to constrain the geometry, mass transfer, and mass loss rates of interaction regions and circumbinary disks in interacting binaries. While magnetic binaries are rare, there are a handful of systems (e.g. $\epsilon$ Lupi, HD\,149277) which are close enough for wind interactions to play a role, and which therefore hold the promise of determining whether and to what degree surface magnetic fields modify these interactions.

\noindent {\bf Massive star binary colliding winds:} \cite{2021arXiv211111552S} describe the utility of Polstar to constrain the geometry of colliding wind binaries. Weak magnetic fields are required to reproduce the gyrosynchrotron radiation detected from colliding wind binaries; while these magnetic fields are too weak to be detected via the Zeeman effect, they may be detectable via the Hanle effect. Furthermore, the rapidly rotating magnetic O-type system Plaskett's Star is a colliding wind binary; a complete understanding of the circumstellar geometry of this system is almost certain to require insights obtained from the study of magnetic stars and non-magnetic colliding wind binaries.

\noindent {\bf Interstellar medium science:} \cite{2021arXiv211108079A} describe several experiments that will investigate the dust polarization properties of the interstellar medium (ISM). Constraints from the ISM project will be a critical element for interpretation of the linear polarization signatures obtained from magnetic stars. Conversely, detailed understanding of the variable intrinsic polarization of magnetic stars will enable these targets to be added to the ISM project. 

\noindent {\bf Protoplanetary disks:} \cite{2021arXiv211106891W} describe the use of Polstar to probe the circumstellar geometry of the protoplanetary disks surrounding Herbig Ae/Be stars in order to determine the nature of the accretion mechanism. Accretion is believed to be magnetospheric in low-mass Herbig stars, but the mechanism unknown for stars more massive than 4~\msun. Comparing accretion signatures of Herbig stars with and without magnetic fields may furthermore provide important insights into the origin of fossil fields, and the reason for the magnetic dichotomy between main sequence stars with and without magnetic fields. 

\section{Summary}\label{sec:summary}

In this white paper we have described how the unique capabilities offered by Polstar will lead to fundamental advances in our understanding of the magnetospheres of hot stars. While the focus of this white paper has been on the capabilities of the Polstar mission, this work is of relevance to any ultraviolet spectropolarimetric mission, such as e.g.\ Arago. 

The high-resolution ultraviolet spectra obtained by Polstar will enable much more precise spectroscopic evaluation of stellar magnetospheres, as compared to the lower-resolution, lower-$S/N$ data available for most stars via the Interstellar Ultraviolet Explorer. Importantly, over half of the none magnetic stars have not a single UV observation; of those that do, less than a third have time-series data adequate for evaluation of the projected magnetospheric geometry and column density across a rotational cycle. 

While surface magnetic field measurements obtained via ground-based visible spectropolarimetry are already available for all stars in the sample, the large number of spectral lines available for multi-line analysis in high-resolution ultraviolet spectra more than compensates for the weaker Zeeman effect at shorter wavelengths, in principle enabling higher-precision magnetic meeasurements to be obtained in the UV as compared to the visible. The full-Stokes capability of Polstar, and expected advantages in the UV over the visible in the amplitude of Stokes $QU$ signals associated with the transverse Zeeman effect, mean that many of the datasets will be optimal for magnetic mapping via full-Stokes Zeeman Doppler Imaging (ZDI). Importantly, the availability of all four Stokes parameters for magnetic inversion breaks degeneracies that can affect maps obtained only in Stokes $IV$.

Polstar will enable measurement of circumstellar magnetic fields, with the projected capabilities of the instrument capable of detecting magnetic signatures originating in the circumstellar environment in a large fraction of known magnetic stars. Strong fields should be detectable via the Zeeman effect (as evaluated using state of the art magnetospheric models), while weak magnetic fields should be detectable via the Hanle effect. 

Both high- and low- resolution linear spectropolarimetry will provide crucial and sensitive constraints on the magnetospheric geometry, enabling degeneracies between rotational axis inclinations and magnetic axis tilt angles to be broken. Importantly, the information available via linear polarization provides geometrical data that cannot be obtained via spectroscopy or photometry alone, as already revealed by the insufficiency of current magnetospheric models to simultaneously reproduce the light curve and polarimetric variation of the key target $\sigma$ Ori E. 

By combining the rich spectroscopic and polarimetric datasets available with Polstar observations, detailed 3D models of of the circumstellar environments of a large number of magnetic hot stars can be compared against constraints on the circumstellar magnetic field, column density, velocity structure, and geometry. This will enable measurement of the escaping and magnetically trapped wind fraction of these stars across a full range of stellar, evolutionary, magnetic, and rotational parameters, thereby providing a crucial test of the expectation that magnetic fields rapidly drain angular momentum and drastically reduce the net mass-loss rates of massive stars. This will provide empirical calibration for evolutionary models incorporating rotation and magnetic fields, which will in turn provide important information for the stellar population synthesis models used to infer the mass and energy budget for the interstellar medium, expectations for the properties of post-main sequence supergiants and supernovae, and the population statistics of stellar remnants. 

\section*{Acknowledgements}

M.E.S. acknowledges financial support from the Annie Jump Cannon Fellowship, supported by the University of Delaware and endowed by the Mount Cuba Astronomical Observatory. A.D.-U. is supported by NASA under award number 80GSFC21M0002. C.E. gratefully acknowledges support for this work provided by NASA through grant number HST-AR-15794.001-A from the Space Telescope Science Institute, which is operated by AURA, Inc., under NASA contract NAS 5-26555. R.I. and C.E. also gratefully acknowledge that this material is based upon work supported by the National Science Foundation under Grant No. AST-2009412. M.C.M.C. acknowledges internal research support from Lockheed Martin Advanced Technology Center. This material is based upon work supported by the National Center for Atmospheric Research, which is a major facility sponsored by the National Science Foundation under Cooperative Agreement No. 1852977. T.dP.A. acknowledges the funding received from the European Research Council (ERC) under the European Union's Horizon 2020 research and innovation programme (ERC Advanced Grant agreement No 742265). Y.N. acknowledges support from the Fonds National de la Recherche Scientifique (Belgium), the European Space Agency (ESA) and the Belgian Federal Science Policy Office (BELSPO) in the framework of the PRODEX Programme (contracts linked to XMM-Newton and Gaia). P.A.S. acknowledges his financial support by the NASA Goddard Space Flight Center to formulate the mission proposal for Polstar.  AuD acknowledges support by NASA through Chandra Award number TM1-22001B issued by the Chandra X-ray Observatory 27 Center, which is oper- ated by the Smithsonian Astrophysical Observatory for and on behalf of NASA under contract NAS8-03060. NS acknowledges support provided by NAWA through grant number PPN/SZN/2020/1/00016/U/DRAFT/00001/U/00001.

\pagebreak

\newpage

\newpage

\bibliography{bib_dat}{}

\end{document}